\documentclass{aa}  
\usepackage{graphicx}
\usepackage{epstopdf}
\usepackage{txfonts}
\usepackage{color}
\def\a{$\alpha$}

\def\caii{{Ca{\sc ii}}}
\def\chisq{$\chi^{2}$}

\def\civ{{\sc{Civ}}$\lambda$1549\/}

\def\cm3{cm$^{-3}$\/}
\def\D{{$\Delta$}}

\def\ergss{ergs s$^{-1}$\/}
\def\fe{{\sc{Fe}}\/}
\def\fe6087{{\sc [Fe vii]}$\lambda$6087\/}

\def\feiiopt{{Fe \sc{ii}}$_{\rm opt}$\/}
\def\feii{{Fe\sc{ii}}\/}

\def\feiia{\rm Fe{\sc ii}$\lambda$5270\/}

\def\gt{$\textgreater$}
\def\gtsima{$\; \buildrel > \over \sim \;$}
\def\gtsim{\lower.5ex\hbox{\gtsima}} 
\def\Gsoft{$\Gamma_{\rm soft}$}

\def\hb{{\sc{H}}$\beta$\/}
\def\hbbc{{\sc{H}}$\beta_{\rm BC}$\/}
\def\hbnc{{\sc{H}}$\beta_{\rm NC}$\/}

\def\heii{He{\sc{ii}}}
\def\heiiopt{{{\sc H}e{\sc ii}}$\lambda$4686\/}

\def\kms{km~s$^{-1}$}
\def\l{$\lambda$}
\def\lt{$\textless$}

\def\lledd{L/L$_{\rm Edd}$}
\def\ltsima{$\; \buildrel < \over \sim \;$}
\def\ltsim{\lower.5ex\hbox{\ltsima}}

\def\mbh{M$_{\rm BH}$\/}

\def\mgi{{Mg\sc{i}}$\lambda$5175\/}

\def\ne{$n_{\rm e}$\/}
\def\nh{$n_{\mathrm{H}}$\/}

\def\o4363{{\sc{[Oiii]}}$\lambda$4363\/}
\def\oii{{\sc [Oii]}$\lambda$3728}
\def\oiii{{\sc [Oiii]}}
\def\oiiiopt{{\sc{[Oiii]}}$\lambda\lambda$4959,5007\/}
\def\oiiia{{\sc [Oiii]}$\lambda$5007}

\def\rfe{$R_{\rm FeII}$}
\def\roiii{$R_{\rm [OIII]}$}

\def\roiii{$R_{\rm [OIII]}$}

\begin{document} 

\title{Highly accreting quasars: The SDSS low-redshift catalog\thanks{Tables 4 and 5 are only available in electronic form at the CDS via anonymous ftp to cdsarc.u-strasbg.fr (130.79.128.5) or via http://cdsweb.u-strasbg.fr/cgi-bin/qcat?J/A+A/}}
\author{C. A. Negrete\inst{1} \and D. Dultzin\inst{2} \and P. Marziani\inst{3} \and D. Esparza\inst{4} \and J. W. Sulentic\inst{5} \and A. del Olmo\inst{5} \and M.~L. Mart\'\i nez-Aldama\inst{5} \and A. Garc\'\i a L\'opez\inst{3,4} \and M. D'Onofrio\inst{6} \and N. Bon\inst{7} \and E. Bon\inst{7}}
\institute{ {CONACyT Research Fellow -- Instituto de Astronom\'{\i}a, UNAM, CDMX 04510, Mexico}
\and {Instituto de Astronom\'{\i}a, UNAM, CDMX 04510, Mexico}
\and{INAF, Osservatorio Astronomico di Padova, IT 35122, Padova, Italy}
\and {Instituto de Radioastronom\'{\i}a y Astrof\'{\i}sica, UNAM, Morelia, Mich. 58089, Mexico}
\and {Instituto de Astrofis\'{\i}ca de Andaluc\'{\i}a, IAA-CSIC, E-18008 Granada, Spain}
\and {Dipartimento di Fisica and Astronomia ``Galileo Galilei'', Univer. Padova, Padova,  Italia}
 \and {Belgrade Astronomical Observatory, Belgrade, Serbia}
}
   \date{}

  \abstract
   {The most highly  accreting quasars are of special interest in studies of the physics of active galactic nuclei (AGNs) and host galaxy evolution. Quasars accreting at high rates ($L/L_{Edd}\sim1$) hold promise for use as `standard candles':  distance indicators detectable at very high redshift.  However, their observational properties are still largely unknown.}  
{We seek to identify a significant number of extreme accretors.  A large sample can clarify the main properties of quasars radiating near $L/L_{Edd}\sim1$ (in this paper they are designated as extreme Population A quasars or simply as extreme accretors) in the \hb\ spectral range for redshift $\lesssim 0.8$.}
   {We use selection criteria derived from four-dimensional Eigenvector 1 (4DE1) studies to identify and analyze spectra for a sample of 334 candidate sources identified  from the SDSS DR7 database. The source spectra were chosen to show a ratio \rfe\ between the FeII emission blend at $\lambda$4570 and \hb, \rfe\ $>$ 1. Composite spectra were analyzed for systematic trends as a function of \feii\ strength, line width, and \oiii\ strength. We introduced  tighter constraints on the signal-to-noise ratio (S/N) and \rfe\ values that allowed us to isolate sources most likely to be extreme accretors.}
   {We provide a database of detailed measurements. Analysis of the data allows us to confirm that \hb\ shows a Lorentzian function with a full width at half maximum (FWHM) of \hb\ $\leq$ 4000 \kms. We find no evidence for a discontinuity  at ~2000 \kms\ in the 4DE1, which could mean that the sources below this FWHM value do not belong to a different AGN class. Systematic \oiii\ blue shifts, as well as a blueshifted component in \hb\ are revealed. We interpret the blueshifts as related to the signature of outflowing gas from the quasar central engine. The FWHM of \hb\ is still affected by the blueshifted emission; however, the effect is non-negligible if the FWHM \hb\ is used as a  ``virial broadening estimator'' (VBE). We emphasize a strong effect of the viewing angle on  \hb\ broadening, deriving a  correction for those sources that shows major disagreement between virial and concordance cosmology luminosity values.   }   
{ The relatively large scatter between concordance cosmology and virial luminosity estimates can be reduced (by an order of magnitude) if a correction  for orientation effects is  included in the FWHM \hb\ value; outflow and sample definition yield relatively minor effects. } 
 
    \keywords{quasars: emission lines -- quasars: supermassive black holes --
                Line: profiles -- quasars: NLSy1 -- quasars: Highly accreting
                }

   \maketitle
%
\section{Introduction}

\label{intro}
\defcitealias{marzianisulentic14}{MS14}
\defcitealias{borosongreen92}{BG92}

With the advent of large databases such as the Sloan Digital Sky Survey (SDSS), we now have access to spectroscopic observations of a large number of galaxies containing active galactic nuclei (AGNs) or quasars. The SDSS Data Release  7 (DR7) provides access to the spectra of more than 100,000 quasars \citep{schneideretal10}. However, as seen throughout this work, for many types of studies it is not enough to have this large amount of data;  careful analysis is also required in order to obtain reliable results.

 Eighteen years ago, \citet{sulenticetal00a} first proposed a parameter space motivated  by the Eigenvector 1 set of correlations \citep[E1, ][hereafter BG92]{borosongreen92} involving an anti-correlation between  the equivalent width (EW) of \feii\ and the  peak intensity of \oiii. Later, \citet{sulenticetal07} proposed an extended parameter  space of four dimensions (4DE1) which included:  1) the full width at half maximum (FWHM) of the broad component (BC) of \hb, FWHM (\hbbc), 2) the ratio of the EWs of \feiiopt\ (\feii\ at 4570\AA) and \hbbc, \rfe\ = EW(\feii)/EW(\hbbc), 3) the photon index of soft X-rays, \Gsoft; and 4)  the centroid at half maximum of \civ, that traces blueshifted  emission of  a  representative high ionization line (HIL). This 4DE1 diagram for quasars is now understood to be mainly driven by Eddington ratio \lledd\   \citep[e.g.,][]{borosongreen92,sulenticetal00a,marzianietal01,boroson02,kuraszkiewiczetal04,ferlandetal09,sunshen15,sulenticetal17,bonetal18}. In practice, in the analysis of optical data, only parameters 1) and 2) are used. In this case, it is customary   to speak about the optical plane (OP) of the 4DE1 parameter space \citep[e.g., ][]{sulenticmarziani15,padovani16,taufikandikaetal16}, and to consider that the data point distribution in the OP defines a quasar main sequence \citep[MS,][and references therein]{sulenticetal11,marzianietal18}.

The work of Sulentic and collaborators  pointed out that the correlations in the OP of the 4DE1 could be referred to as a surrogate of a Hertzprung-Russel (H-R) diagram for quasars. In addition, they proposed two main populations based on the optical plane (FWHM(\hb) vs. \rfe): Population A (Pop A) for quasars with FWHM(\hb) \lt\ 4000 \kms\ and Population B (Pop B) including sources  with FWHM(\hb) \gt\ 4000 \kms\ (e.g., Fig. \ref{fig:E1}). The phenomenological study of \citet{sulenticetal02}   showed that the broad \hb\ profiles of  Pop. A objects are well-modeled with Lorentzian profiles, while those of Pop. B can be described with Gaussians. There are many other spectral characteristics that differentiate objects within the same population, especially in Pop. A \citep[see Figure 2 of ][and e.g., \citealt{craccoetal16}, and references therein]{sulenticetal02}. For this reason, the OP  was divided into bins with \D FWHM(\hb) = 4000\kms\ and \D\rfe = 0.5. This created  bins A1, A2, A3, and A4 defined as \rfe\ increases, and the B1, B1+, and B1++  bins defined as   FWHM(\hb) increases \citep[see Figure 1 of ][]{sulenticetal02}. Thus, spectra belonging to the same bin are expected to have very similar characteristics (e.g., line profiles and UV line ratios). The MS organizes the diverse quasar properties and makes it possible to identify quasars in different accretion states \citep{sulenticetal14}.    In this work we present in particular those  which are characterized by having \rfe\ \gt\ 1 (i.e., belonging to the A3 and A4 bins), most likely associated with high accretion rates \citep[e.g.,][]{sulenticetal14a,duetal16a}. 

These sources (which we indicate in the following as xA, acronym for extreme Pop. A)  have an importance that goes beyond their extreme properties in the 4DE1 context (highest \rfe,  largest \Gsoft, highest \civ\ blueshifts); they are the most efficient radiators per unit black hole mass, radiating at Eddington ratios of order unity.\footnote{The exact values depend on black hole mass scaling and bolometric corrections, both of which are uncertain.}
In other words, radiation forces are maximized with respect to gravity \citep{ferlandetal09}. These xA sources frequently show evidence of large blueshifts in their high-ionization broad \citep[typically \civ, see e.g.,][]{corbinboroson96,willsetal95,richardsetal02,sulenticetal07,marzianietal16} and narrow lines \citep[typically \oiiiopt, e.g.,][]{zamanovetal02,komossaetal08}. These sources should display the most evident feedback effects on their hosts, especially at high L.  

\citet[][hereafter \citetalias{marzianisulentic14}]{marzianisulentic14} proposed that xA quasars are observationally defined  by simple criteria (in the visual domain, \rfe $\ge$ 1 is a sufficient condition; more details are given in Sect.\ \ref{sec:sample}) and radiate at \lledd $\rightarrow$ 1.  These are the most luminous quasars (at a fixed \mbh) that exist over a 5-dex range in luminosity that is predictable, at least in principle,  from line width measurements only: L $\propto$ ($\delta$ v)$^4$, where $\delta $v  is a suitable ``virial broadening estimator'' (VBE).   The virial luminosity equation is a restatement of the virial theorem for constant luminosity-to-mass ratio, and  assumes the rigorous validity of the scaling relation r$ \propto$ L$^{1/2}$. In the case of xA quasars, its validity is observationally verified by the consistency of the emission line spectrum of xA sources over a wide luminosity range. A deviation from the scaling law would imply a change in ionization parameter, and hence in the emission line ratios that are used for the xA definition.  Similar considerations apply to the spectral energy distribution (SED) parameters entering into the virial luminosity equation as written by \citepalias[][ see also Sect.  \ref{sec:cosmo}]{marzianisulentic14}. The consistency in observational parameters hints at  replicable structure and dynamics.  
This and other recent attempts to link the quasar luminosity to the velocity dispersion \citep{lafrancaetal14,wangetal14a} echo analogous attempts done for galaxies \citep[e.g.,][]{faberjackson76,tullyfisher77}.  Spheroidal galaxies are however non-homologous systems \citep[e.g.,][and references therein]{donofrioetal17}, and the assumption of a constant luminosity-to-mass ratio is inappropriate.  

Accretion theory supports the empirical findings of \citetalias{marzianisulentic14} on xA sources.  First,  \lledd\ $\rightarrow$ 1 (up to a few times the Eddington luminosity)  is a  physically motivated condition.  The accretion flow remains optically thick  which allows for the possibility that the radiation pressure  ``fattens'' it.  However, when the mass accretion rate becomes super-Eddington, the emitted radiation is advected toward the black hole, so that the source luminosity increases only with the logarithm of accretion rate. In other words, the radiative efficiency of the accretion process is expected to decrease, yielding an asymptotic behavior of the luminosity as a function of the mass-accretion rate \citep{abramowiczetal88,mineshigeetal00,wataraietal00}. 
In observational terms, the luminosity-to-black hole mass ratio (L/\mbh $\propto$ \lledd) should tend  toward a well-defined value. The resulting ``slim'' accretion disk is expected to emit a steep soft and hard X-ray spectrum,  with hard X-ray photon index (computed between 2 and 20 KeV) converging toward $\Gamma_\mathrm{hard} \approx 2.5$ \citep{wangetal13}.  

The prospect of a practical application of virial luminosity estimates raises the issue of the availability of a VBE.  Major observational constraints from \civ\  indicate   high-ionization wind in Pop. A, even at relatively low luminosity  \citep[e.g.,][]{sulenticetal07,richardsetal11}.  The blueshifted component of the \oiiiopt\ lines  traces  winds  which are most likely of nuclear origin \citep{marzianietal16} and may become extremely powerful at high luminosities \citep[e.g.,][]{netzeretal04,zakamskaetal16,bischettietal17,marzianietal17,vietrietal18}. However, previous studies have indicated that part of the broad line emitting regions remains ``virialized.'' A virialized low-ionization broad emission line region is present even at the highest quasar luminosities, and coexists with high ionization winds \citep{negreteetal12,shen16,sulenticetal17}. Could this also be true for  xA sources?   

The identification of a reliable VBE is clearly a necessary condition to link the luminosity to the virial broadening.  The width of the HI Balmer line  \hb\ is considered one of the most reliable VBEs by several studies \citep[][and references therein]{vestergaardpeterson06,shenliu12,shen13,marzianietal17}, and the obvious question is whether or not the \hb\ line can also be considered as such in xA sources. Low-$z$\ xA sources show evidence of systematic blueshifts of the \oiiiopt\ lines \citep{zamanovetal02,komossaetal08,zhangetal11,marzianietal16}, indicating radiative or mechanical feedback from the AGN. Is the \hb\ profile   also affected? While there are several  studies of \oiiiopt\ over a very broad range of redshifts that are able to partially resolve the emitting regions  \citep[e.g., ][and references therein]{canodiazetal12,harrisonetal14,carnianietal15,crescietal15}, the broad component of \hb\ has not been considered in detail and in some cases has even been misinterpreted.  

In addition, it is believed that the width of \hb\ is affected by the viewing angle. Evidence of an orientation effect for radio loud (RL) quasars has grown  since the early work of \citet{willsbrowne86}. \citet{zamfiretal08} deduced a narrowing of the \hb\ profile by  a factor almost two, if lobe- and core-dominated RL  were compared. More recent work also suggests extreme orientation for blazars \citep{decarlietal11}. The issue is of utmost importance since \mbh\ estimates are so greatly influenced by orientation effects, if the line is emitted in a flattened configuration \citep[e.g.,][]{jarvismclure06}. In some special cases the \hb\ line width can be considered an orientation indicator \citep{punslyzhang10}, but the FWHM of \hb\ depends, in addition to the viewing angle, on \mbh\ and \lledd\ \citep{nicastro00,marzianietal01}. A strong dependence on \mbh\ trivially arises from the virial relation which implies \mbh $\propto $FWHM$^{1/2}$, while the dependence on \lledd\ is more complex and less understood; it may  come from the balance between radiation forces and gravity, which   in turn affects the distance of the line-emitting region from the central black hole. Therefore, at present there is no established way to estimate the viewing angle from optical spectroscopy for individual radio quiet quasars. We infer a significant effect from the analogy with the RL quasars. 

xA sources are found at both low-$z$\ and high-$z$.  The low-$z$ counterpart offers the advantage of high-S/N data coverage of the \hb\ spectral range needed for the identification of xA sources in the OP of the 4DE1. In the present work, we therefore use a sample of quasars based on the SDSS DR7 \citep{shenetal11} to identify the extreme accretors at low redshift ($z \lesssim$ 0.8) in the OP of the 4DE1 parameter space. Objects with  \rfe\ \gt\ 1 (regardless of their FWHM) as measured in this paper are considered extreme accretors.  In a parallel work, we are investigating xA sources at $2 \lesssim z \lesssim$ 2.6\ \citep[][accepted]{martinez-aldamaetal18}. 

In Section \ref{sec:sample} we describe the steps that we follow for the selection of the sample.  Section \ref{sec:specan} describes the spectral fit measurements. In Section \ref{sec:method} we explain the methodology used for population allocation, and how to obtain the spectral components for each object in the sample. Section \ref{sec:species}  describes the catalog of measurements that is the immediate result of this paper. Section \ref{results}  analyzes the results with a focus  on \hbbc\ and \oiii\ based on the individual measurements of the spectral components. In Section \ref{sec:discussion}, we discuss first the relation of xA sources to other quasar classes, and the meaning of the outflow tracers for feedback.  We subsequently analyze  the effect of orientation and outflows on the virial luminosity estimates. To this aim we develop a method that allows for an estimate of the viewing angle for each individual quasar. Based on the work of \citetalias{marzianisulentic14}, we also suggest the use of xA quasars as ``Eddington standard candles''. The conclusions and a summary of the paper are reported in Section \ref{sec:conclusion}. 

\section{Sample selection}
\label{sec:sample}

In order to obtain a sample that is  representative of low-redshift extreme quasars, we use the Sloan Digital Sky Survey Data Release 7 as a basis.\footnote{http://skyserver.sdss.org/dr7}

\subsection{Initial selection}
\label{sec:initialselection}

The quasar sample presented by \citet{shenetal11} consists in 105,783 spectra taken from the SDSS DR7 data base. For a detailed analysis of the quasars with high accretion rates, we need to select only those spectra with good quality that meet the criteria described by the E1 parameter space. Initially we used the following filters:
\begin{enumerate}
\item We selected quasars with $z$ \lt\ 0.8 to cover the range around \hb\ and include the \feii\ blends around 4570 and 5260\AA. As a first approximation, we use the $z$\ provided by the SDSS. With this criteria, we selected 19,451 spectra. We detected 103 spectra with an erroneous $z$ identification  which were also eliminated because they are noisy (Table 4 online\footnote{Vizier link to Table 4}). The measurement of $z$ is described in Section \ref{sec:rf}. 
\item  For the selected spectra, we carried out automatic measurements using the IRAF task \textit{splot}, to estimate the signal-to-noise ratio (S/N) around 5100\AA. We took three windows of 50\AA\ to avoid effects of local noise. Those with S/N \lt\ 15 (76.5\% of this sub sample), are mostly pure noise. Spectra with S/N between 15 and 20 are still very bad, and we can vaguely guess where the emission lines would be. 
In this way, we find that only those spectra with S/N \gt\ 20 have the minimum quality necessary to make reliable measurements. These represents only 14.1\% of the objects of the sample at item 1 with 2,734 spectra. This illustrates the importance of considering only good-quality spectra, especially when performing automatic measurements. However, as we see in the following sections, even considering only such spectra, the automatic measurements can introduce up to 30\% additional error on \rfe.  
\item The  criterion that we use to isolate extreme quasars is  \rfe\ \gt\ 1. In order to determine this ratio, we perform automatic measurements on the objects selected as described in item (2) of this list of filters. First we normalized all the spectra by the value of their continuum at 5100 \AA. We then took an approximate measurement of the EW of \feii\ and \hb\ in the ranges 4435-4686 and 4776-4946\AA,\ respectively \citepalias{borosongreen92}. We selected 468 objects with \rfe\ \gt\ 1. At this point, we further rejected 134  spectra that were either noisy or intermediate type (Sy 1.5), that is, the \hb\ broad component  (\hbbc) was weak compared to its narrow component, which is usually very intense.

\end{enumerate}

Our final selection is a sample consisting of 334 high-quality spectra. The OP of the E1 using this sample is shown in Fig. \ref{fig:E1}.

\subsection{Spectral-type assignment using automatic measurements}
\label{sec:automes}

Within the automatic measurements described in section \ref{sec:initialselection}, we also estimate the FWHM of \hb\ considering Lorentzian and Gaussian profiles to separate Pops. A and B, respectively (see Sect. \ref{intro}). We consider that objects with FWHM \lt\ 4000 \kms\ measured with a Lorentzian profile belong to Pop. A. The remaining spectra were classified as Pop. B objects.    In total we counted 211 objects  that   belong to the A3 and A4 bins. There is an overlap in the FWHM(\hb) measured with both profiles (Lorentzian and Gaussian). We found 41 objects with FWHM(\hb)$_{\rm Lorentz}$ \lt\ 4000 \kms\ which also have FWHM(\hb)$_{\rm Gauss}$ \gt\ 4000 \kms. For these cases we chose the measurements with Lorentz profile assigning these objects to Pop. A.  The remaining 82 objects, with a FWHM(\hb)$_{\rm Gauss}$ \gt\ 4000 \kms, belong to Pop. B, that is, these quasars would be in principle, B3 and B4 objects.  These sources are relatively rare, to the point that in previous studies of low-$z$ quasars, almost no objects with  \rfe\ \gt\ 1 and FWHM \gt\ 4000 \kms\ were found \citep[for B2, and none at all for B3 or B4,][]{zamfiretal10}. If xA sources are $\sim 10$\%\ of optically selected samples based on the SDSS \citep{zamfiretal10}, then they are roughly $\sim 2.5$\%\ of all type-1 quasars. Nonetheless, they are relatively important for understanding the nature of the \hb\ broadening. 

One of the purposes of this work is to study the systematic differences between each population, which help us to characterize and isolate the extreme accretors. For this purpose, as a first approximation, and in order to increase the S/N, we made average spectra of each of the bins A3, A4, B3, and B4. However, we realized that this separation is not enough, because within each bin, there are evident spectral differences, such as the intensities of \feii\ and \oiiiopt, as well as the width of the lines. In the objects of our sample, we can see a wide variation in the \oiiiopt\ intensities. To characterize the \oiii\ emission with respect to \hb\, we define the parameter \roiii\ = \oiiia/\hb\ as the ratio of the peak intensities of both lines. The range of values of \roiii\ goes from zero to  more than ten in the most extreme cases.  Therefore, each bin was separated into four sub-bins, taking into account a \D FWHM = 2000 \kms\ and an \roiii\ either greater or less than one. In this way, we have isolated the Narrow Line Seyfert 1s (NLSy1s) of our sample which, by definition, should have FWHM of the broad components of   less than 2000 \kms\  \citep{osterbrockpogge85,pogge00}.

\subsubsection{Automatic measurement bias}
\label{sec:automesbias}

Using the \textit{specfit} task of IRAF \citep{kriss94}, and following the methodology described in Section 3.1, we fit the spectral components to these average spectra. From the measurements of these fits, we find that some spectral characteristics are different from the originally assigned population. The most important difference is that for the average spectra of the A3 and B3 bins; the measured \rfe\  was lower than 1. 
The same happens with some average spectra of the sub-bins in A4 and B4, where the \rfe\ was lower than 1.5.

The principal reason for the differences between the individual automatic measurements and the average spectrum, is that there are some objects that do not actually have the spectral characteristics of the extreme accretors. For example:
\begin{itemize}
\item We found 32 sources that have a strong contribution from the host galaxy.  
There are more objects with smaller contributions (Sections \ref{sec:method}, \ref{database}).
\item Objects that are at the limit of \rfe\ = 1.
\item Objects with a strong contribution of the \feii\ template, that widens  the red wing of \hb, or a significant contribution of an \hb\ blueshifted component. Both components artificially increase the FWHM(\hb), if the FWHM is measured with the automatic technique.
\end{itemize}

In order to study the spectral differences in our initial sample of 334 objects in more detail, as well as to limit our sample to those very reliably defined extreme quasars, we decided to analyze them individually with a maximum-likelihood multicomponent-fitting software.

\section{Spectral analysis}
\label{sec:specan}

\subsection{Methodology: multi-component fitting}
\label{sec:method}

The \textit{specfit} task of IRAF allows us to simultaneously fit all the components present in the spectrum: the underlying continuum, the \feii\ pseudo-continuum, and the lines, whether in emission or, if necessary, in absorption. \textit{Specfit} minimizes the \chisq\ to find the best fit. The value of the \chisq\ reflects the difference between the original spectrum and the components fit; it tends to be larger in objects where the fit does not accurately reproduce the original spectrum. 

The steps we followed to accomplish identification, deblending, and measurement of the emission lines in each object are the following:
\begin{description}
\item[ The continuum -- ] In the optical range for the extreme quasars, \feii\ is especially strong, so we must look for continuum windows where \feii\ emission is minimal. We adopted a single power law to describe it using the continuum windows around 4430, 4760, and 5100 \AA\ (see, e.g., \citealt{francisetal91}). Only for  two objects, J131549.46+062047.8 and J150813.02+484710.6, was it necessary to use a broken power law with two different slopes to reproduce the \feii\ template at the blue side of \hb.
\item[\feii\ template -- ] We used the semi-empirical template by \citet{marzianietal09}, obtained from a high-resolution spectrum of I Zw 1, with a model of the \feii\ emission computed by a photoionization code in the range of \hb. This template adequately reproduces the \feii\  emission in the optical range for the vast majority of our sample spectra. For the most extreme \feii\ emitters, such as J165252.67+265001.9, this template does not reproduce the \feii\ features. It seems that a more detailed analysis of the \feii\ emission such as the one of \citet{kovacevicetal10} is needed. However, this analysis is beyond the scope of this paper.
\item[ \hb\ broad component profile --] Initially, we took as a basis the output of the automatic measurements that gave us the FWHM estimates, in order to fit the \hb\ line profile according to the E1. That is, we use a Lorentzian profile for objects with FWHM \lt\ 4000 \kms, and a Gaussian one for those with  FWHM \gt\ 4000 \kms. In the most extreme quasars with \rfe $\gg 1$, the FWHM is  widened due to the presence of two components in \hb, and the FWHM is artificially increased  if the measurement is automatic. In these objects, \feii\ is extremely strong, and/or a very intense blueshifted component of \hb\ appears (e.g., J131150.53+192053.1). When we performed a second individual profile fitting (see below) we reassigned a Lorentzian instead of a Gaussian for these sources.
\item[\oiiiopt\ --] We fitted this doublet with two Gaussians, considering the ratio of theoretical intensities of 1:3 \citep{dimitrijevicetal07}, the same FWHM, and the same line shift. We call this \oiiiopt\ component the ``narrow'' or ``core'' component.  We found some extreme cases with no detectable \oiiiopt\ emission (e.g., J085557.12+561534.7).

\end{description}

Apart from these four features, in some cases it was necessary to add other emission lines. In some cases the extra emission lines are strong and therefore obvious. When emission lines are weak  we identified them in the residuals of the fit. These extra emission lines are:

\begin{description}
\item[ \hb\ narrow component --]  This component, if present, is very evident in population B, however, we can also observe it in Pop. A objects. We model it consistently with the core component of \oiii, with a Gaussian profile and  the same FWHM. We used this emission line to define the restframe in the spectra where we measure it.

\item[ \hb\ blueshifted component  -- ] In some spectra we detected an extra component in the blue wing of \hb. When this component is weak, we can see it in the residuals of the fit. When strong, it appears as though the \hb\ line base is shifted to the blue. This component is likely to be associated with non-virialized outflows in quasars with high accretion rates \citep[e.g.,][]{coatmanetal16}. We fitted this feature with a blueshifted symmetric Gaussian.

\item[ \oiiiopt\ semi broad component -- ]  This second component -- added to the ``core'' or narrow component described above -- of the \oiii\ doublet is also associated with outflows \citep [e.g., ][]{zamanovetal02,zhangetal11}. It is characterized as being wider than the main component of \oiii, and is generally shifted to the blue. In some cases, we find this component with blue shifts up to 2000 \kms\ (e.g., J120226.75-012915.2).

\item[ \heiiopt --] In some cases we detect this component as residual emission of the fit; we fitted it with a Gaussian component. In some spectra it is not clear whether the extra emission around 4685-95 \AA\ is actually \heii\ or an \feii\ feature.

\item[Spectrum of the quasar host galaxy --]  We find absorption lines characteristic of the host galaxy, such as \mgi\ and \feiia\ which appear especially significant for 32 objects in our sample. The \hb\  absorption due to the host may create  the appearance of  the  broad \hb\  profile as double peaked, or let the \hbnc\ disappear altogether. 
Outside of the fitting range, we detect the \caii\ K and H bands (at 3969 and 3934 \AA).  
These remaining 32 objects were initially fitted along with the author sources. However, we then considered this sample apart because none of the objects are true extreme accretors. A detailed analysis of these sources will be described in Bon et al., in preparation.

\end{description}

\begin{figure}
\includegraphics[scale=0.55]{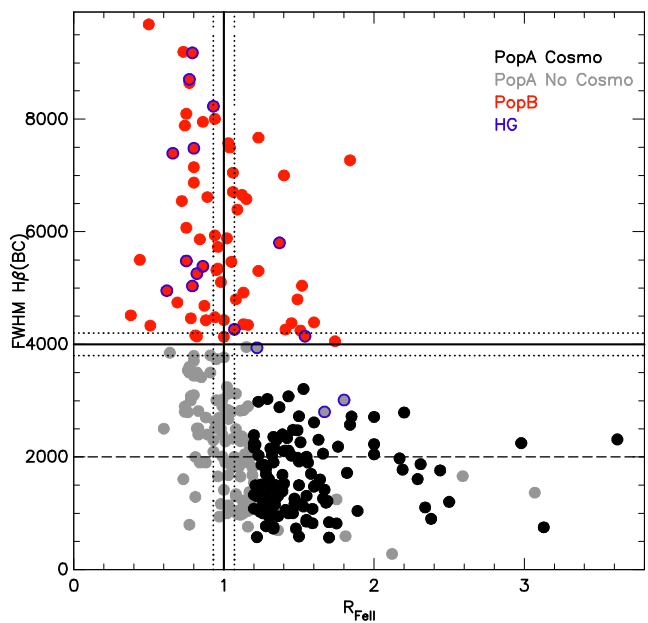}
\caption{Optical plane of the 4DE1 parameter space, FWHM(\hb) vs. \rfe, for the present sample. Black dots indicate sources chosen for the ``Cosmo'' sample, i.e., with \rfe$\ge1.2$, and gray dots indicate the remaining Pop. A sources. Population B sources are indicated in red. A blue contour identifies sources with weak host galaxy contaminations in their spectrum (labeled as HG). Vertical line separates the objects with \rfe\ \textgreater\ 1. This criteria is used to identify the extreme accretors. The filled horizontal line separates populations A and B, according to \citet{sulenticetal00a}. Dotted lines indicate the typical error associated to \rfe\ and the FWHM(\hb). Dashed horizontal line delimits the region of  NLSy1s (FWHM(\hb) \textless\ 2000 \kms).
\label{fig:E1}}
\end{figure}

Figure \ref{fig:E1} illustrates the location of the samples in the OP of the 4DE1 parameter space. The 32 sources that show strong contamination from the host galaxy (circled blue in Fig. \ref{fig:E1}) are not considered in the present work,; the analysis described in the sections therefore takes into account 302 spectra out of the 334 previously selected. The individual fits done with \textit{Specfit} are shown in Fig. 2 (online\footnote{Link to Figure 2.}).

\begin{sidewaystable*}
\setcounter{table}{0}
\scriptsize
\caption{Sample properties. \label{tab:sampleprop}}
\begin{tabular}{lcccccccccccccccccccccccc}
\hline\hline       
{Bin$^{a}$}  & {\# sources$^{b}$} & \multicolumn{3}{c}{Mag$_{G}$}& & \multicolumn{2}{c}{Log \l L$_{\lambda}$(5100)} &  & \multicolumn{2}{c}{z}  & {S/N} & \multicolumn{2}{c}{FWHM(\kms)}  &  {\rfe} &  {\D} &  {\roiii}&  {\D}  & & \multicolumn{6}{c}{EW (\AA)}  \\
\cline{3-5} \cline{7-8} \cline{10-11}\cline{13-14}  \cline{20-25} 
  &    &   {min}&   {max} &  {ave}&  &   {min}&   {max} &  &  {min}&   {max} &  &  {\hbbc} &  {\D}&  {}&   {}& {}&  {}&  {}&  {\hb$_{total}$}&  {\D}&  {\oiii$_{total}$}&  {\D}&  {\hb$_{Blue/BC}$}&  {\D} \\
(1)&(2)&(3)&(4)&(5)&&(6)&(7)&&(8)&(9)&(10)&(11)&(12)&(13)&(14)&(15)&(16)&&(17)&(18)&(19)&(20)&(21)&(22)
\\
\hline\\
PopA  & 211 & 15.5 & 19.4 & 18.0 &    & 44.06 & 46.13 &    & 0.08994 & 0.71646 & 28 & 1949 & 52 & 1.45 & 0.28 & 1.05 & 0.12 &    & 39.5 & 1.0 & 9.8 & 1.2 & 0.18 & 0.09\\
PopB  & 91 & 15.7 & 19.7 & 18.3 &    & 44.12 & 46.81 &    & 0.15184 & 0.77136 & 25 & 5887 & 189 & 0.91 & 0.09 & 2.24 & 0.11 &    & 29.7 & 1.3 & 9.0 & 0.8 &    &  \\
\hline 
A2b0  & 14 & 16.4 & 19.4 & 18.2 &    & 44.35 & 45.51 &    & 0.27820 & 0.46197 & 67 & 2778 & 35 & 0.75 & 0.06 & 0.55 & 0.03 &    & 48.1 & 0.5 & 5.7 & 0.4 &    &  \\
A2b1  & 25 & 16.4 & 19.3 & 18.4 &    & 44.13 & 45.11 &    & 0.09158 & 0.40778 & 62 & 2899 & 12 & 0.71 & 0.05 & 1.70 & 0.15 &    & 39.6 & 0.9 & 13.0 & 2.5 &    &  \\
A2n0  &  3 (4)  & 17.0 & 18.9 & 18.3 &    & 44.44 & 45.78 &    & 0.29954 & 0.51711 & 46 & 1557 & 35 & 0.78 & 0.13 & 0.57 & 0.03 &    & 39.0 & 1.0 & 10.2 & 0.3 &    &  \\
A2n1  &  3 (5)  & 18.0 & 19.2 & 18.6 &    & 44.36 & 44.80 &    & 0.19696 & 0.30477 & 47 & 1392 & 43 & 0.85 & 0.23 & 1.89 & 0.04 &    & 33.8 & 0.5 & 21.6 & 0.6 &    &  \\
A3b0  & 39 & 15.5 & 19.4 & 17.7 &    & 44.30 & 46.12 &    & 0.15038 & 0.68950 & 76 & 2303 & 26 & 1.10 & 0.05 & 0.32 & 0.04 &    & 47.9 & 0.5 & 4.6 & 0.8 &    &  \\
A3b1  & 9 & 16.4 & 18.9 & 18.3 &    & 44.24 & 45.61 &    & 0.17522 & 0.46621 & 56 & 2022 & 13 & 1.13 & 0.79 & 1.45 & 0.32 &    & 36.0 & 1.1 & 10.8 & 1.4 &    &  \\
A3n0  & 69 & 16.2 & 19.4 & 17.9 &    & 44.06 & 45.86 &    & 0.08994 & 0.63026 & 92 & 1621 & 71 & 1.16 & 0.08 & 0.36 & 0.14 &    & 44.3 & 1.4 & 6.6 & 0.8 & 0.06 & 0.10\\
A3n1  & 12 & 17.6 & 19.1 & 18.4 &    & 44.14 & 45.14 &    & 0.20842 & 0.41191 & 75 & 1349 & 12 & 1.09 & 0.07 & 1.68 & 0.06 &    & 33.0 & 1.1 & 22.0 & 2.3 & 0.02 & 0.15\\
A4b0  & 9 & 16.8 & 18.7 & 17.6 &    & 44.57 & 46.18 &    & 0.26492 & 0.71646 & 60 & 2054 & 79 & 1.81 & 0.55 &    &    &    & 36.4 & 1.6 & 2.4 & 0.4 & 0.12 & 0.11\\
A4b1  &  1 (3)  & 16.9 & 19.4 & 18.2 &    & 44.24 & 44.64 &    & 0.17942 & 0.30640 & 56 & 3237 & 136 & 1.51 & 0.12 & 1.26 & 0.07 &    & 49.0 & 1.3 & 8.1 & 0.8 & 0.31 & 0.05\\
A4n0  & 23 & 17.1 & 19.3 & 17.9 &    & 44.18 & 45.30 &    & 0.12736 & 0.49682 & 74 & 1145 & 17 & 1.62 & 0.16 & 0.30 & 0.08 &    & 37.4 & 1.0 & 5.2 & 0.8 & 0.15 & 0.11\\
A4n1  &  1 (5)  & 16.7 & 18.7 & 18.0 &    & 44.17 & 44.51 &    & 0.13464 & 0.27647 & 61 & 1188 & 32 & 1.65 & 0.18 & 2.31 & 0.06 &    & 29.4 & 1.0 & 14.7 & 0.7 & 0.16 & 0.06\\
A5b  &  3 (4)  & 17.4 & 18.1 & 17.7 &    & 44.98 & 45.59 &    & 0.35038 & 0.58198 & 45 & 2286 & 18 & 2.37 & 0.33 &    &    &    & 41.9 & 0.9 &    &    & 0.24 & 0.06\\
A5n  & 9 & 16.7 & 19.1 & 18.0 &    & 44.47 & 45.93 &    & 0.25868 & 0.56391 & 51 & 1294 & 13 & 2.33 & 0.05 & 0.22 & 0.09 &    & 32.3 & 1.0 & 2.9 & 0.8 & 0.17 & 0.11\\
Pec  &  3 (5)  & 16.6 & 18.6 & 17.8 &    & 44.25 & 45.43 &    & 0.15080 & 0.40004 & 40 & 2106 & 54 & 2.83 & 0.05 &    &    &    & 44.6 & 1.0 &    &    & 0.38 & 0.05\\
B2b  & 9 & 18.0 & 19.5 & 18.8 &    & 44.39 & 44.92 &    & 0.18646 & 0.46889 & 64 & 6999 & 13 & 0.55 & 0.06 & 1.83 & 0.20 &    & 34.2 & 0.8 & 8.0 & 1.2 &    &  \\
B2n  & 19 & 15.8 & 19.4 & 18.3 &    & 44.23 & 46.81 &    & 0.18637 & 0.77136 & 71 & 4705 & 12 & 0.74 & 0.10 & 1.37 & 0.15 &    & 26.0 & 0.9 & 6.7 & 1.0 &    &  \\
B3b  & 9 & 18.0 & 19.2 & 18.7 &    & 44.21 & 44.91 &    & 0.17688 & 0.35636 & 55 & 6227 & 154 & 1.07 & 0.27 & 2.04 & 0.05 &    & 28.0 & 0.9 & 12.2 & 0.7 &    &  \\
B3n  & 14 & 17.2 & 19.7 & 18.3 &    & 44.12 & 45.65 &    & 0.15184 & 0.54423 & 58 & 4697 & 74 & 1.16 & 0.09 & 1.76 & 0.04 &    & 31.1 & 0.9 & 9.7 & 0.7 &    &  \\
NA  & 16 & 15.7 & 19.7 & 18.4 &    & 44.33 & 46.69 &    & 0.20789 & 0.60325 & 69 & 6806 & 386 & 1.05 & 0.18 & 4.20 & 0.31 &    & 29.1 & 2.4 & 8.1 & 0.3 &    &  \\
\hline
 \hline
\end{tabular}
\end{sidewaystable*}

The first two rows of Table \ref{tab:sampleprop} present the general properties of the sample: column (1) shows population and spectral type (ST) assignment (see Sect. \ref{sec:composite}), column (2) number of sources, columns (3-5) minimum, maximum, and average $g$\ magnitude for each population and ST, columns (6-9) minimum and maximum luminosity, and redshift, column (10)  average S/N, columns (11-12) FWHM(\hbbc) and its uncertainty, columns (13-16) \rfe\ and \roiii\ and its uncertainties, and columns (17-22) EW of \hb$_{total}$ ($BC + NC + Blue$), \oiii$_{total}$ ($NC + SB$), and EW(\hb$_{Blue/BC}$) with its uncertainties.
The wide majority of sources has $44 \lesssim \log L \lesssim 45$, with a tail in the luminosity range $45 \lesssim \log L \lesssim 46$. The majority of the redshifts are $\lesssim 0.4$, with a tail reaching $z \approx 0.8$. 

 Only 187 of  the 236 initially selected Pop. A sources were found to have \rfe $\ge 1$ after
the individual fitting.  Restricting our attention to the xA sources, Pop. B sources are prudentially kept separated from the cosmo sample described below, even if a minority of them meet the selection criteria.  We define a ``cosmo'' sample with a stronger restriction to \rfe $\ge 1+ 2\delta $\rfe$ = 1.2 $, where $\delta $\rfe\ is the uncertainty associated with the \rfe\ measurement at a $1-\sigma$\ confidence value. In this way, we should exclude 95\%\ of the sources that are not true xA sources, but are misplaced due to measurement uncertainties (some true xA objects will also be  lost because they are brought to the region \rfe $\lesssim 1 $). The cosmo sample includes 117 objects with FWHM $\le$ 4000 \kms.

\subsection{Full profile measurements}

The full profiles of \hb\ (BC+BLUE) and \oiii\ (narrow + semi broad) were reconstructed adding the two components isolated through the \textit{ specfit} analysis. The full profiles are helpful for the definition of width and shift parameters (as described in Sect. \ref{database}) that are not dependent on the \textit{specfit decomposition}. As shown in Sect. \ref{sec:method}, the decomposition of the \oiii\ semi-broad and narrow components is especially difficult.

\subsection{Computation of luminosity and accretion parameters}

The bolometric luminosity is given by $L = C \lambda L_\lambda (5100 \mathrm{\AA})$, where we assume C $\approx$ 12.17  as   bolometric correction for the   luminosity at 5100\AA\  $\lambda L_\lambda (5100 \mathrm{\AA})$  \citep{richardsetal06}. The Eddington ratio \lledd, the ratio between the bolometric luminosity and the Eddington luminosity $L_{\rm Edd} \approx 1.3 \cdot 10^{46} (10^{8} M_{\rm BH}/M_\odot)$ \ergss , is then  computed by using the mass relation derived by \citet{vestergaardpeterson06}.  In this relation we enter the FWHM of the broad component of \hb\ only.

\section{Immediate results: the database of spectral measurements}
\label{sec:species}

\addtocounter{figure}{1}
\begin{figure}
\includegraphics[scale=0.5]{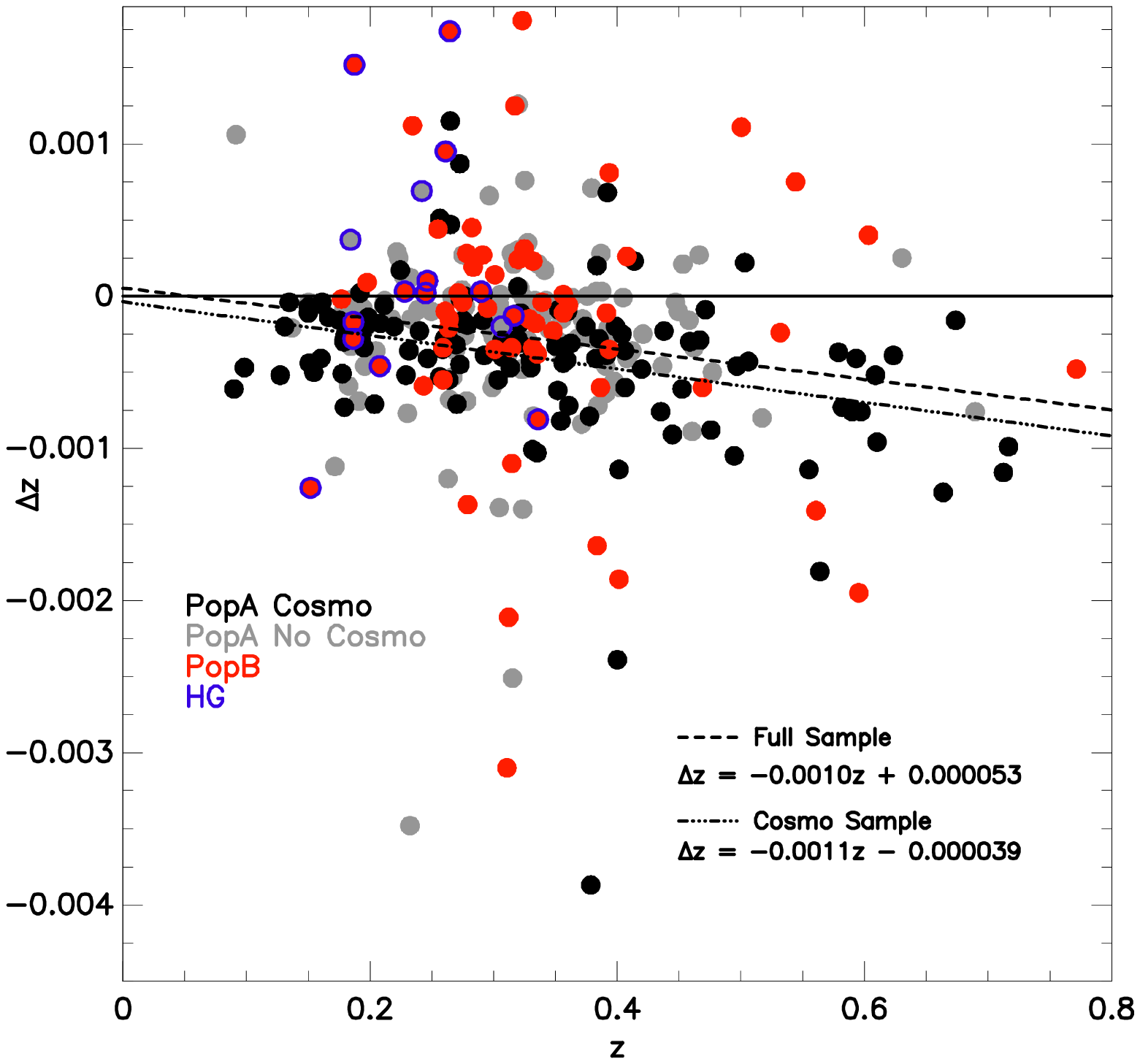}
\caption{Bias in redshift estimates $\Delta z$\ as a function of $z$\ given by the SDSS database.  Meaning of symbols is as for Fig. \ref{fig:E1}.  
\label{fig:Dz}}
\end{figure}

\subsection{Rest frame}
\label{sec:rf}

The redshift spectral correction was initially done  using the $z$\ given by the SDSS database. However, it is well known that this estimate can be biased \citep{hewettwild10}. We made an independent estimation of the rest frame using the \hb\ narrow component, if present;  if unavailable, we used the broad component (especially for Pop. A objects).  We find discrepancies between the $z$\ given by the SDSS and our estimation (this work estimation - TW), spanning from \D z = z$_{SDSS}$ - z$_{TW}$ = 0.0038 to  -0.0047, with an average value of \D z = -0.0006 (equivalently to -3 \AA\ around \hb; Figure \ref{fig:Dz}). We believe there are two main reasons for these systematic differences. Firstly, the rest frame is usually estimated using the average $z$\ derived from narrow lines such as \oiii. For highly accreting objects, blueshifts of \oiii\ have been observed reaching up to 2000 \kms\ (e.g., SDSSJ030000.00-080356.9).  Secondly,  Fig.  \ref{fig:Dz}  shows that above $z \approx 0.35$, the  majority of sources belong to the cosmo sample. They are the ones expected to show  blueshifts of the highest amplitude. There could also be  a luminosity effect, as sources in the redshift range $0.4 \lesssim z \lesssim 0.7$\   belong to the high-luminosity tail of the sample.  The distributions for \hb\ and \oii\ agree if the effective wavelength of the \oii\ is assumed consistent with the low-density case ($\lambda_{\rm eff} \approx 3728.8 $\AA\ in vacuum). In columns 2 and 3 of Table \ref{tab:indmeas} we report the value of $z$ estimated in this work and its uncertainty, respectively.

\subsection{ Spectral parameters of broad and narrow emission lines}
\label{database}

The  measurements of the individual spectral fits, along with other properties, are reported in Table 5 (online\footnote{Vizier link to Table 5}). The headers of the online table columns are described in Table \ref{tab:indmeas}, and are as follows: 

\begin{description}
\item[1  --]  SDSS DR7 designation. 

\item[2 -- 5 ] Redshift. As mentioned, the rest frame of the sample was estimated using the narrow component of \hb, for the cases where this component was measured.  For the objects where no narrow component was found, we used the broad component, except for objects when there was a weak contribution of the host galaxy (labeled as HG in the table). The \hb\ absorption line from the host galaxy affects the \hbnc. In this case we used \oiiia\ since in a companion work we have shown that there is good agreement between \oiiia\ and redshift derived from the host galaxy (Bon et al., in preparation).  The SDSS redshift was taken directly from the header of each spectrum. 

\item[6 -- 13]  S/N, Continuum at 5100 \AA\ in the rest-frame measured from the fitted power law, and the value used to normalize the continua in the automatic measurements, along with their associated errors. The real continuum flux at 5100\AA\ (in units of  erg  s$^{-1}$ cm$^ {-2}$\, \AA$^{-1}$) is the multiplication of the C(5100) $\times$ Norm $\times 10^{-17}$.  Columns 11 -- 12 provide the values of the spectral index $\alpha$, and its associated error. Column 13 labels 16 sources for which a weak contamination by the host galaxy had been detected. Their contribution  cannot be reliably estimated  because it is small ($ < 50$\%).

\item[14 -- 22] \hbbc\  flux, EW, shift, and FWHM, with associated errors. Column 22 yields the \hb\ line profile shape: G=Gaussian, L=Lorentzian.

\item[23 -- 30] Flux and EW, shift relative to the restframe, and FWHM of the blueshifted component of  \hb, along with associated errors.

\item[31 -- 34] Flux and EW of \feii\ in the range 4435 -- 4685 \AA, with errors.

\item[35]  Assigned population: A or B, according with the E1 formalism.  

\item[36 -- 49]  Asymmetry index (AI), kurtosis index (KI), and centroids  of \hb, with respective uncertainties. 
The centroid at fraction $x$ of the peak intensity is given by
\begin{equation}
c (x) = \frac{\lambda_{\rm R(x)}+\lambda_{\rm B(x)} - 2\lambda_0}{2\lambda_0} c
.\end{equation}
Values are reported for x=0.25,0.5,0.75,0.9. Asymmetry index is only different from zero for objects with a blueshifted component. The asymmetry index at one quarter is defined as twice the centroid using the peak wavelength (in practice the c(0.9) is used as a proxy) as a reference, that is,

\begin{equation}
{\rm AI} =  \frac{\lambda_{\rm R(1/4)}+\lambda_{\rm B(1/4)}-2\rm \lambda_P}{\rm \lambda_P}
.\end{equation}

The kurtosis index is defined as the ratio between the line widths at three quarters and one quarter of fractional intensity:

\begin{equation}
{\rm KI} = \frac{\lambda_{\rm R(3/4)}-\lambda_{\rm B(3/4)}}{\lambda_{\rm R(1/4)}-\lambda_{\rm B(1/4)}}
.\end{equation}

\item[50 -- 55] Flux, shift relative to the restframe, and FWHM of \heiiopt\ broad component, with errors.

\item[56 -- 63] Flux, EW, shift relative to the restframe, and FWHM of \hbnc\ , with errors.

\item[64 -- 79] Flux, EW, shift relative to the restframe, and FWHM of \oiiia\ and of the semi-broad component, with errors.

 \end{description}

\subsection{Error sources}
\label{sec:errs}

 The {\em specfit} analysis builds a model of the spectrum that implies an a priori-assumption on the emission line components.  Formal uncertainties (as provided by the fitting routine) are relatively small and include the effect of the line blending. The continuum placement is the number one source of uncertainty, especially for the semi-broad or weak narrow component of \oiii, and \heiiopt: a small change of continuum level can easily change the line fluxes by a factor of approximately two. Continuum placement may be ambiguous in the case of A4 and B4 because of the extremely strong \feii\ emission that obliterates the regions almost free of line emission.

\begin{table*}
\caption{Header description of the online table with individual measurements. \label{tab:indmeas}}
\scriptsize
\centering
\begin{tabular}{lllll}
\hline \hline\\
{Column} & {Identifier} & {Type} & {Units}& {Description}  \\
\hline
1       &       SDSS    &       CHAR    &       NULL    &       SDSS Object Name    \\
\\
2       &       $z$     &       FLOAT   &       NULL    &       $z$ considered in this work, measured using the \hbnc\ or \oiiia\ line (see text).     \\
3       &       $z$\_ERR        &       FLOAT   &       NULL    &       $z$ (This work) error       \\
4       &       $z$SDSS &       FLOAT   &       NULL    &       $z$ given by the SDSS database    \\
5       &       $z$SDSS\_ERR    &       FLOAT   &       NULL    &       $z$ (SDSS) error    \\
\\
6       &       SN      &       FLOAT   &       NULL    &       S/N ratio measured around 5100 \AA        \\
7       &       C5100   &       FLOAT   &       $10^{-17}$ ergs\,s$^{-1} $cm$^{-2}$ \AA$^{-1}$   &       Continuum Flux at 5100 \AA      \\
8       &       C5100\_ERR      &       FLOAT   &       $10^{-17}$ ergs\,s$^{-1}$ cm$^{-2}$ \AA$^{-1}$    &       Continuum Flux at 5100 \AA\ error       \\
9       &       N5100   &       FLOAT   &       NULL    &       Continuum Normalization at 5100 \AA       \\
10      &       N5100\_ERR      &       FLOAT   &       NULL    &       Continuum Normalization at 5100 \AA\ error        \\
11      &       ALPHA   &       FLOAT   &       NULL    &       Power Law Index - \a\     \\
12      &       ALPHA\_ERR      &       FLOAT   &       NULL    &       Power Law Index - \a\ error   \\
13      &       FAINT\_HG       &       INTEGER &       NULL    &       Faint contribution of the HG (9 objects)      \\
\\
14      &       FLUX\_HBBC      &       FLOAT   &       $10^{-17}$ ergs\,s$^{-1}$ cm$^{-2} $      &       \hbbc\ Line Flux        \\
15      &       FLUX\_HBBC\_ERR &       FLOAT   &       $10^{-17}$ ergs\,s$^{-1}$ cm$^{-2} $      &       \hbbc\ Line Flux error  \\
16      &       EW\_HBBC        &       FLOAT   &       \AA     &       Rest-frame Equivalent Width of \hbbc       \\
17      &       EW\_HBBC\_ERR   &       FLOAT   &       \AA     &       Rest-frame Equivalent Width of \hbbc\ error        \\
18      &       SHIFT\_HBBC     &       FLOAT   &       \kms    &       \hbbc\ shift with respect to the Rest-frame    \\
19      &       SHIFT\_HBBC\_ERR        &       FLOAT   &       \kms    &       \hbbc\ shift with respect to the Rest-frame error      \\
20      &       FWHM\_HBBC      &       FLOAT   &       \kms    &       \hbbc\ Full Width at Half Maximum      \\
21      &       FWHM\_HBBC\_ERR &       FLOAT   &       \kms    &       \hbbc\ Full Width at Half Maximum error        \\
22      &       HB\_PROFILE     &       CHAR    &       NULL    &       \hb\ Line Profile. G = Gaussian, L = Lorentzian      \\
\\
23      &       FLUX\_HBBLUE    &       FLOAT   &       $10^{-17}$ ergs\,s$^{-1}$ cm$^{-2} $      &       \hb\ BLUE Flux  \\
24      &       FLUX\_HBBLUE\_ERR       &       FLOAT   &       $10^{-17}$ ergs\,s$^{-1}$ cm$^{-2} $       &       \hb\ BLUE Flux error    \\
25      &       EW\_HBBLUE      &       FLOAT   &       \AA     &       Rest-frame Equivalent Width of \hb\ BLUE   \\
26      &       EW\_HBBLUE\_ERR &       FLOAT   &       \AA     &       Rest-frame Equivalent Width of \hb\ BLUE error     \\
27      &       SHIFT\_HBBLUE   &       FLOAT   &       \kms    &       \hb\ BLUE shift with respect to the Rest-frame       \\
28      &       SHIFT\_HBBLUE\_ERR      &       FLOAT   &       \kms    &       \hb\ BLUE shift with respect to the Rest-frame error \\
29      &       FWHM\_HBBLUE    &       FLOAT   &       \kms    &       \hb\ BLUE Full Width at Half Maximum \\
30      &       FWHM\_HBBLUE\_ERR       &       FLOAT   &       \kms    &       \hb\ BLUE Full Width at Half Maximum error   \\
\\
31      &       FLUX\_FEII      &       FLOAT   &       $10^{-17}$ ergs\,s$^{-1}$ cm$^{-2} $      &       \feiiopt\ Flux  \\
32      &       FLUX\_FEII\_ERR &       FLOAT   &       $10^{-17}$ ergs\,s$^{-1}$ cm$^{-2} $      &       \feiiopt\ Flux error    \\
33      &       EW\_FEII        &       FLOAT   &       \AA     &       Rest-frame equivalent width of \feiiopt    \\
34      &       EW\_FEII\_ERR   &       FLOAT   &       \AA     &       Rest-frame equivalent width of \feiiopt\ error     \\
\\
35      &       POP     &       CHAR    &       NULL    &       Population designation     \\
36      &       RFeII & FLOAT   &       NULL    &       \rfe\   \\
37      &       RFEII\_ERR &    FLOAT   &       NULL    &       \rfe\   error\\
\\
38      &       AI\_HB  &       FLOAT   &       NULL    &       \hb\ Asymetry (only objects with \hb\ BLUE        \\
39      &       AI\_HB\_ERR     &       FLOAT   &       NULL    &       \hb\ Asymetry error  \\
40      &       KURT    &       FLOAT   &       NULL    &       Kurtosis        \\
41      &       KURT\_ERR       &       FLOAT   &       NULL    &       Kurtosis error   \\
42      &       C010    &       FLOAT   &       \kms    &       \hb\ centroid at 0.10 of the line intensity   \\
43      &       C010\_ERR       &       FLOAT   &       \kms    &       \hb\ centroid at 0.10 of the line intensity error    \\
44      &       C025    &       FLOAT   &       \kms    &       \hb\ centroid at 0.25 of the line intensity   \\
45      &       C025\_ERR       &       FLOAT   &       \kms    &       \hb\ centroid at 0.25 of the line intensity error    \\
46      &       C050    &       FLOAT   &       \kms    &       \hb\ centroid at 0.50 of the line intensity   \\
47      &       C050\_ERR       &       FLOAT   &       \kms    &       \hb\ centroid at 0.50 of the line intensity error    \\
48      &       C075    &       FLOAT   &       \kms    &       \hb\ centroid at 0.75 of the line intensity   \\
49      &       C075\_ERR       &       FLOAT   &       \kms    &       \hb\ centroid at 0.75 of the line intensity error    \\
50      &       C090    &       FLOAT   &       \kms    &       \hb\ centroid at 0.90 of the line intensity   \\
51      &       C090\_ERR       &       FLOAT   &       \kms    &       \hb\ centroid at 0.90 of the line intensity error    \\
\\      
52      &       FLUX\_HEII      &       FLOAT   &       $10^{-17}$ ergs\,s$^{-1}$ cm$^{-2} $      &       \heii\ Line Flux        \\
53      &       FLUX\_HEII\_ERR &       FLOAT   &       $10^{-17}$ ergs\,s$^{-1}$ cm$^{-2} $      &       \heii\ Line Flux error  \\
54      &       SHIFT\_HEII     &       FLOAT   &       \kms    &       \heii\ shift with respect to the rest frame    \\
55      &       SHIFT\_HEII\_ERR        &       FLOAT   &       \kms    &       \heii\ shift with respect to the rest frame error      \\
56      &       FWHM\_HEII      &       FLOAT   &       \kms    &       \heii\ Full Width at Half Maximum      \\
57      &       FWHM\_HEII\_ERR &       FLOAT   &       \kms    &       \heii\ Full Width at Half Maximum error        \\
\\
58      &    FLUX\_HBNC &       FLOAT   &       $10^{-17}$ ergs\,s$^{-1}$ cm$^{-2} $      &       \hbnc\ Line Flux        \\
59      &       FLUX\_HBNC\_ERR &       FLOAT   &       $10^{-17}$ ergs\,s$^{-1}$ cm$^{-2} $      &       \hbnc\ Line Flux error  \\
60      &       EW\_HBNC        &       FLOAT   &       \AA     &       Rest-frame Equivalent Width of \hbnc       \\
61      &       EW\_HBNC\_ERR   &       FLOAT   &       \AA     &       Rest-frame Equivalent Width of \hbnc\ error        \\
62      &       SHIFT\_HBNC     &       FLOAT   &       \kms    &       \hbnc\ shift with respect to the Rest-frame    \\
63      &       SHIFT\_HBNC\_ERR        &       FLOAT   &       \kms    &       \hbnc\ shift with respect to the Rest-frame error      \\
64      &       FWHM\_HBNC      &       FLOAT   &       \kms    &       \hbnc\ Full Width at Half Maximum      \\
65      &       FWHM\_HBNC\_ERR &       FLOAT   &       \kms    &       \hbnc\ Full Width at Half Maximum error        \\
\hline\\
\end{tabular}
\end{table*}

\clearpage
\addtocounter{table}{-1}
\begin{table*}\scriptsize
\caption{Header description (cont.) \label{tab:indmeas1}}
\centering
\begin{tabular}{lllll}
\hline \hline\\
{Column} & {Identifier} & {Type} & {Units}& {Description}  \\
\hline
\\
66      &       FLUX\_OIII      &       FLOAT   &       $10^{-17}$ ergs\,s$^{-1}$ cm$^{-2} $      &       \oiiia\ Line Flux       \\
67      &       FLUX\_OIII\_ERR &       FLOAT   &       $10^{-17}$ ergs\,s$^{-1}$ cm$^{-2} $      &       \oiiia\ Line Flux error \\
68      &       EW\_OIII        &       FLOAT   &       \AA     &       Rest-frame Equivalent Width of \oiiia\     \\
69      &       EW\_OIII\_ERR   &       FLOAT   &       \AA     &       Rest-frame Equivalent Width of \oiiia\ error       \\
70      &       SHIFT\_OIII     &       FLOAT   &       \kms    &       \oiiia\ shift with respect to the Rest-frame    \\
71      &       SHIFT\_OIII\_ERR        &       FLOAT   &       \kms    &       \oiiia\ shift with respect to the Rest-frame error      \\
72      &       FWHM\_OIII      &       FLOAT   &       \kms    &       \oiiia\ Full Width at Half Maximum      \\
73      &       FWHM\_OIII\_ERR &       FLOAT   &       \kms    &       \oiiia\ Full Width at Half Maximum error        \\
\\
74      &       FLUX\_OIIISB    &       FLOAT   &       $10^{-17}$ ergs\,s$^{-1}$ cm$^{-2} $      &       \oiiia\ Semi Broad Line Flux    \\
75      &       FLUX\_OIIISB\_ERR       &       FLOAT   &       $10^{-17}$ ergs\,s$^{-1}$ cm$^{-2} $       &       \oiiia\ Semi Broad Line Flux error      \\
76      &       EW\_OIIISB      &       FLOAT   &       \AA     &       Rest-frame Equivalent Width of \oiiia\ Semi Broad  \\
77      &       EW\_OIIISB\_ERR &       FLOAT   &       \AA     &       Rest-frame Equivalent Width of \oiiia\ Semi Broad error    \\
78      &       SHIFT\_OIIISB   &       FLOAT   &       \kms    &       \oiiia\ SB shift with respect to the rest frame \\
79      &       SHIFT\_OIIISB\_ERR      &       FLOAT   &       \kms    &       \oiiia\ SB shift with respect to the rest frame error   \\
80      &       FWHM\_OIIISB    &       FLOAT   &       \kms    &       \oiiia\ SB Full Width at Half Maximum   \\
81      &       FWHM\_OIIISB\_ERR       &       FLOAT   &       \kms    &       \oiiia\ SB Full Width at Half Maximum error     \\
\\
82      &       LOG\_MBH        &       FLOAT   &       NULL    &       Logarithmic Black Hole Mass in solar masses \\
83      &       LOG\_MBH\_ERR   &       FLOAT   &       NULL    &       Logarithmic Black Hole Mass error   \\
84      &       LOG\_L\_BOL     &       FLOAT   &       NULL    &       Logarithmic Bolometric Luminosity   \\
85      &       LOG\_L\_BOL\_ERR        &       FLOAT   &       NULL    &       Logarithmic Bolometric Luminosity error     \\
86      &       L/L\_EDD        &       FLOAT   &       NULL    &       Eddington ratio   \\
87      &       L/L\_EDD\_ERR   &       FLOAT   &       NULL    &       Eddington ratio error     \\
\hline\\
\end{tabular}
\end{table*}

To further assess measurement errors we considered the composite spectra for bins A3n0 and A3b0 (see the description of the sub-bins in Sect. \ref{sed:blend}) which are practically noiseless (S/N $\approx 200$). We added random Gaussian noise to obtain (1) S/N $\approx 20$ which is the lowest S/N in our sample, and (2) S/N $\approx 40$\ which roughly corresponds to the highest S/N, and repeated the {\em specfit} analysis at least several hundred times. The {\em specfit}  simulations that included Gaussian noise (with a different random pattern for each simulation) were computed leaving all parameters free to vary and also giving a random offset to the initial values of  several  of them (e.g., continuum level, \hbbc\ line flux, continuum slope). 
 
\begin{table} 
 \caption{Uncertainty estimates for selected parameters. \label{tab:mc}}
\tabcolsep=3pt
\begin{tabular}{lccccc}
 \hline \hline
Parameter &  Average  & $\sigma$ & Median & SIQR$_\mathrm{l}$ & SIQR$_\mathrm{u}$\\
\hline
 \multicolumn{6}{c}{A3n0} \\ 
\hline
 FWHM \hbbc\          &   1480  &  160  &  1480  &    75  &  85     \\
 Intensity \hbbc     &  40.75  & 3.54  & 41.74  &  2.17  & 1.33    \\
 Intensity \oiii\ NC  &  2.98  & 0.89  & 2.85  &  0.44  & 0.45    \\
 Intensity \oiii\ SB  &   3.49$^{1}$     &0.73$^{1}$  &  3.44  & 0.47   & 0.50   \\
 Shift  \oiii\ SB         &   -390  &  230  &  -350  &  65      &   60    \\
$ I_{\lambda,0}$     & 15.77   &0.98  &  15.58 &  0.45  &  0.65   \\
 $a$     & 1.73    &0.04   &  1.72  &  0.02  &  0.025  \\ 
  \hline
 \multicolumn{6}{c}{A3b0}\\ \hline
 FWHM \hbbc\             & 2365 & 120  &     2365  &  75   &   70     \\
 Intensity \hbbc        & 49.65 & 2.05 &    49.89 & 1.04 & 0.87    \\
 Intensity \oiii\ NC    & 2.84 & 1.02 &    2.88 & 0.87 & 0.78    \\
 Intensity \oiii\ SB    & 1.65 & 1.16 &    1.60 & 0.88 & 0.89    \\
 Shift  \oiii\ SB       &  -790 & 920   &     -630  &  300  &   290  \\
$ I_{\lambda,0}$        & 14.90 & 0.52 &     14.89 &  0.32 &   0.32  \\
 $a$       & 1.712 & 0.023 &     1.71  & 0.013 &  0.015   \\
\hline
\end{tabular}
\tablefoot{FWHM \hbbc\ is in \kms; intensities are computed on spectra with continuum normalized to 1 at 5100 \AA. $ I_{\lambda,0}$\ and $a$ are the   intensity at 1000 \AA\ and the index of the power law $I(\lambda) = I_{\lambda,0} (\lambda/1000)^{a}$. $^{1}$: from the $F$ distribution within $1 \le F \le F(2\sigma)$.   }
\end{table}

Table \ref{tab:mc} lists the average and standard deviation $\sigma$ along with the median and the lower and upper semi-interquartile ranges (SIQR) of the parameter distributions (under the restriction $1 \le F \le F(3\sigma)$; there is practically no difference with $1 \le F \le F(2\sigma)$\ save for one case identified in the table).  Uncertainties in the \hbbc\ parameters (FWHM and intensity) are modest, around 10\%\ for the A3n0 case, and 5\%\ for  A3b0. Continuum placement and shape are also well-defined  within uncertainties of a few percent.  The difference between the lower and upper SIQR indicates that in some cases the parameter distribution is skewed.   Significant uncertainties are associated with the measurements of the \oiii\ NC and SB (we have chosen the composites built for  \oiii/\hbnc$\lesssim 1$), but these are nonetheless usually within $30$\%\ (Table \ref{tab:mc}). These components become detectable if they are $\approx$ 5\%\ of the \hbbc\ intensity. Errors in fluxes and detection limits are then expected to scale with the inverse of the square root of the S/N:  $\delta I \sim \sqrt{20/\mathrm{(S/N)}}\delta I_{20}$, where $\delta I_{20}$ are the uncertainties for S/N$\approx$20\  in Table \ref{tab:mc}. 

The \hb\ BLUE component shows a large range of intensities within the spectral bin; this means that a median composite is unlikely to be representative of the sample. From the simulations for the  A3b0 bin with S/N$\approx$20 we derive that \hb\ BLUE with a strength approximately equal to one tenth of  \hbbc\ should be detectable at a $2 \sigma$\ confidence level.

\section{Results}
\label{results}

\begin{figure}
\includegraphics[scale=0.27]{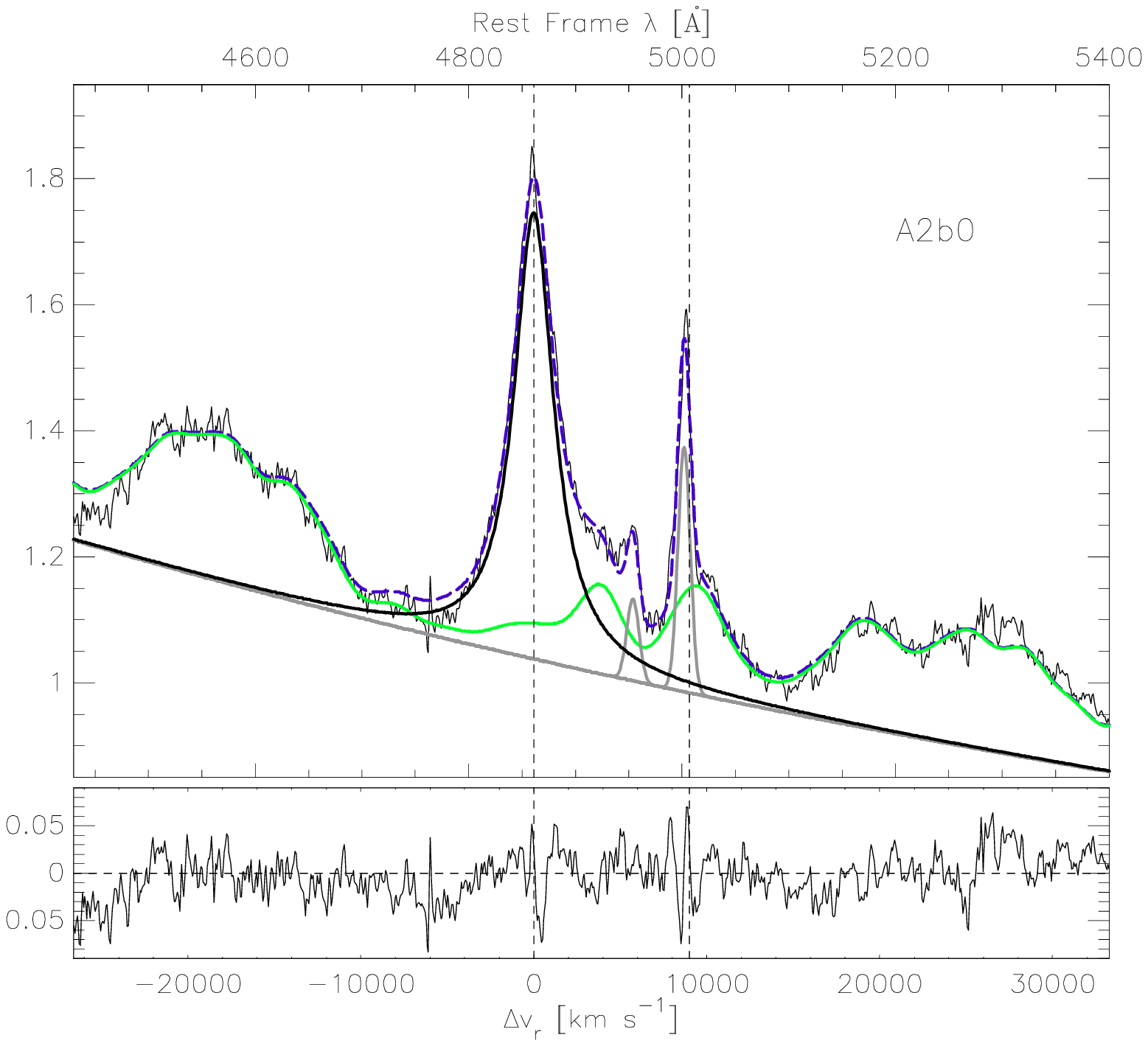}
\includegraphics[scale=0.27]{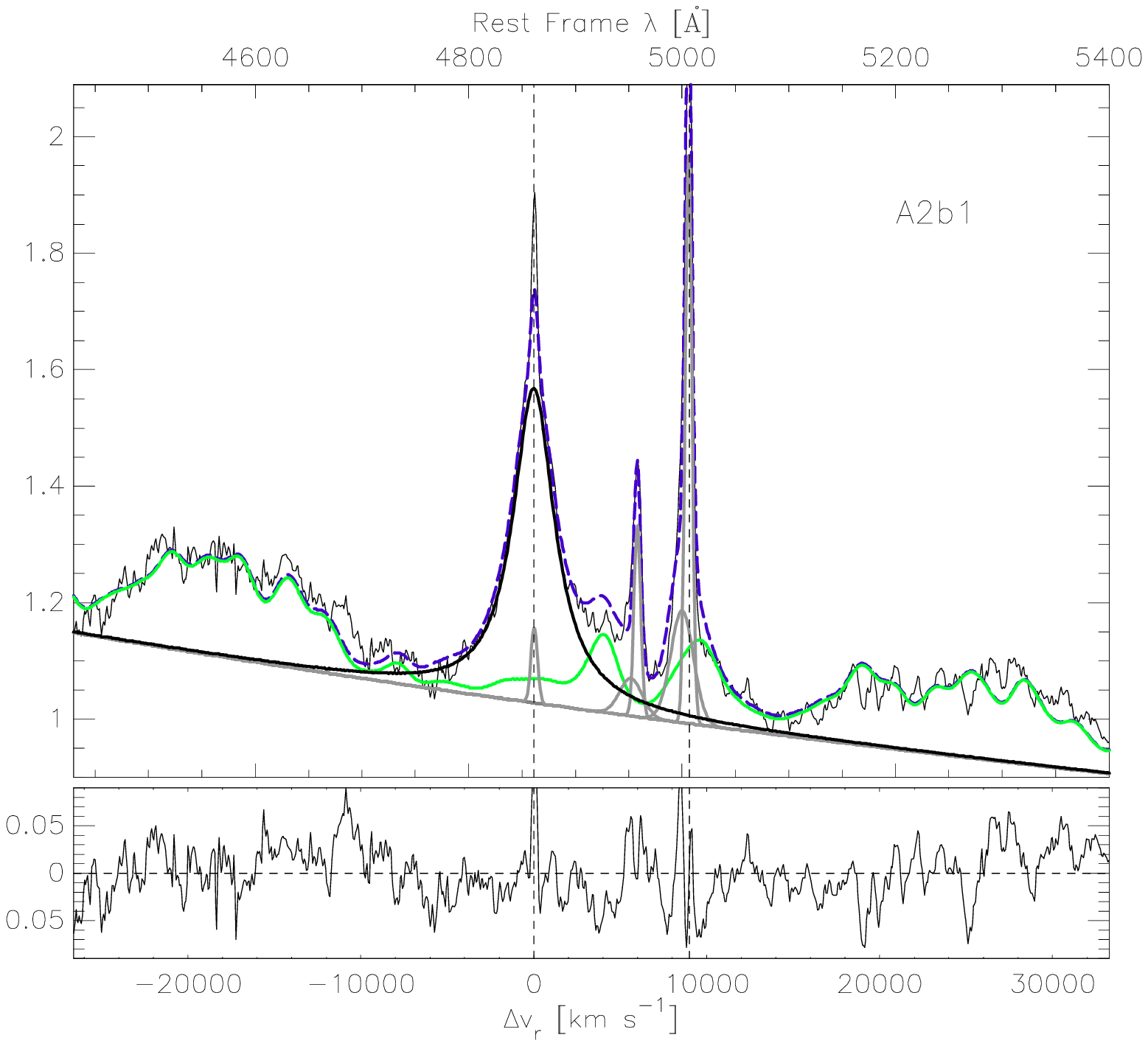}

\includegraphics[scale=0.27]{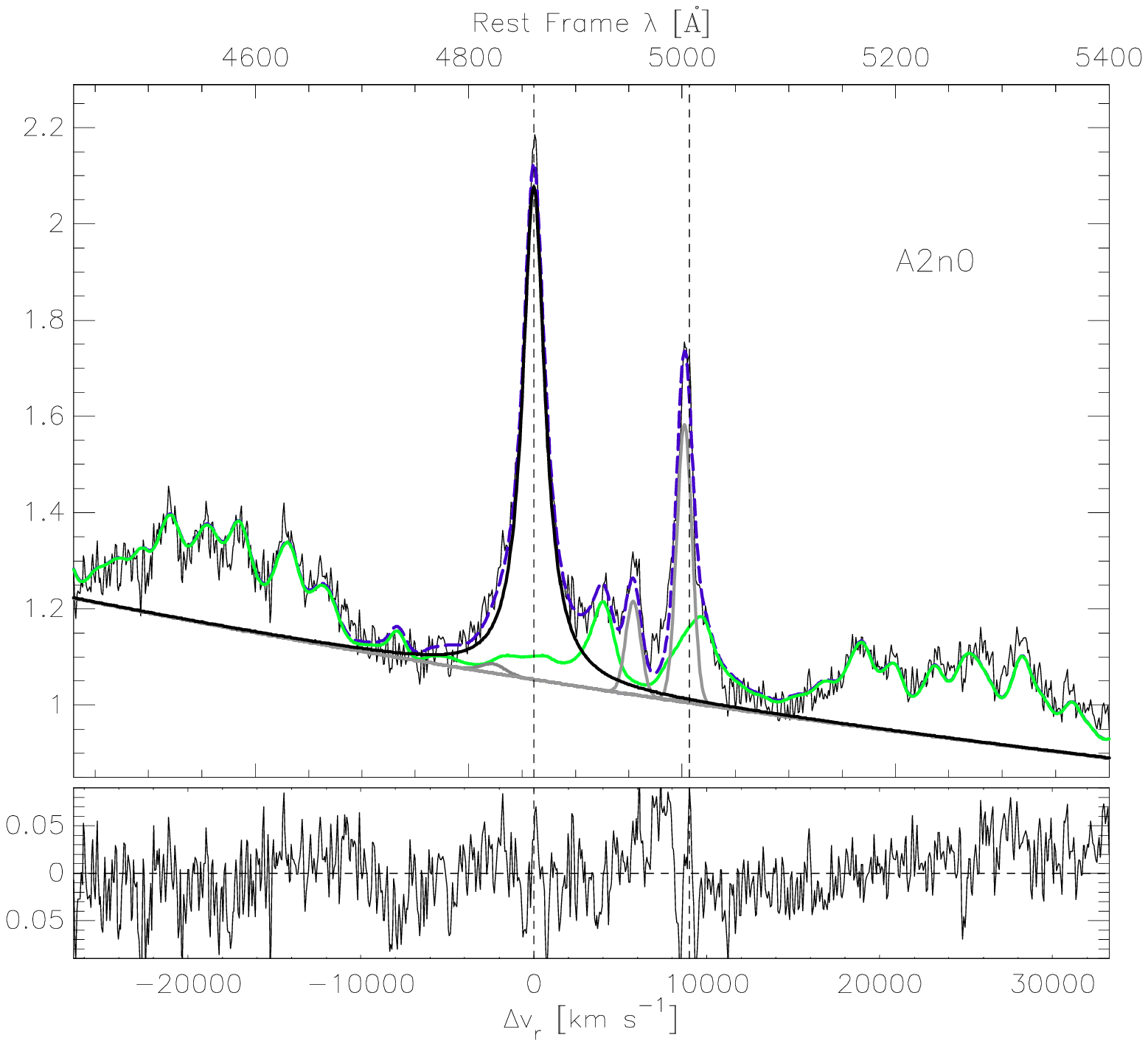}
\includegraphics[scale=0.27]{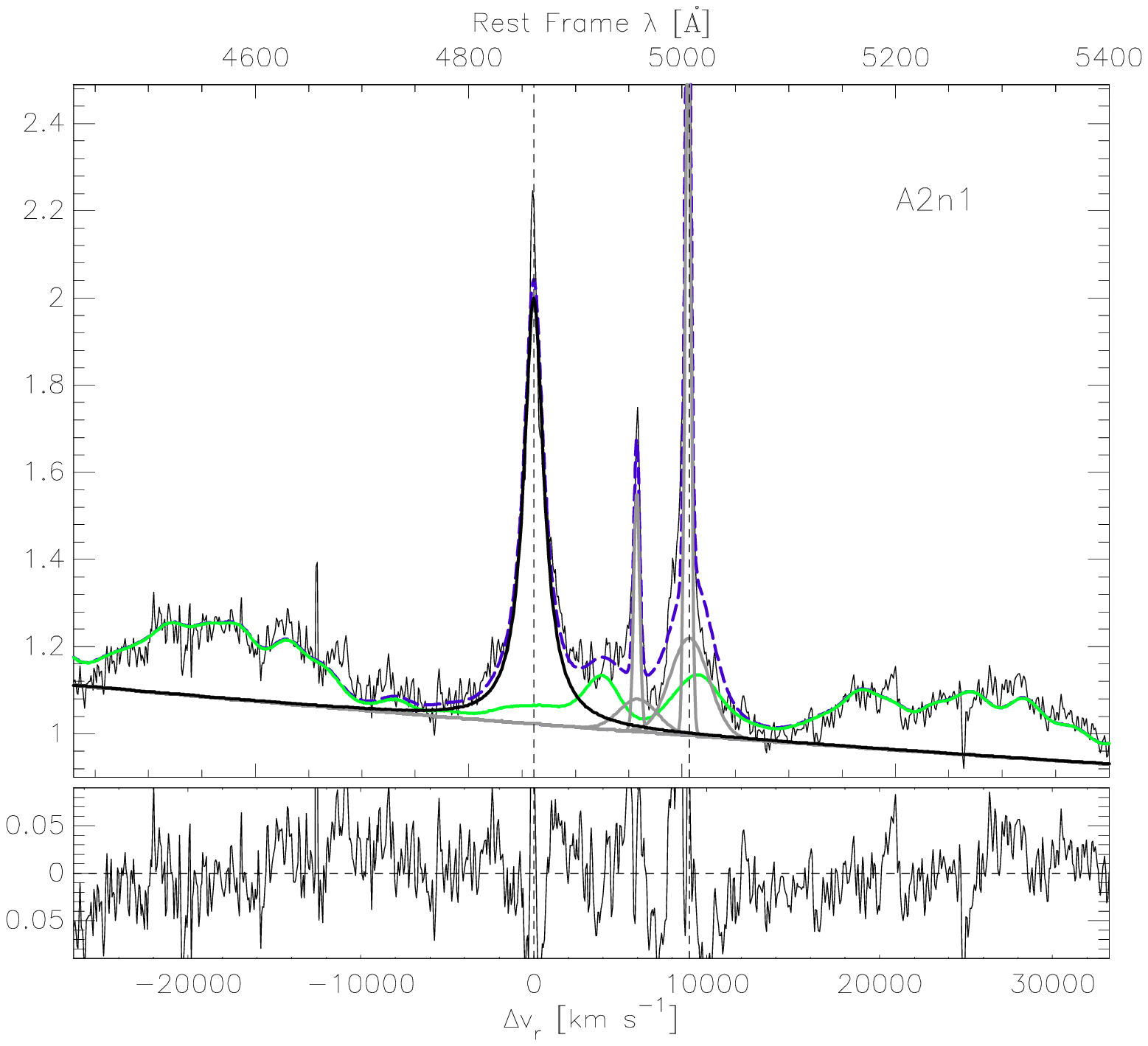}
\caption{Fits of the spectral type composite spectrum A2 for four spectral sub-types: A2b0: broad line and weak \oiii; A2b1: broad line and strong \oiii; A2n0: narrow line and weak \oiii; A2n1: narrow line and strong \oiii.  The original spectrum is shown superimposed to the adopted continuum (thick black line) and to \feii\ emission (thin green line). The \hbbc\ fitting functions are shown in black and gray (the blue shifted component). The long-dashed purple lines are the fits. Dashed vertical lines identify the rest-frame wavelengths of \hb\ and \oiii. The lower panels show the residuals to the fits. 
\label{fig:fitsA2} }
\end{figure}

\begin{figure}
\includegraphics[scale=0.27]{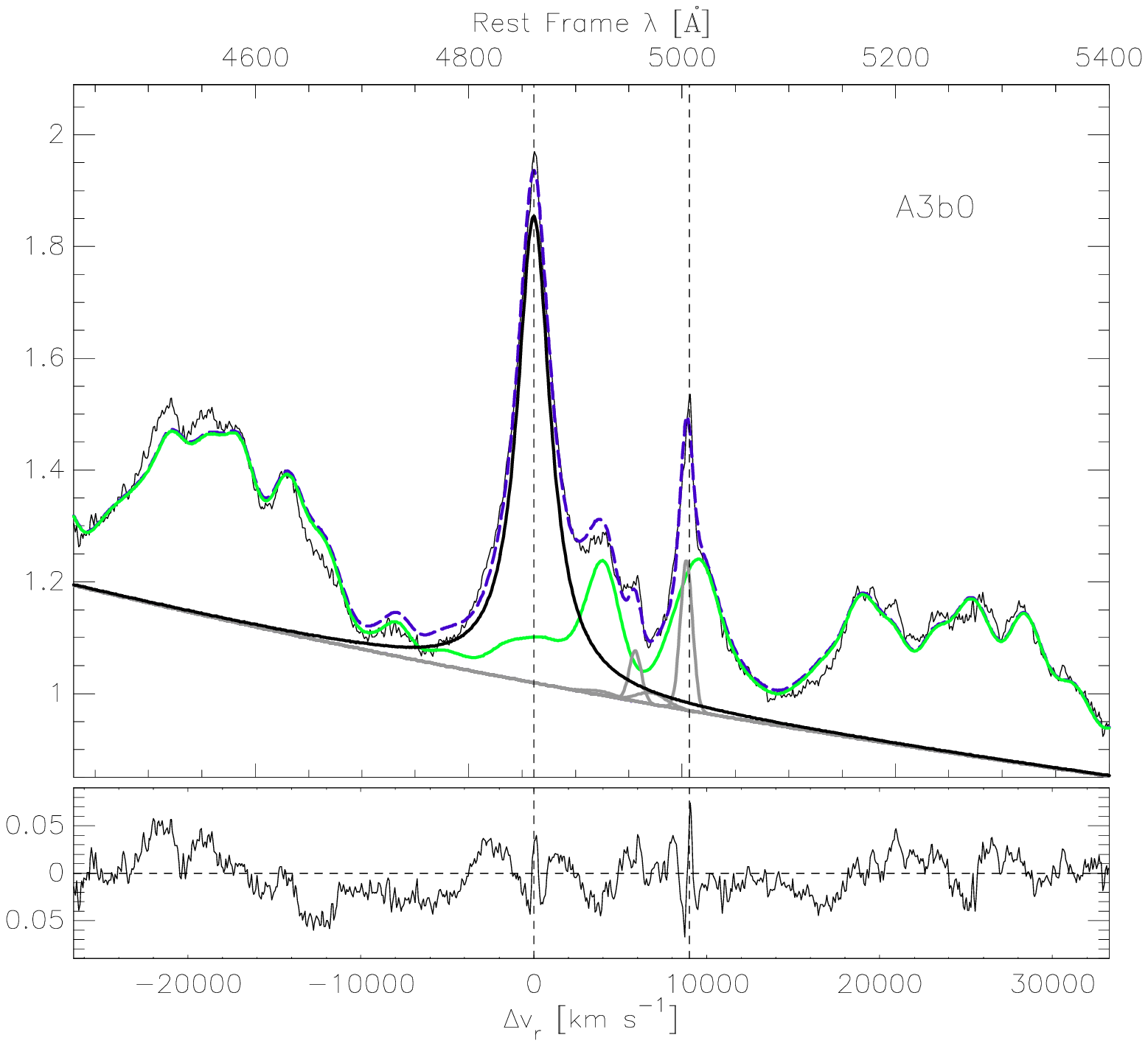}
\includegraphics[scale=0.27]{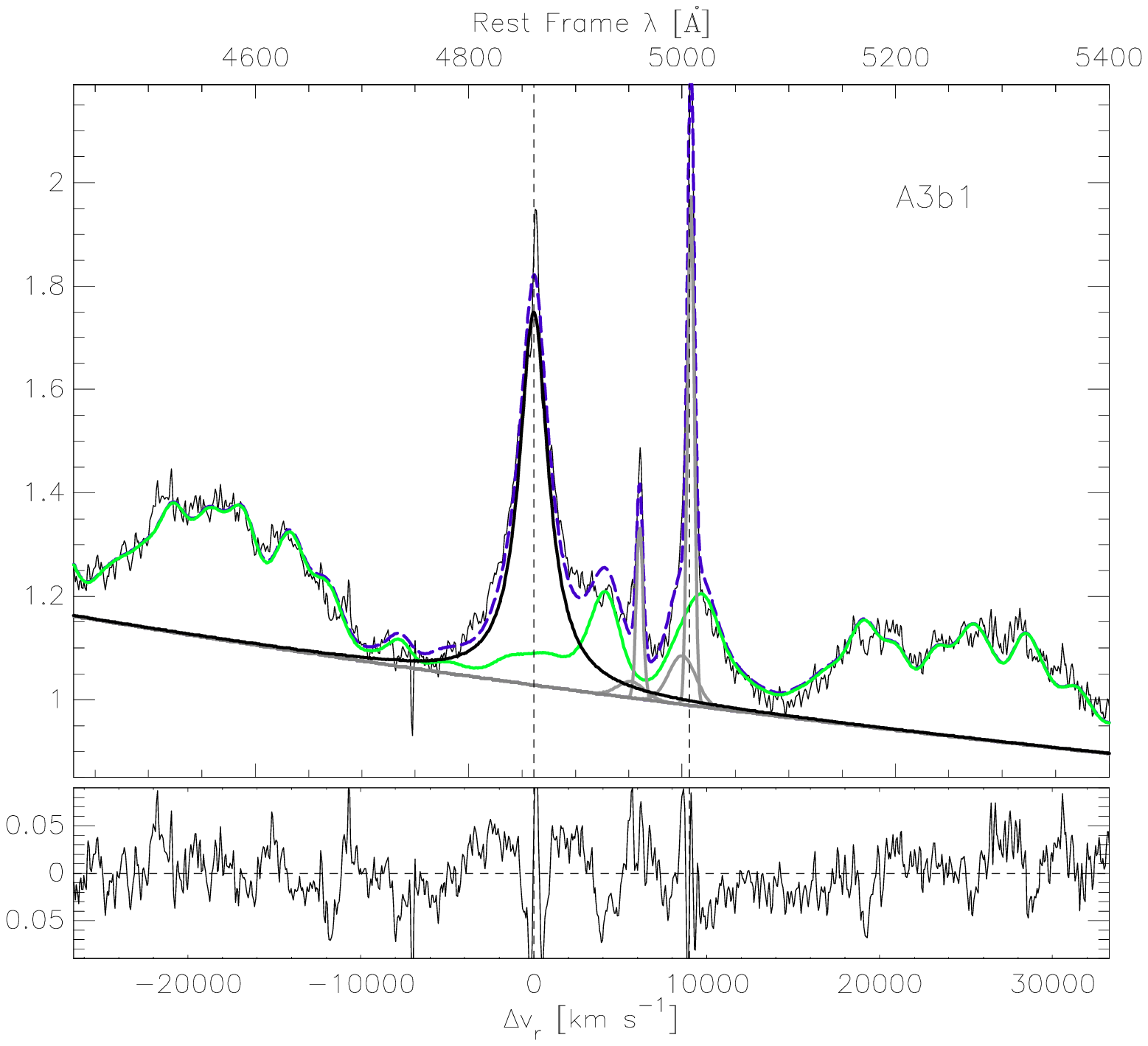}

\includegraphics[scale=0.27]{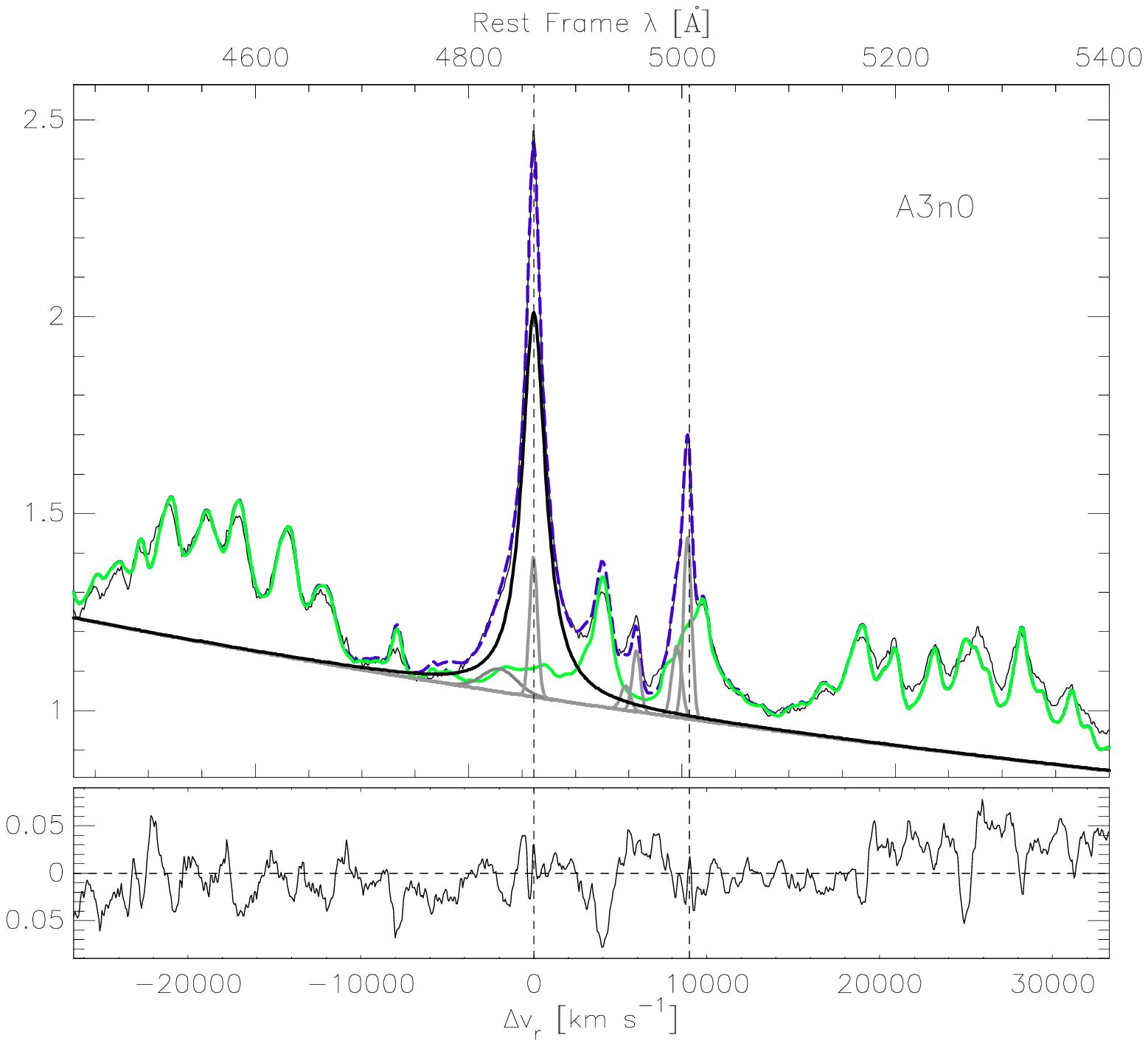}
\includegraphics[scale=0.27]{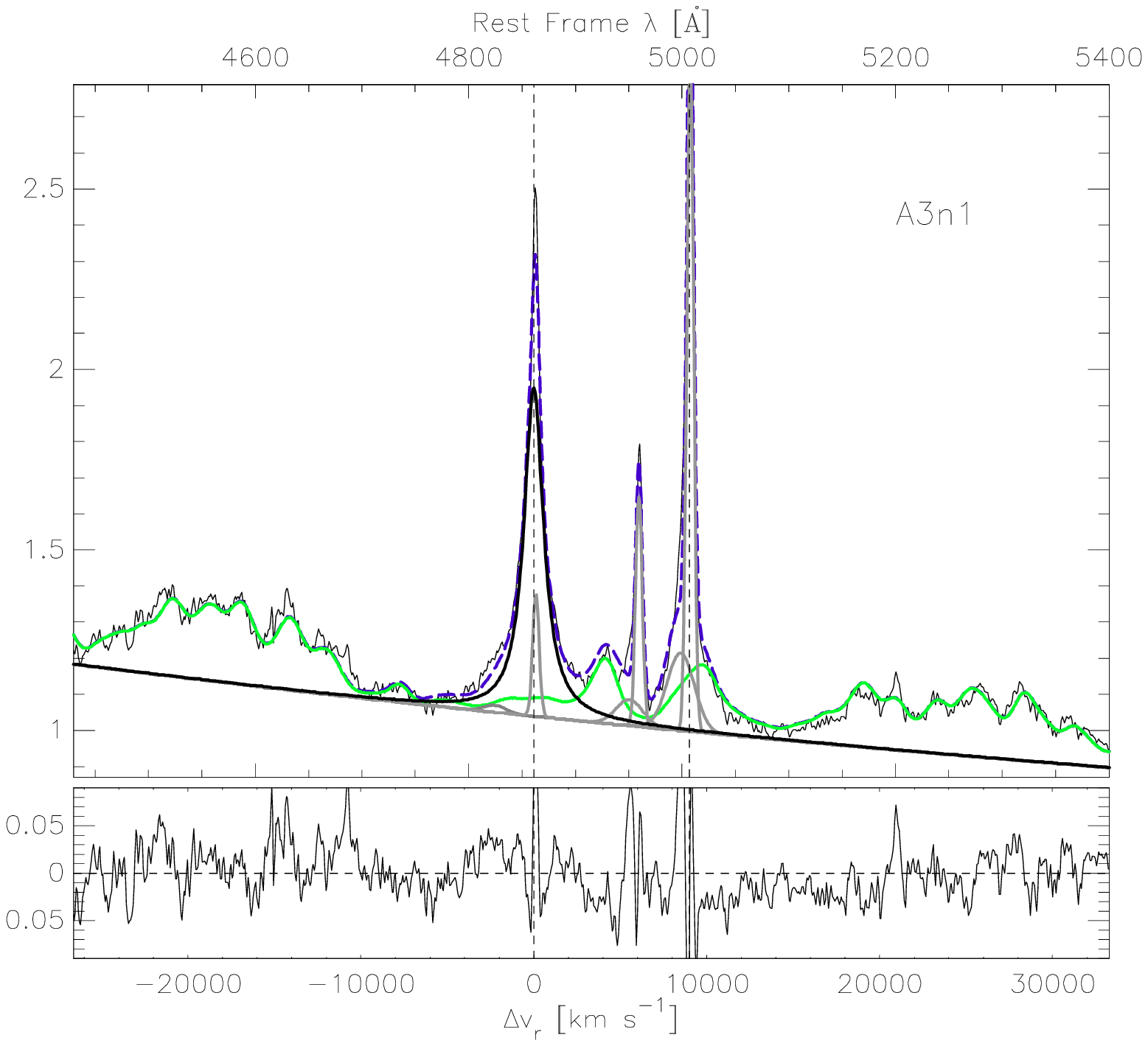}
\caption{As in Fig. \ref{fig:fitsA2}, but for spectral type A3.
\label{fig:fitsA3} }
\end{figure}

\begin{figure}
\includegraphics[scale=0.27]{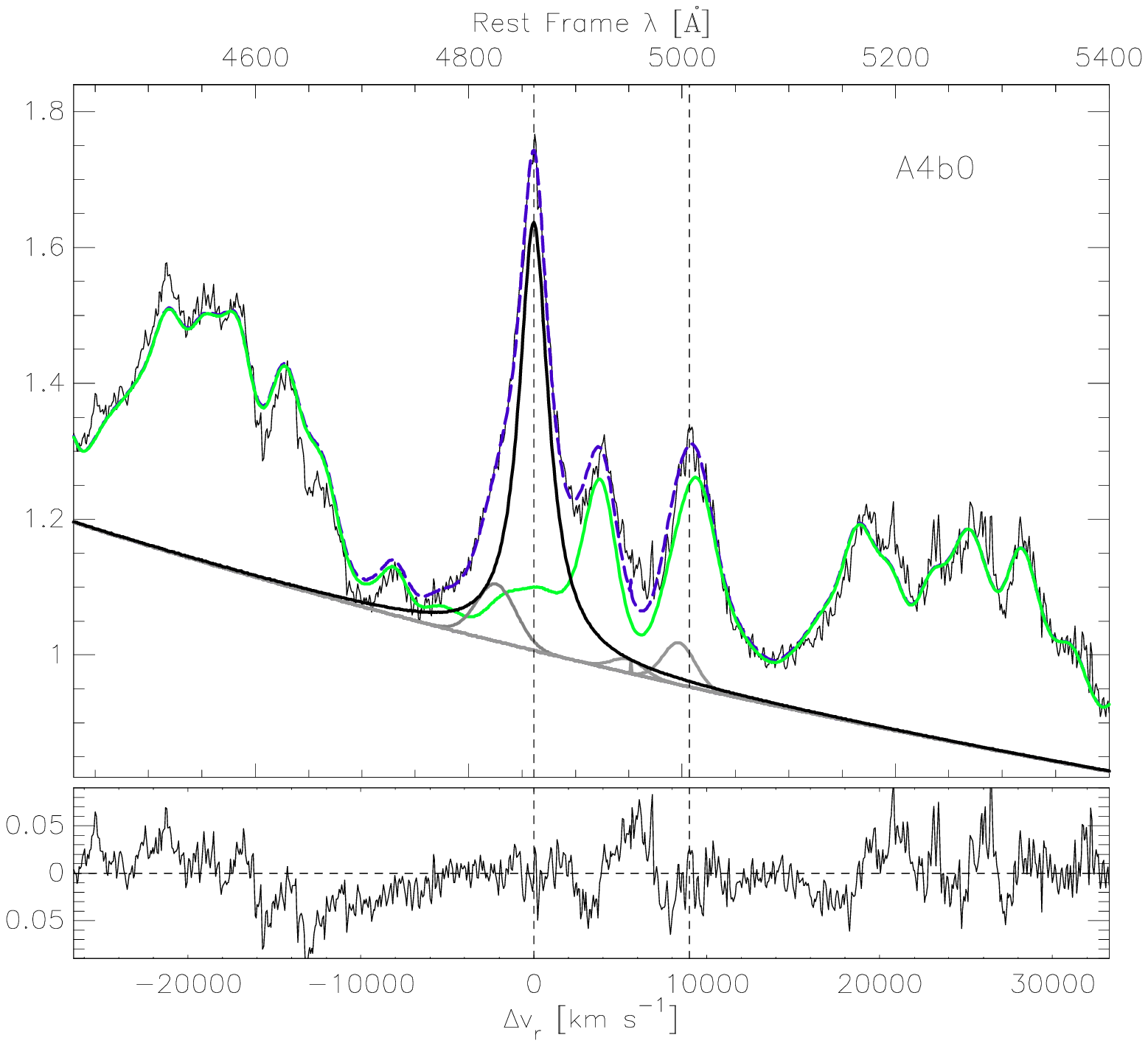}
\includegraphics[scale=0.27]{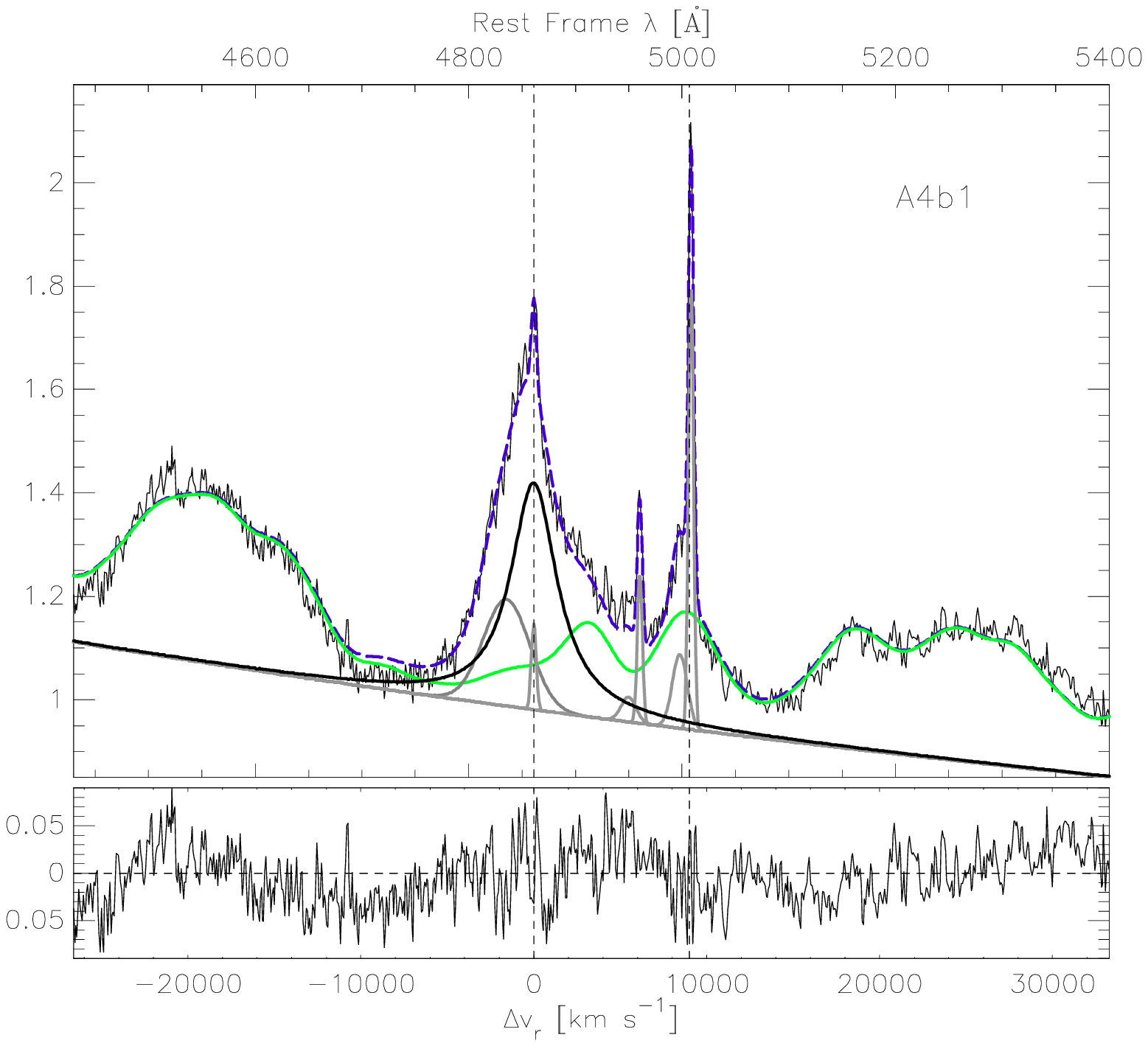}

\includegraphics[scale=0.27]{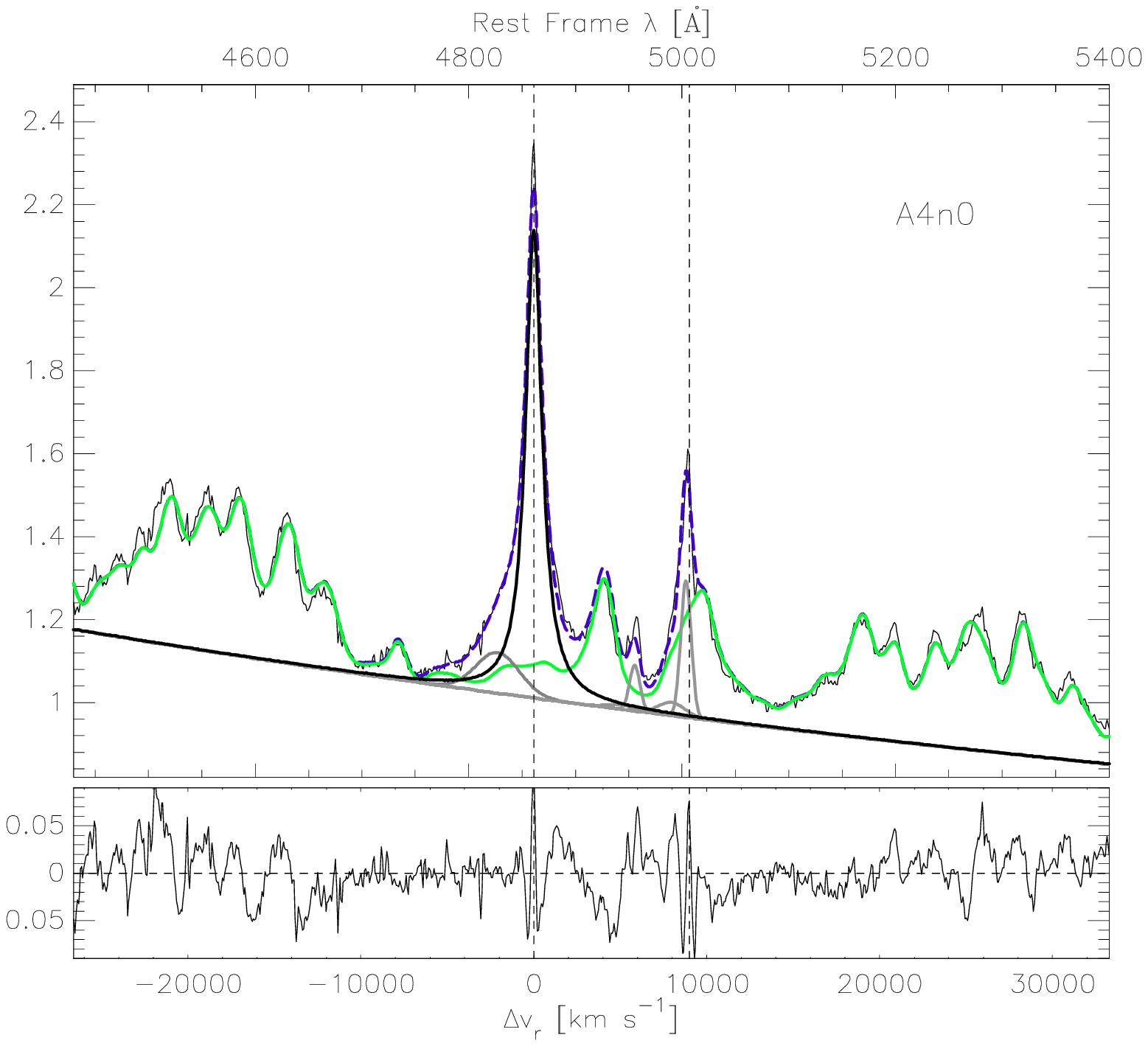}
\includegraphics[scale=0.27]{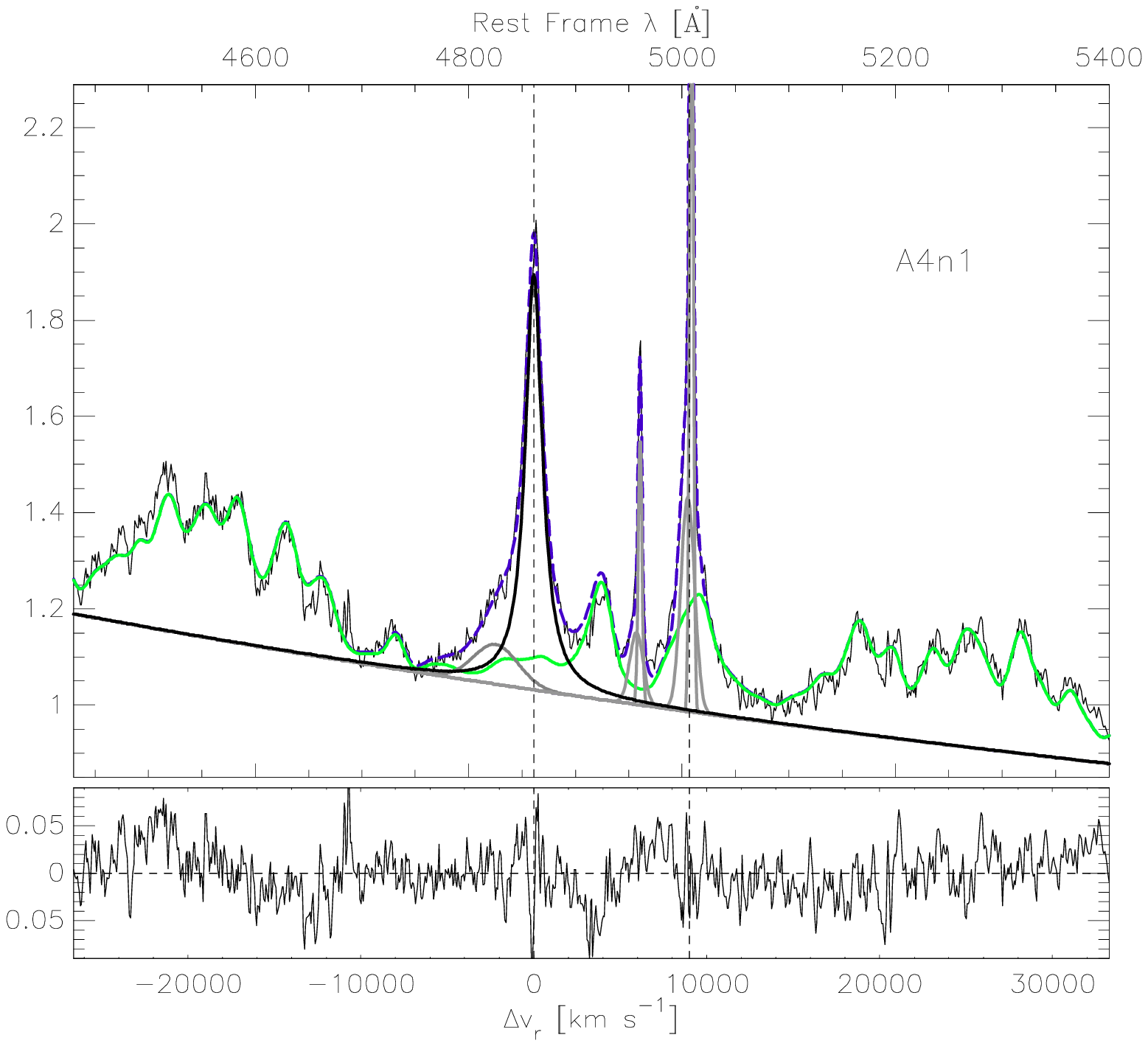}
\caption{As in Fig. \ref{fig:fitsA2}, but for spectral type A4.
\label{fig:fitsA4} }
\end{figure}

\begin{figure}
\includegraphics[scale=0.27]{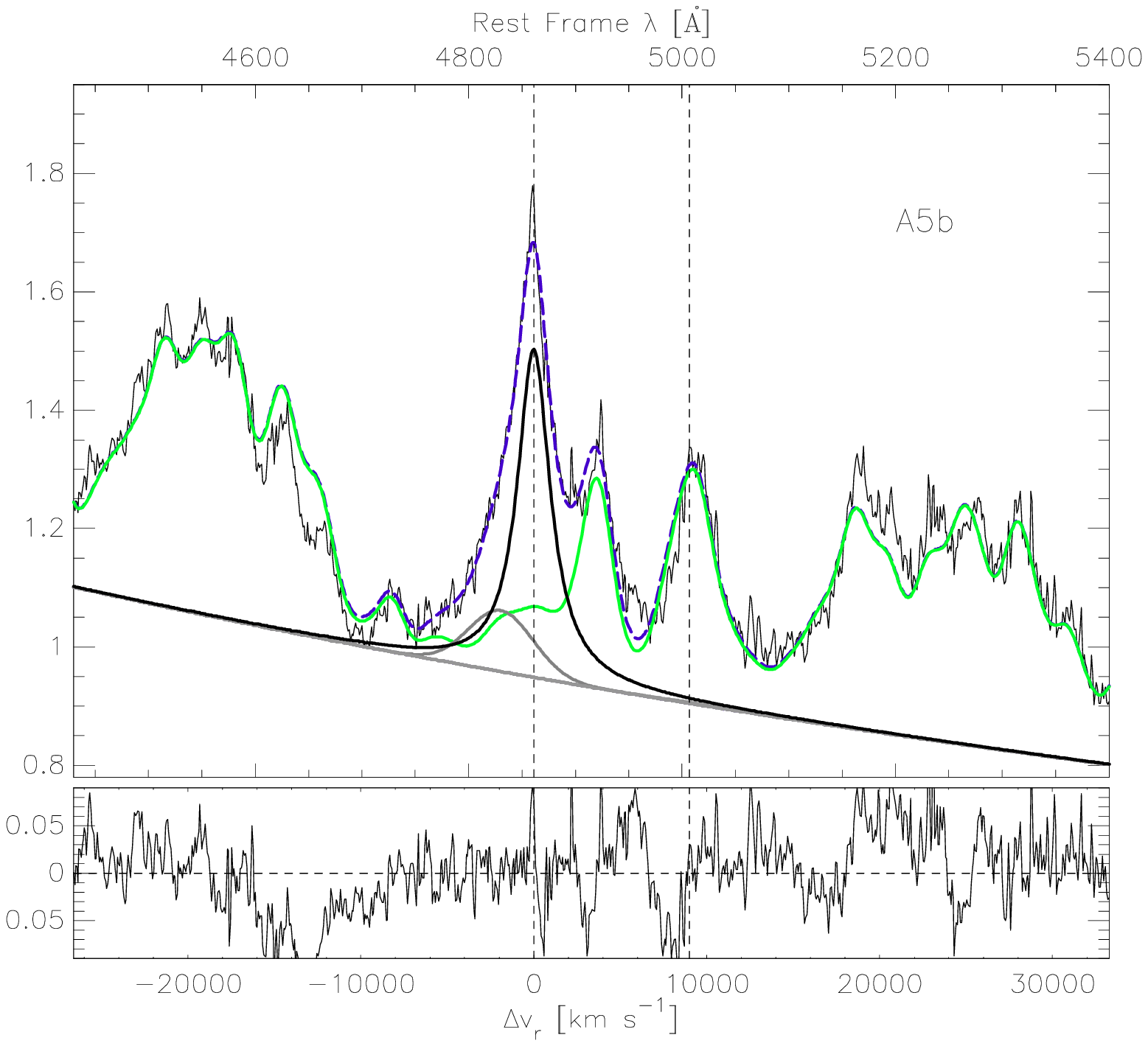}
\includegraphics[scale=0.27]{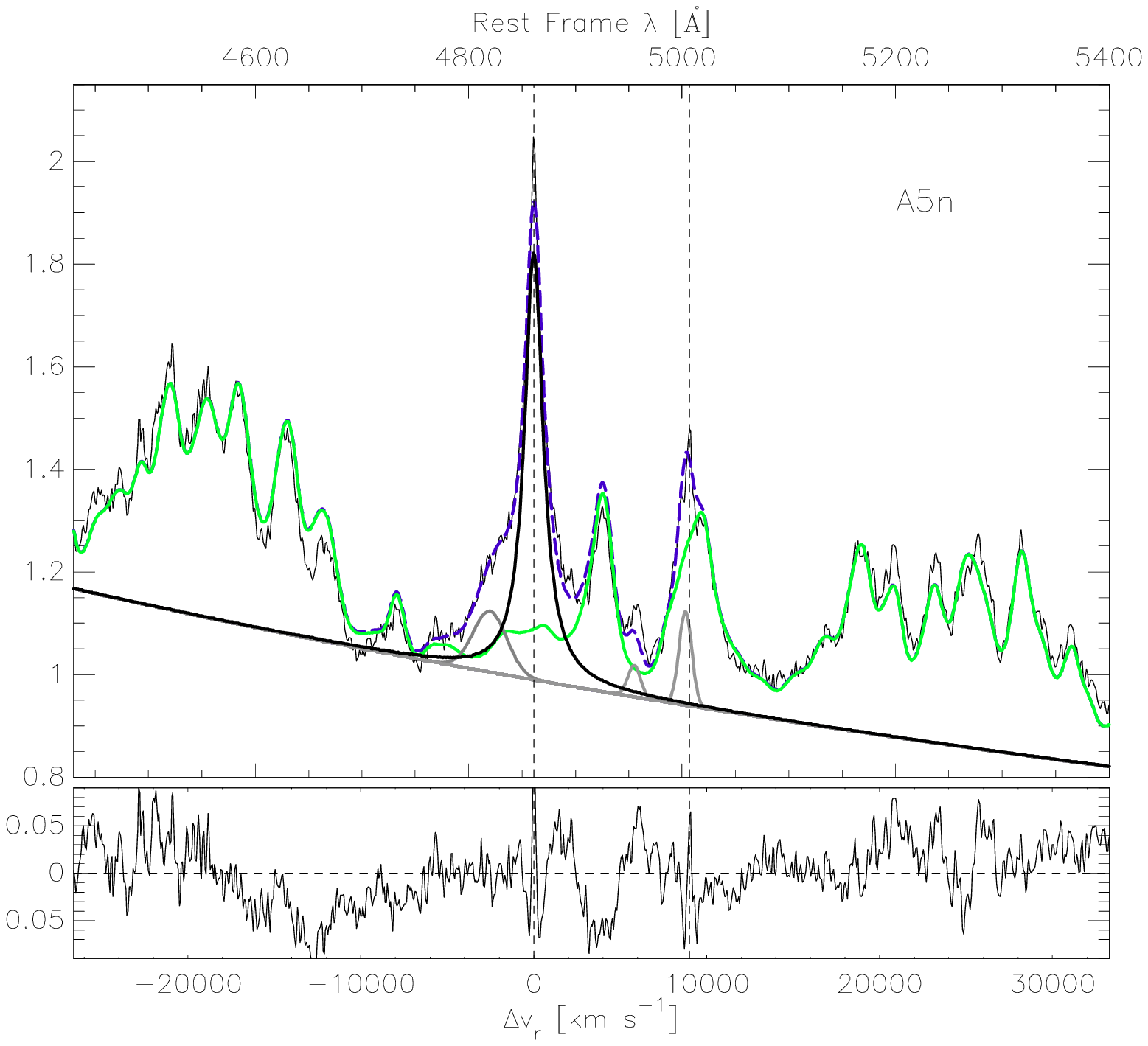}

\includegraphics[scale=0.27]{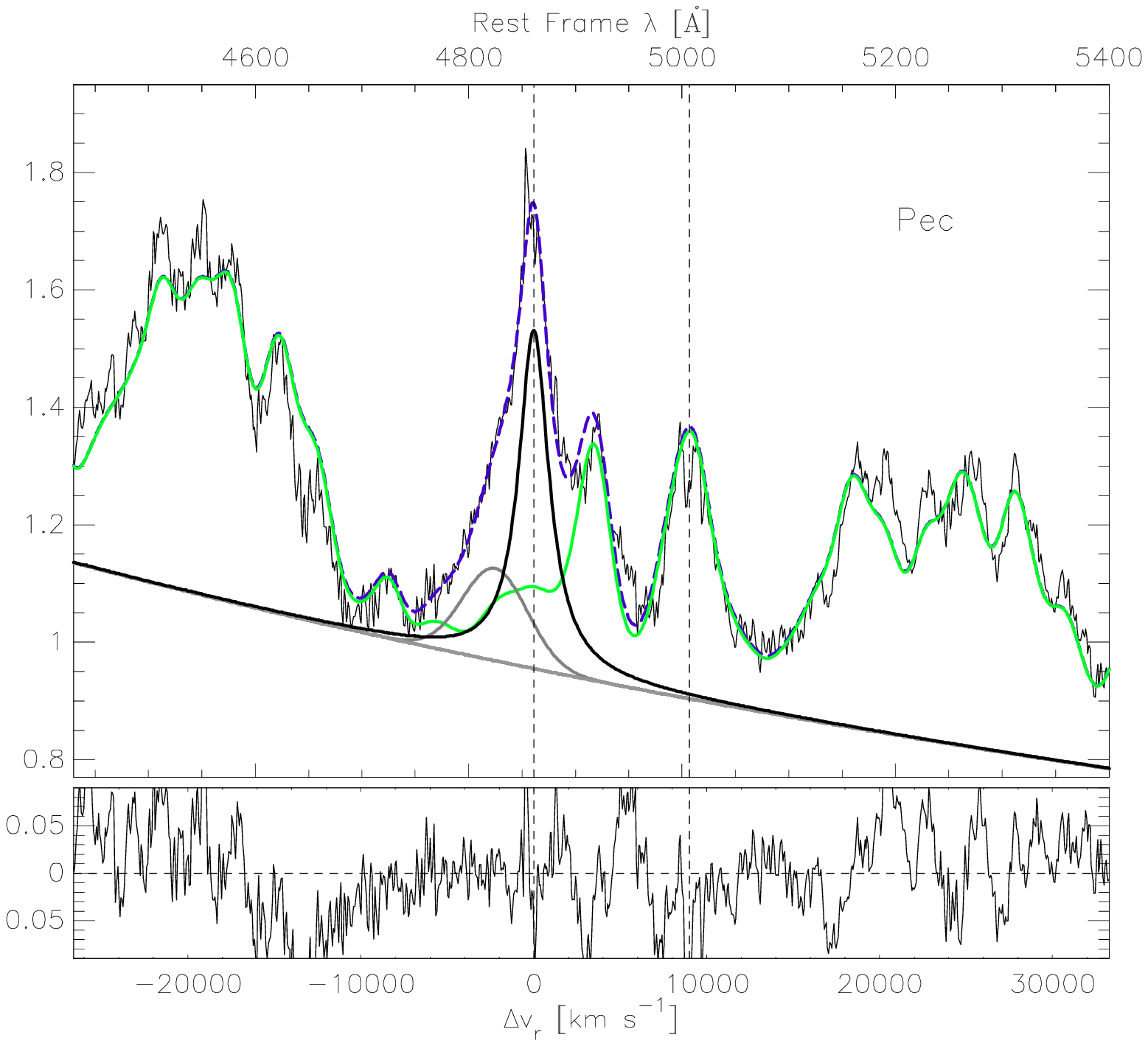}
\caption{As in Fig. \ref{fig:fitsA2}, but for spectral type A5 and peculiar objects.
\label{fig:fitsA5} }
\end{figure}

\subsection{Systematic line profile change as a function of FWHM}

\subsubsection{Composite spectra: qualitative trends}
\label{sec:composite}

Composite spectra were built from averages in four sub-bins that split the original spectral types of \citet{sulenticetal02} in four: narrower ``n'' (lower 2000 \kms\ range in FWHM \hb), broader ``b'' (upper 2000 \kms\ range in FWHM \hb), weak ``0'' and strong ``1'' \oiiiopt. Table \ref{tab:sampleprop} presents the properties of the composite spectra. As described in Sect. \ref{sec:method}, column (1) gives the Population and the ST assigned. In this column, the labels Pec and NA are for ``Peculiar'' and ``Not Assigned'' objects. Peculiar objects are defined as those with an unusual \rfe\ \gt\ 2, while Not Assigned objects belong to bins B3 and B4; most of these, however, show a host galaxy contribution. On the other hand, column (2) gives the total number of objects for each sub-bin. When the total number is $\lesssim$ 5 objects we used only the highest-quality spectra. For this reason we give the total number of objects in the bin in parentheses, while the number of objects that we use for the composite spectra is out of parentheses. We note that for two cases we use only a representative spectrum. Additionally, column (10) now gives the S/N for the composite spectra.

Figures \ref{fig:fitsA2}, \ref{fig:fitsA3}, and \ref{fig:fitsA4} show the composite spectra in the region of \hb, along with the principal components included in the {\em specfit } analysis for spectral types A2, A3 and A4, respectively.  
The similarity of the sub-bin spectra is striking: \feii\ emission is consistently fit by a scaled and broadened template, and Lorentzian functions provide  satisfactory fits in all cases. This is also true for the A2 spectral type, even if  A2 sources do not satisfy the criteria of xA. Progressing from A2 to A5 (Fig. \ref{fig:fitsA5}), there is  evidence of a growing relevance of excess blueshifted emission for both \hb\ and \oiiia. In A2, the \hb\ line is fit by symmetric profiles, while for bins A4 and A5 (Fig. \ref{fig:fitsA5}, where A5 are the objects with 2 < \rfe < 3, see column 13 of Table \ref{tab:sampleprop}), there is an obvious emission hump on the blue side of \hb, the hump being most prominent in spectral type A5. We note  that the interpretation of the A5 composite taken out of the MS context would be rather ambiguous, as there is no strong evidence of an actual Lorentzian-like shape (we return to this issue in Sect.  \ref{sed:blend}).
In line with the argument of a continuous sequence, we adopt the interpretation Lorentzian + BLUE that is consistent with the previous spectral types. 

The semi-broad blueshifted component of \oiii\ is also increasing in prominence, and is stronger in the case of the ``1'' composites with  strong \oiiiopt\ emission. However, in the most extreme \feii\ emitters, \oiiiopt\ almost disappears, in line with the trend seen for the ``0'' composites from A2 toward stronger \feii\ emitters: the ``0'' composites for A2 still show significant \oiiia\ narrow component which becomes much fainter in A3, barely detectable in A4, and absent in A5. 

\subsubsection{NLSy1s as part of Population A}
\label{sed:blend}

The sub-spectral types involving the narrower half of the FWHM range of each spectral type (label ``n'') in Figs. \ref{fig:fitsA2}, \ref{fig:fitsA3}, \ref{fig:fitsA4}, and \ref{fig:fitsA5} show that the \hb\ profile is well-fit by a Loretzian function. The Lorentzian-like profiles are also appropriate for the spectral sub-type including sources with 2000 \kms $<$ FWHM(\hb) $<$ 4000 \kms\ (label ``b'' in Figs. \ref{fig:fitsA2}, \ref{fig:fitsA3}, \ref{fig:fitsA4}, and \ref{fig:fitsA5}), in agreement with early and more recent  results  \citep{veroncettyetal01,sulenticetal02,craccoetal16}. There is no evidence of discontinuous properties  corresponding to the 2000 \kms\ FWHM limit defining NLSy1s.

\begin{figure}
\includegraphics[scale=0.27]{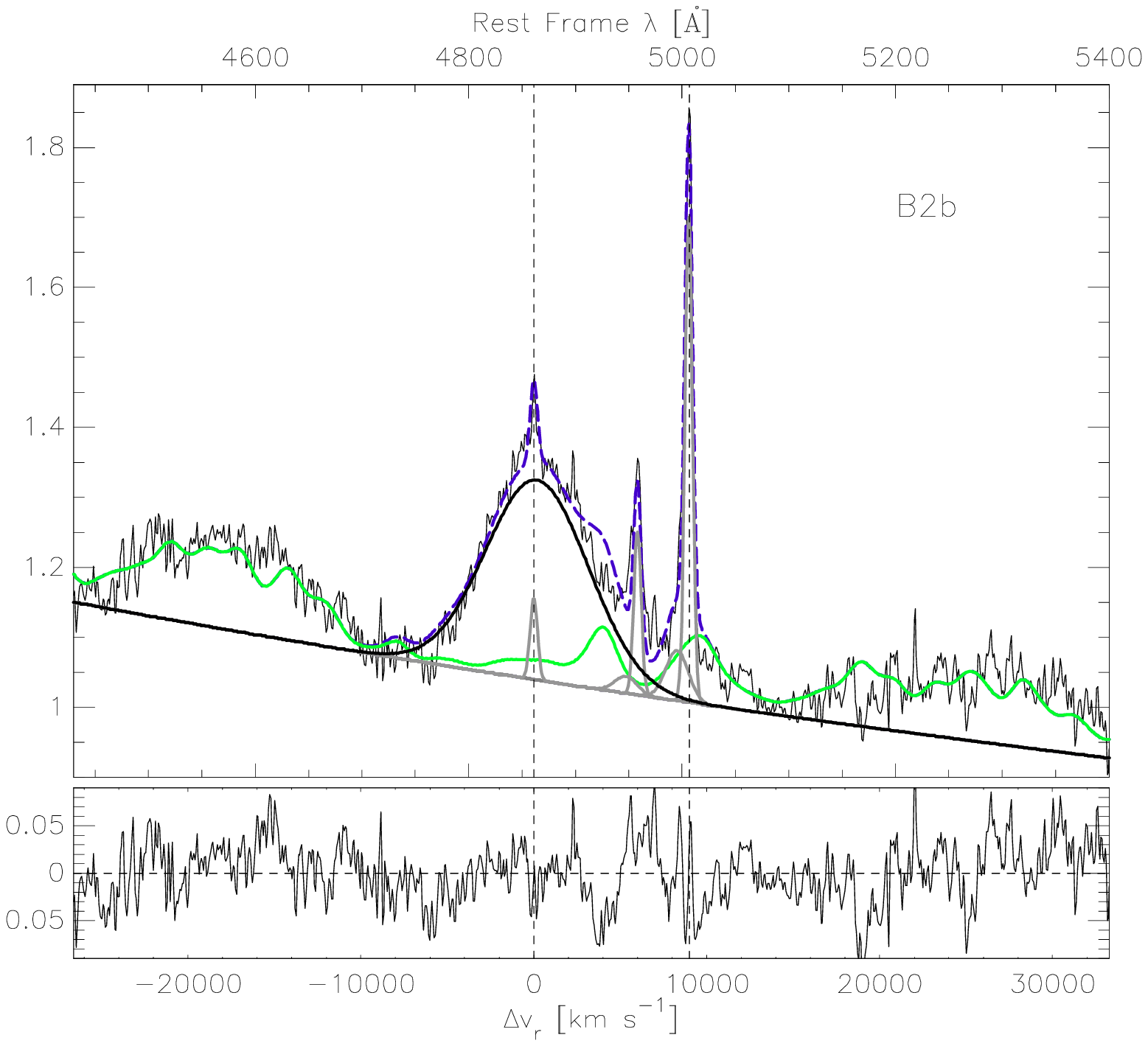}
\includegraphics[scale=0.27]{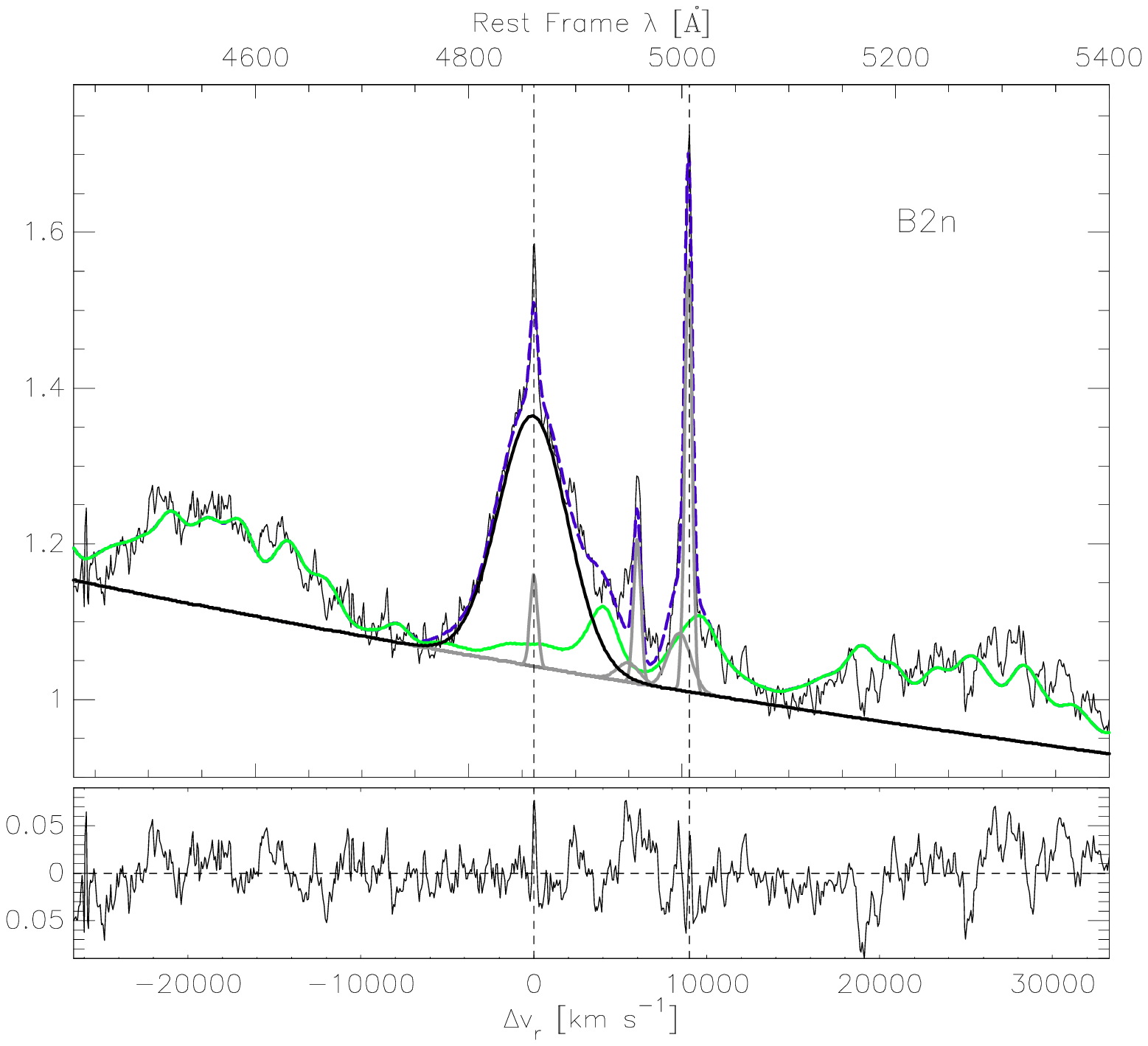}

\includegraphics[scale=0.27]{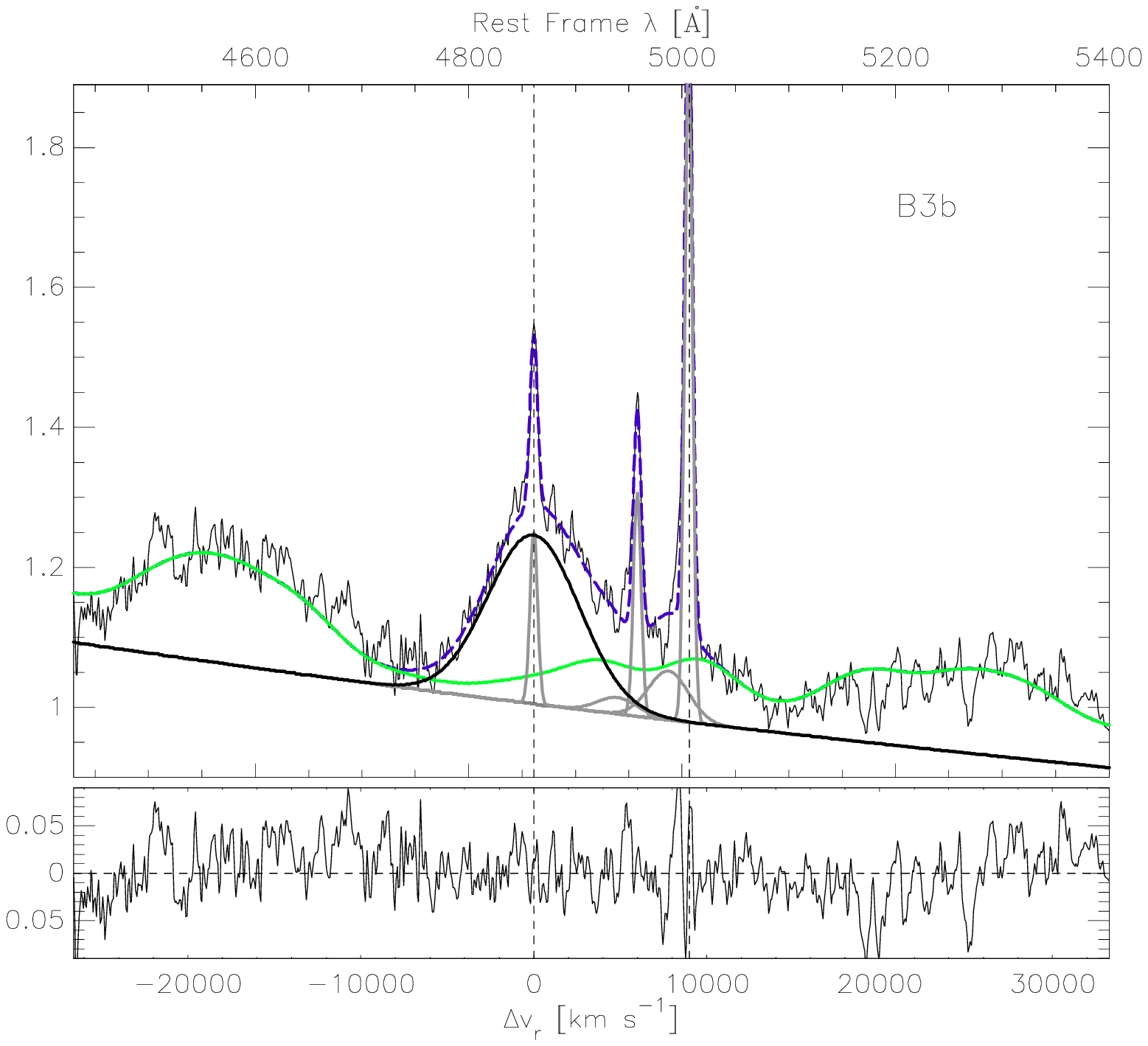}
\includegraphics[scale=0.27]{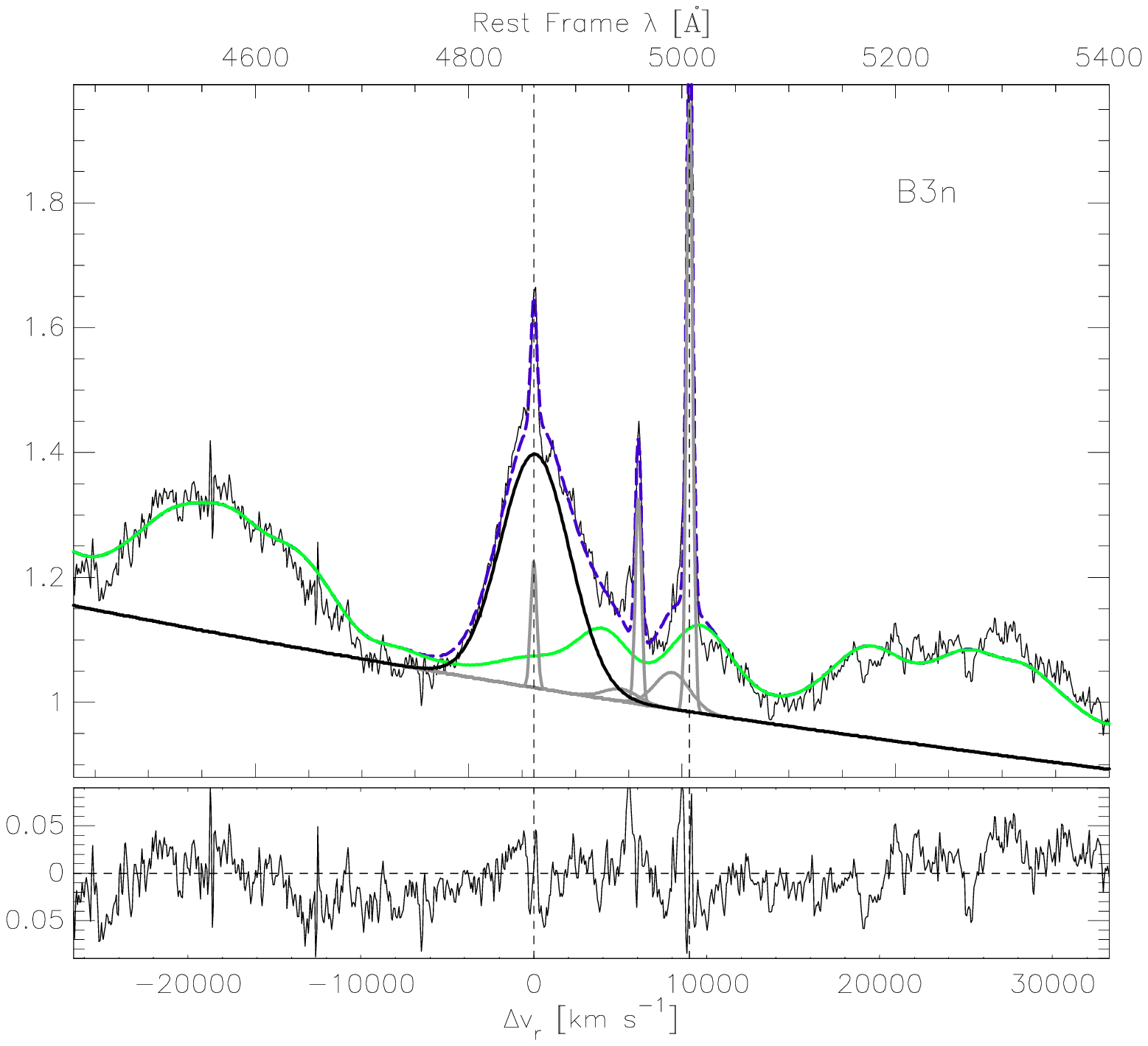}
\caption{As in Fig. \ref{fig:fitsA2}, but for Pop. B. An individual Gaussian is fit in place of the  Lorentzian  used for Pop. A.
\label{fig:fitsB} }
\end{figure}

\subsubsection{Population B: still xA sources?}

Figure \ref{fig:fitsB} shows that in regards to the composite for Pop. B (spectral types B2 and B3) the best fit can be achieved with a Gaussian function. In this figure, `n' signifies a FWHM(\hbbc) between 4000 and 6000 \kms, and `b' a FWHM(\hbbc) between 6000 and 8000 \kms. Also, we noticed that 90\% of the individual spectra show strong \oiiiopt\ emission.

The difference in the line profile of A and B spectral types is striking, and especially so if the A2 and A3 composites are compared to B2 and B3: in A2, where the moderate  \feii\ emission allows for an easier visual evaluation of the \hb\ profile,   a Lorentzian shape is clearly indicated. 

A key aspect is the presence of a redward asymmetry in the \hb\ line profiles which is a defining feature of Pop. B. In the present context, the lines are modeled with a Gaussian core component (the \hbbc) with no (or small) peak shift with respect  to rest frame, plus no additional redshifted very broad component: no VBC is needed to achieve a satisfactory fit. This may have gone unnoticed because of the rarity of B3 and B4 sources, as mentioned in Sect.  \ref{sec:automes}.

\subsection{Selection criteria: ``true'' extreme quasars}

\begin{figure}
\includegraphics[scale=0.4]{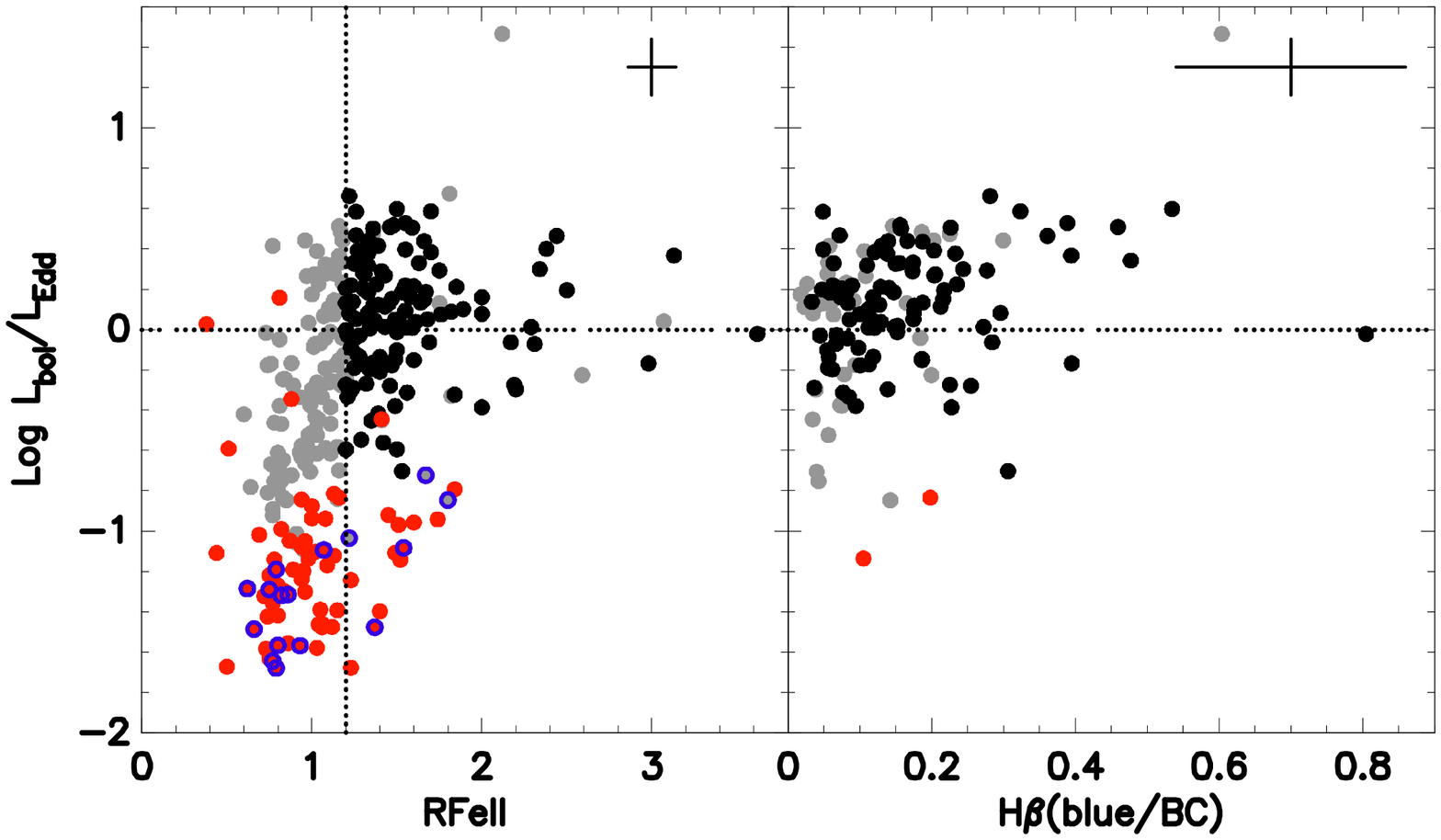}
\caption{Behavior of  Eddington ratio as a function of \rfe\ (left panel) and fraction of \hb\ BLUE intensity (right). The black crosses are the average errors. Meaning of symbols is as in Fig. \ref{fig:E1}.
\label{fig:BlueRels} }
\end{figure}

Figure \ref{fig:BlueRels} shows the dependence of \rfe\ and of the relative intensity of the \hb\ BLUE with respect to \hbbc\ as a function of \lledd.  There is a  dependence of \rfe\ on \lledd\ (the Pearson's correlation coefficient $r\approx$ 0.494 is significant at a confidence level $\approx1-10^{-17}$ for 302 objects) supporting the idea that the empirical selection criterion \rfe$\ge 1$ is indeed setting a lower limit on \lledd. Only a minority of the cosmo sample objects  show \lledd$\lesssim$1.  Population A sources below the limit \rfe$=1$ may include some xA sources placed there due to observational errors, or as true high radiators if the condition \rfe$>$1 is sufficient but not necessary:  some A2 and even A1 are xAs according to \citet{duetal14}. The main feature differentiating A2 from A3 is, apart from the \rfe\ value, the presence of a detectable \hb\ BLUE in A3, A4, and A5.  This feature is missing in the A2 composites and the right panel of Fig. \ref{fig:BlueRels} shows that \hb\ BLUE is  associated with  $\log$ \lledd $\gtrsim $ -0,2, and is detected in just a few sources in the rest of Pop. A outside of the cosmo sample, and only  in some cases in Pop. B. 

\subsection{The blended nature of the  \hb\ emission profile in xA sources}
\label{sec:blend}

\begin{figure}
\includegraphics[scale=0.34]{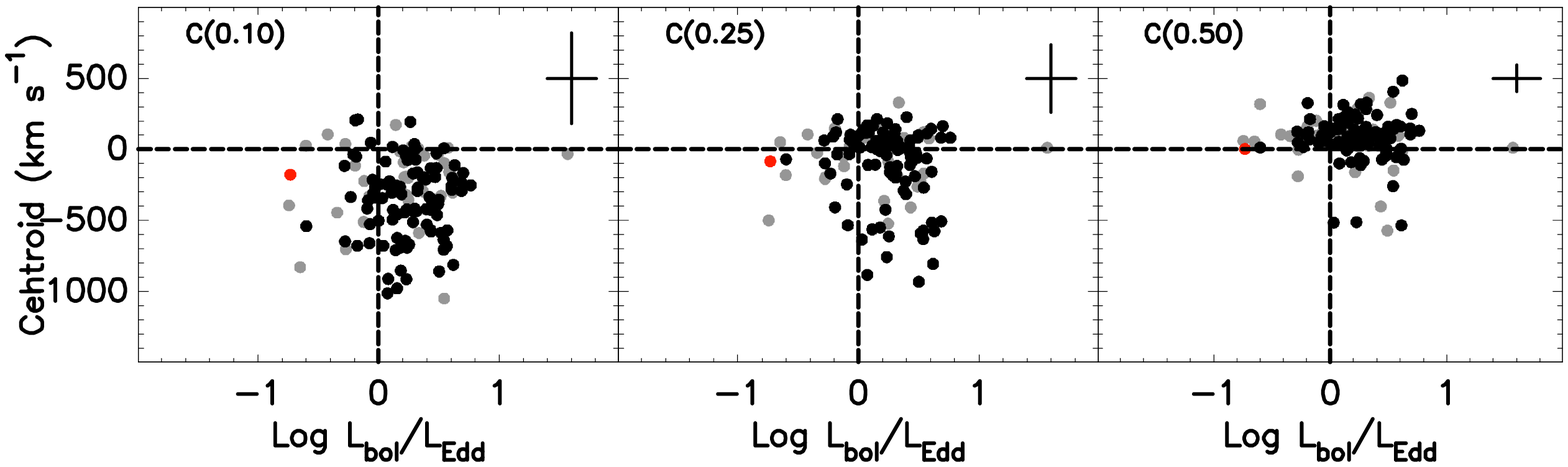}

\includegraphics[scale=0.34]{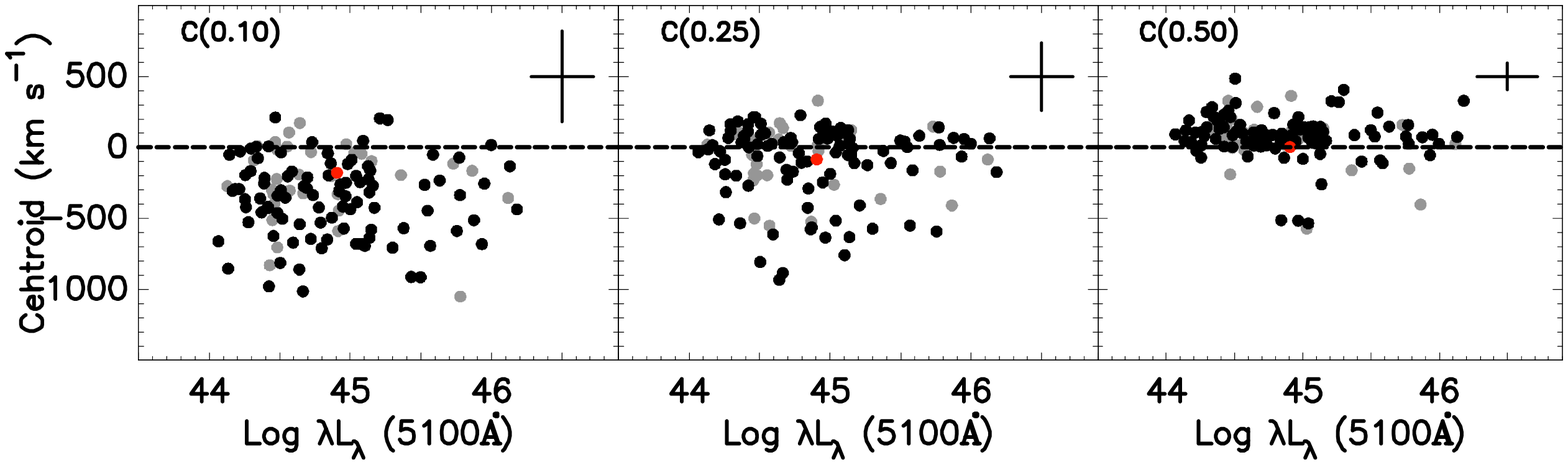}

\includegraphics[scale=0.34]{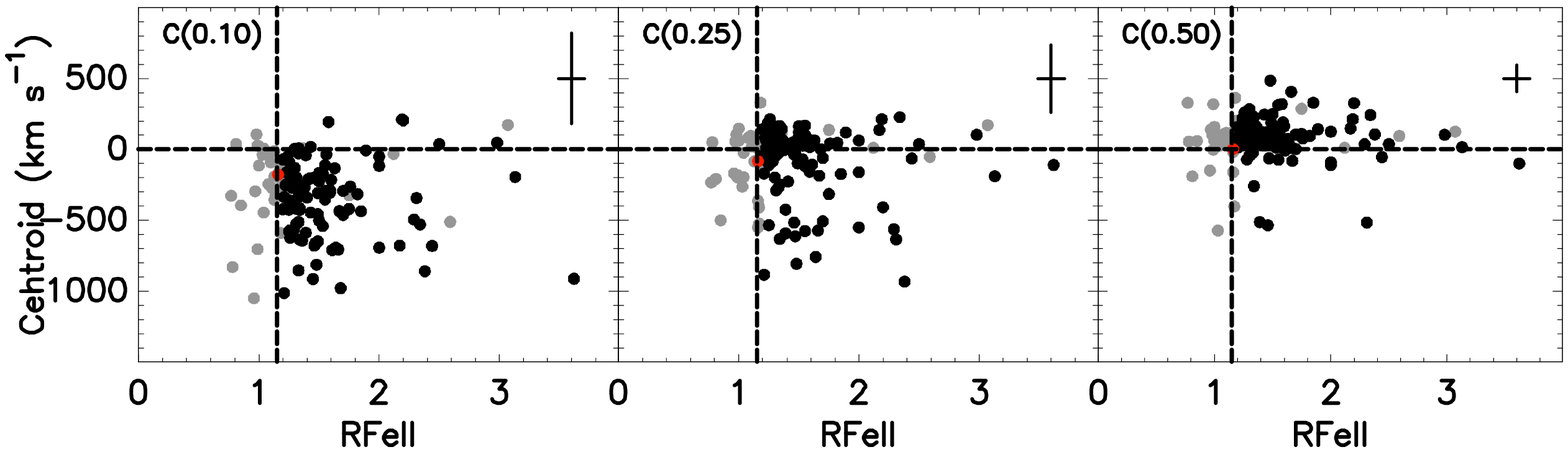}

\includegraphics[scale=0.34]{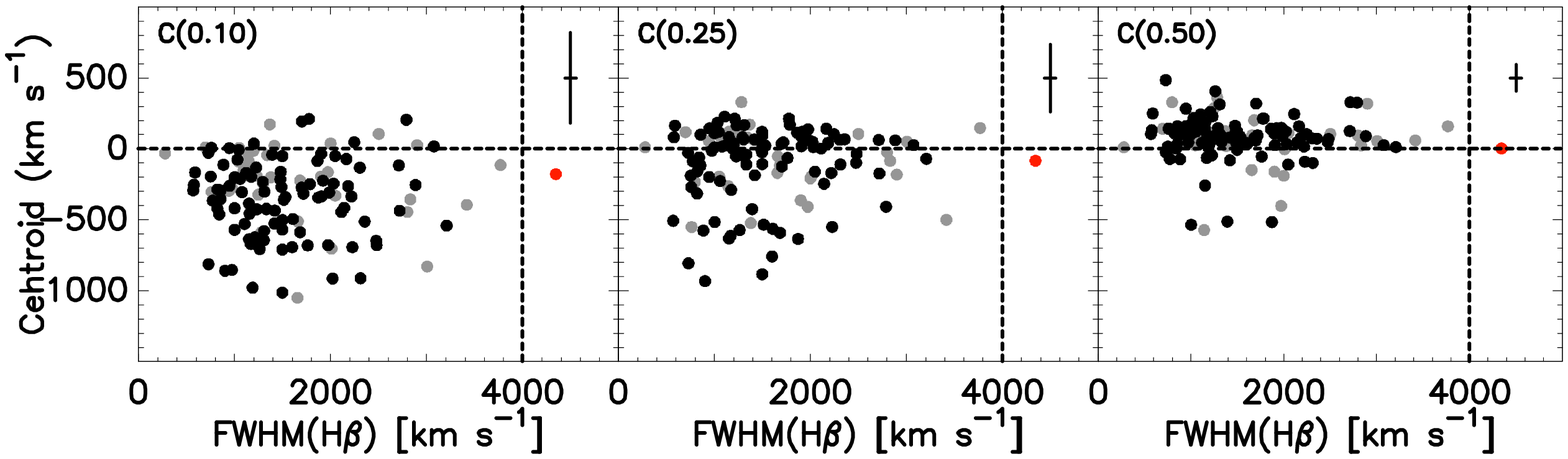}
\caption{Dependence of \hb$_{BC+BLUE}$ centroids c(0.10), c(0.25) and c(0.50) on Eddington ratio (top row), luminosity at 5100 \AA\ (second row from top), and the 4DE1 parameters \rfe\ and FWHM(\hb) (third and fourth rows, respectively). We note that almost all objects with \hb$_{BLUE}$ are Pop. A sources. Meaning of symbols is as in Fig. \ref{fig:E1}. 
\label{fig:CentRels} }
\end{figure}

The \hb\ BLUE component is most likely associated with an outflow which produces the prominent blueshifted emission more clearly observed in the \civ\ profiles \citep[e.g.,][and references therein]{sulenticetal17,coatmanetal16}. The \hb\ blueshifted emission is barely detectable with respect to \hbbc, implying that for \hb\ BLUE, the intensity ratio \civ/\hb\ can be expected to be $\gg 1$. The \civ/\hb\ ratio is very high, most likely above 20, implying a very high ionization level (ionization parameter $\log U \gtrsim -1$) for a moderate-density gas (hydrogen density \footnote{In a fully ionized medium the electron density \ne\ $\approx$ 1.2 \nh. We prefer to adopt a definition based on \nh\ because it is the one employed in CLOUDY computations.} $\log n_{\rm H} = 9$ [\cm3], following CLOUDY simulations. On the full profile, it is necessary to compute the centroid at one tenth or at one quarter fractional intensity to detect the effect of \hb\ BLUE. There is no systematic blueshift at half maximum: the distribution of the data points scatters around zero in  the rightmost panels of Fig. \ref{fig:CentRels}. The c(1/2) values predominantly  show a small redward displacement ($\sim 100 - 200$ \kms; i.e., $\sim 1.5 - 3 \AA$) on average.    
Figure \ref{fig:CentRels} shows the dependence of the centroids at 0.1, 0.25, and 0.5 intensity of \hb$_{BC+BLUE}$, as a function of several parameters (\lledd, luminosity,  \rfe,  FWHM \hb):  there is no strong dependence of the c(1/10) blueshift on FWHM \hb,  Eddington ratios or \rfe.  We  rather see a segregation effect: large blueshifts of c(1/10) and c(1/4) are  predominant in the cosmo sample  at \lledd $\gtrsim 1$ and \rfe $\gtrsim 1$.

\subsection{The narrow and semi-broad components of \oiiiopt}
\label{nsb}

The interpretation of the \oiiiopt\ profile in terms of  a narrow (or core) component (NC) and of a semi-broad component (SB) displaced toward the blue is now an established practice \citep[e.g., ][]{zhangetal11,pengetal14,craccoetal16,bischettietal17}.  Figure 1 of \citet{marzianietal16} shows a mock profile where the NC and SB are added to build the full \oiii\ profile, on which centroids can be measured, as done for \hb.  

At one end we find the symmetric core component to have a typical line width $\lesssim$ 600 \kms; in most cases the core component is superimposed to broader emission  (the SB component) skewing the line base of \oiiiopt\ toward the blue. On the one hand, only the semi-broad component is present.  These are the cases in which the \oiiiopt\ blueshift is largest, even if it is measured at line peak.

\begin{figure}
\includegraphics[scale=0.4]{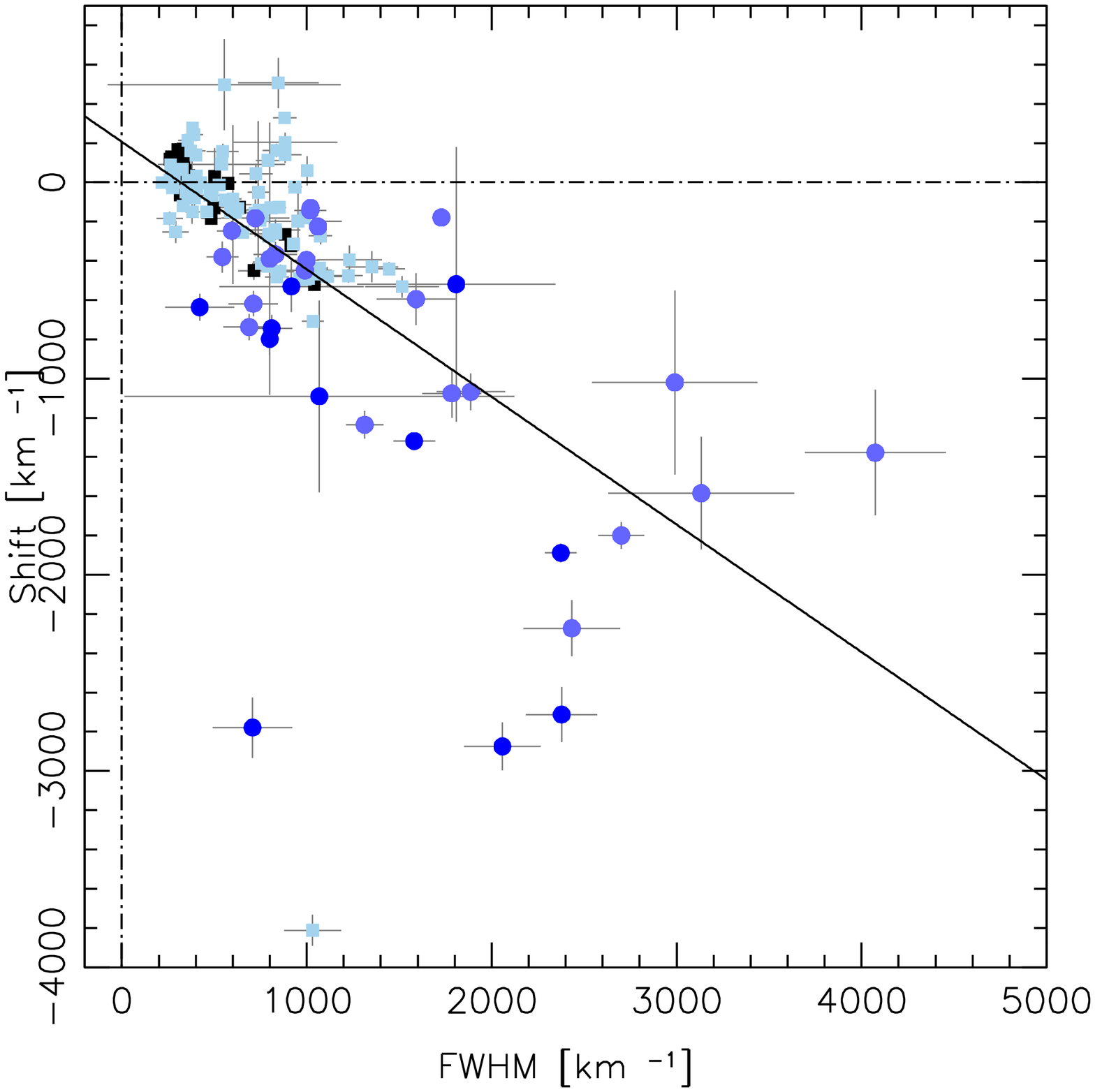}
\caption{Line shifts vs. FWHM for the Pop. A objects. For the objects in which both \oiii\ NC and SB were fitted, the black squares represents the NC while the SB component is in light blue circles. The sources fitted with a single NC are plotted in pale blue squares, while the ones with only a SB component are represented by darker blue circles.  
\label{fig:fwhm_o3shift_histo} }
\end{figure}

Restricting the attention to the Pop. A objects, in Fig. \ref{fig:fwhm_o3shift_histo} we can see that the distribution of the narrow (black and pale blue squares) and SB (light and dark circles) component  
is overlapping in both FWHM and shifts. The vast majority of \oiiiopt\ profiles show blueshift or no shift; the SB in our sample almost always shows a blueshift large enough to be considered a blue outlier (light and dark circles of Fig. \ref{fig:fwhm_o3shift_histo}, with blueshift amplitude larger than 250 \kms). Several NCs are so broad and shifted that they overlap with the FWHM/shift domain of the SB. The narrowest profiles (FWHM $\lesssim $ 600 \kms) all cluster around 0 \kms\ shift. Figure \ref{fig:fwhm_o3shift_histo} shows that the identification of the NC and SB is blurred. This is not  very relevant to the physical interpretation, as long as NC, SB, and composite profiles explained as due to the superposition of NC plus SB  are considered all at once. Figure \ref{fig:fwhm_o3shift_histo} shows the behavior of the shift versus FWHM for four groups of data points: \oiii\ NC only (pale blue squares), \oiii\ NC (black squares) and \oiii\ SB (light blue circles) in case both are detected, and SB only (dark blue circles). Shifts and FWHM are strongly correlated, with $r \approx$ --0.72, and a probability $P \sim 10^{-17}$\ of a chance correlation. The best fitting line equation  obtained with the unweighted least squares method is: $v_{\rm Peak} \approx (-0.650 \pm 0.054) {\rm FWHM} + (206 \pm 62)$ \kms. Analogous  correlations have been found for \civ\ in Pop. A sources \citep{coatmanetal16,sulenticetal17}.

The behavior of  \oiii\ resembles that of \hb, albeit in a somewhat less-ordered fashion: there is no relation with luminosity, and large shifts at one tenth of the total intensity are possible for relatively large \lledd\ ($\log$ \lledd $\gtrsim -0.5$). It is not surprising to find that the largest shifts are found when the semi-broad  component dominates and that the majority of data points follow a correlation between FWHM and c(0.1),  since this correlation is a reformulation of the shift-FWHM correlation of Fig. \ref{fig:fwhm_o3shift_histo}. There is no dependence on \rfe, but one has to consider that the sources of our sample are all with \rfe $\gtrsim 0.5$, and that a large fraction of them show large blueshifts above 200 \kms, which are rare in the general population of quasars (a few percent, \citealt{zamanovetal02}).

\subsection{The \oiiiopt\ emission and its relation to \hb}

\begin{figure}
\includegraphics[scale=0.8]{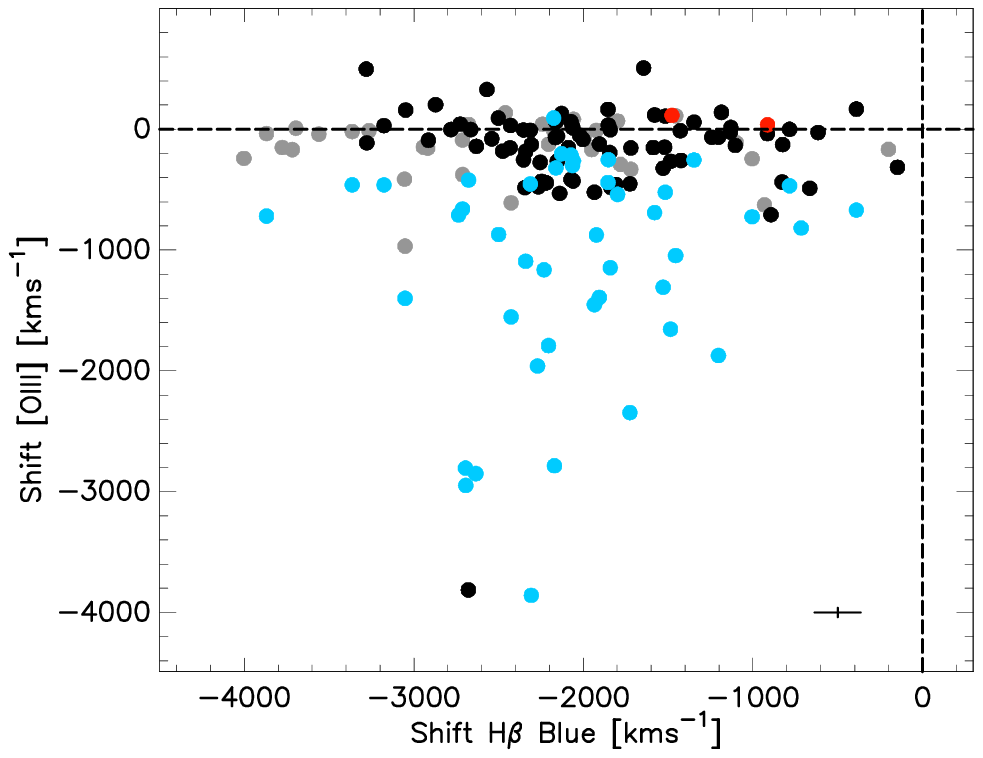}
\caption{Line shifts of \hb\ BLUE vs. shift of \oiii. Meaning of symbols is as for Fig. \ref{fig:E1}, with the addition of pale blue dots that represent sources whose \oiii\ peak shift emission is due to the semi-broad component. Dotted lines are zero shifts. 
\label{fig:OIII_shift_shift} }
\end{figure}

One of the main results of the present investigation is the detection of a blueward excess in the \hb\ profile. The detection has been made possible by the high S/N of the spectra selected for our sample, and is not surprising for the reasons mentioned in Sect. \ref{sec:blend}. 

The relation between \hb\ BLUE and \oiii\ shift is not tight, and it is not expected to be so: the shifts measured on \oiii\ are influenced by aperture, and by the intrinsic extension of the narrow line region (NLR). The results shown in Fig. \ref{fig:OIII_shift_shift} imply that there are several sources that show no significant \oiii\ blueshifts even if \hb\ BLUE is detected and observed with large shifts. If the sources with only \oiii\ SB are considered, then in the wide majority of cases there is a significant blueshifted \oiii, giving a fan-like shape (with the vertex at 0) to the distribution of the data point. 
If we consider \oiii\ profiles and \hb\ BLUE showing blueshifts larger than 250 \kms, the correlation coefficient is 0.05.    

These considerations are  confirmed by the analysis of the centroids at one tenth, one quarter, and half fractional intensity. At one tenth, we observe a similar data point distribution to that in Fig. \ref{fig:OIII_shift_shift}: apart from one outlier, the distribution also suggests that c(1/10) \oiii $\approx $ 0.5 c(1/10) \hb.  This distribution suggests that, at least in part, the \hb\ BLUE might be associated with a NLR outflow. Spatially resolved observations with integral-field spectroscopic units  are needed to ascertain the nature of the relation between the \oiii\ and \hb\ blueshifts.

\section{Discussion}
\label{sec:discussion}

\subsection{xA sources: relation to other AGN classes}

\subsubsection{Relation to NLSy1s}

NLSy1s are an  ill-defined class. The previous analysis shows that there is no clear boundary at 2000 \kms\ in terms of line-profile shapes. In addition, NLSy1s cover all the range of \feii, from \rfe\ $\approx 0$ to very high values exceeding  \rfe\ $=2${. }Sources  in spectral bins A1 and A3/A4 are expected to be different, as we found a strong dependence of physical properties on \rfe, and specifically on \lledd\ which is most likely the main physical parameter at the origin of the spectroscopic diversity in low-$z$ samples \citep[e.g.,][]{borosongreen92,sulenticetal00a,kuraszkiewiczetal04,sunshen15}. As a corollary, it follows that comparing  samples of NLSy1 to broader type-1 AGNs is an approach that is bound to yield misleading results, since the broader type-1 AGNs include sources that are homologous to NLSy1 up to FWHM $\approx$ 4000 \kms\ \citep[e.g.,][]{negreteetal12}. 

\subsubsection{xA sources and SEAMBHs}

\citet{duetal16a} introduced the notion of the fundamental plane of super-Eddington accreting massive black holes (SEAMBHs) defined by a bivariate correlation between the parameter $\dot{{\mathcal M}} = \frac{ \dot{M_{\rm BH}}c^{2}}{L_\mathrm{Edd}}$\,, that is, the dimensionless accretion rate $\dot{m} = \frac{\eta \dot{M}c^{2}}{L_\mathrm{Edd}}$\   for $\eta = 1$ \citep{duetal15}, the Eddington ratio, and the   observational parameters \rfe\ and ratio  FWHM/$\sigma$ of \hb, where $\sigma$\ is the velocity dispersion. The fundamental plane can then be written as two linear relations   between $\log  \dot{\cal M}$\ and $ L/L_{\rm Edd}$ 
versus  $\approx \alpha  + \beta  \frac{\rm FWHM}{\sigma}+\gamma R_{\rm Fe},$ where $\alpha, \beta, \gamma$ are  reported by \citet{duetal16a}. The identification criteria included in the fundamental plane are consistent with the ones derived from the E1 approach (\lledd\ and $\dot{{\mathcal M}}$\ increase as the profiles become Lorentzian-like, and \rfe\ becomes higher).

The cosmo sample satisfies the condition \rfe $\ge 1.2$ by definition. The \hb\ profiles of Pop. A sources are Lorentzian, implying that the ratio $\frac{\rm FWHM}{\sigma} \rightarrow 0$.  This is not actually occurring because the wings of a Lorentzian profile cannot be detected beyond a limit set by the spectrum S/N.  If the detection limits are between 4800 and 4950 \AA\ (appropriate for the typical S/N $\approx 26$\ of our sample), then FWHM/$\sigma \approx 1.2$\ for a pure Gaussian of FWHM$\approx$ 1850 \kms, implying $\dot{{\mathcal M}} \gtrsim 10^{3}$,  $\log $ \lledd\ $\approx 0.3$. The values of  Eddington ratio and $\dot{{\mathcal M}}$ derived from the fundamental plane equation are large enough to qualify the xA sources of the cosmo sample as SEAMBHs. 
 
Assuming  detection between 4500  and 5200 \AA,  one obtains that the Pop. A sources of the cosmo sample  with typical  FWHM/$\sigma \approx 0.5$\ (including the \hb\ BLUE in addition to the Gaussian) and \rfe$\approx$1.56  have Eddington ratio $\log $ \lledd $\approx 0.9$ and a dimensionless accretion rate of $\dot{{\mathcal M}} \sim 10^{4}$. The fundamental plane of \citet{duetal16a}  is not able to consistently consider purely Lorentzian profiles and very high \rfe. There is also a problem with the reliability of the   FWHM/$\sigma$ parameter: its value depends on the line width if the summation range used for the computation of the $\sigma$ is kept constant: the ratio changes from 0.79 to 0.51 for FWHM $\approx$ 4000 \kms\ and 1860 \kms, respectively,  assuming that the line is detected from 4500 to 5200 \AA.

\begin{figure}
\includegraphics[scale=0.45]{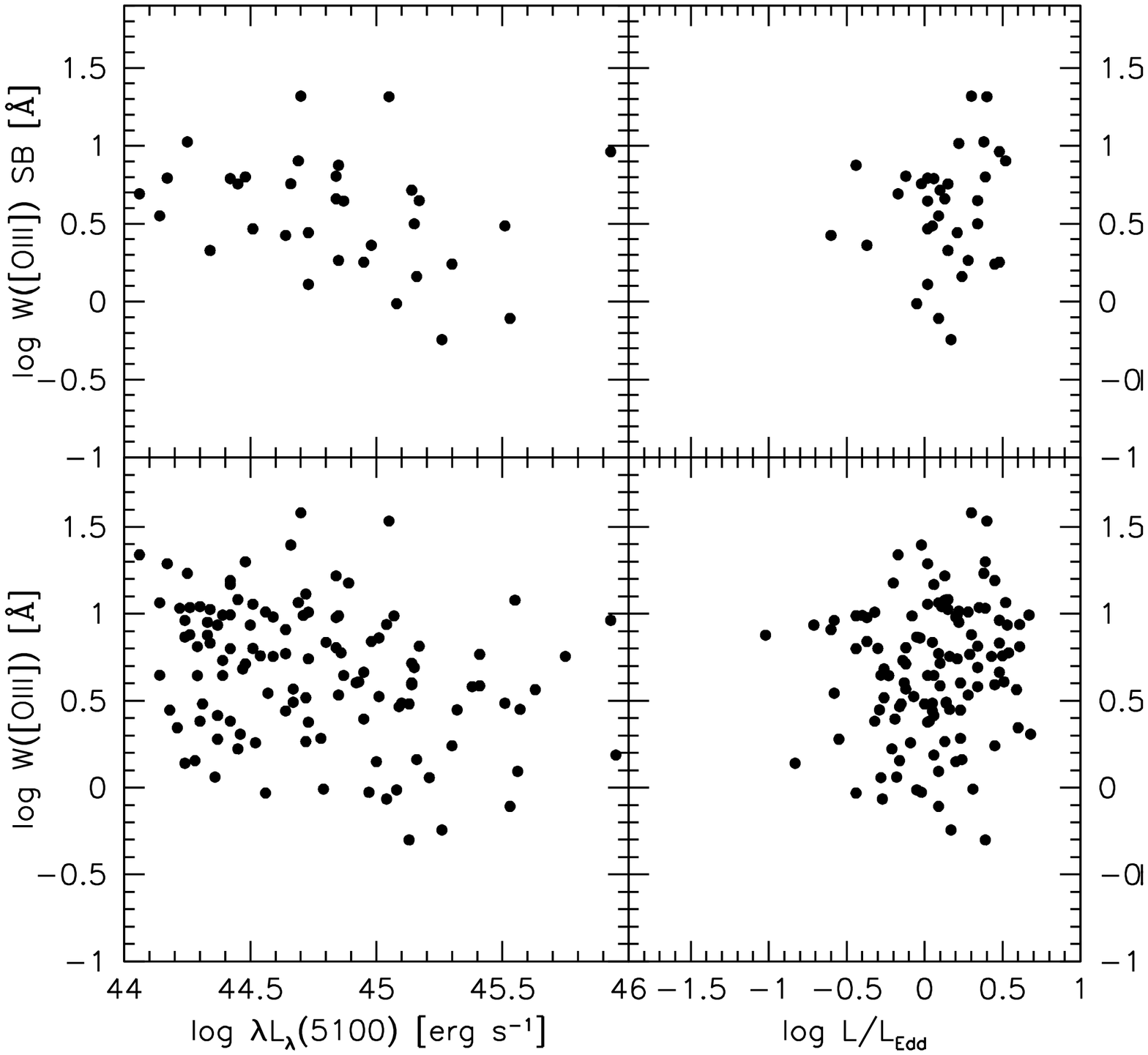}
\caption{Relation W \oiii\ semi-broad component (top) and total (SB+NC) W \oiii\  (bottom) vs. luminosity (left) and \lledd\ (right) for the cosmo sample. \label{fig:EWO3SB_Lbol} }
\end{figure}

\subsection{A Baldwin effect in  \oiii?}

Figure \ref{fig:EWO3SB_Lbol} shows that W(\oiii) is not dependent on Eddington ratio. This remains true for all sources in the cosmo sample as well as in sources in which only the SB component is detected.  A dependence on Eddington ratio is traced by the systematic trends observed along the entire quasar MS: extreme cases imply a change by a factor $\approx 100$ in equivalent width, from A3/A4 to B1+/B1++. In A3 and A4 sources, the \oiii\ EW may become so low that it renders the line undetectable ($\lesssim 1 $ \AA). In the cosmo sample, the \lledd\ covers a relatively restricted range. 

\begin{figure}
\includegraphics[scale=0.45]{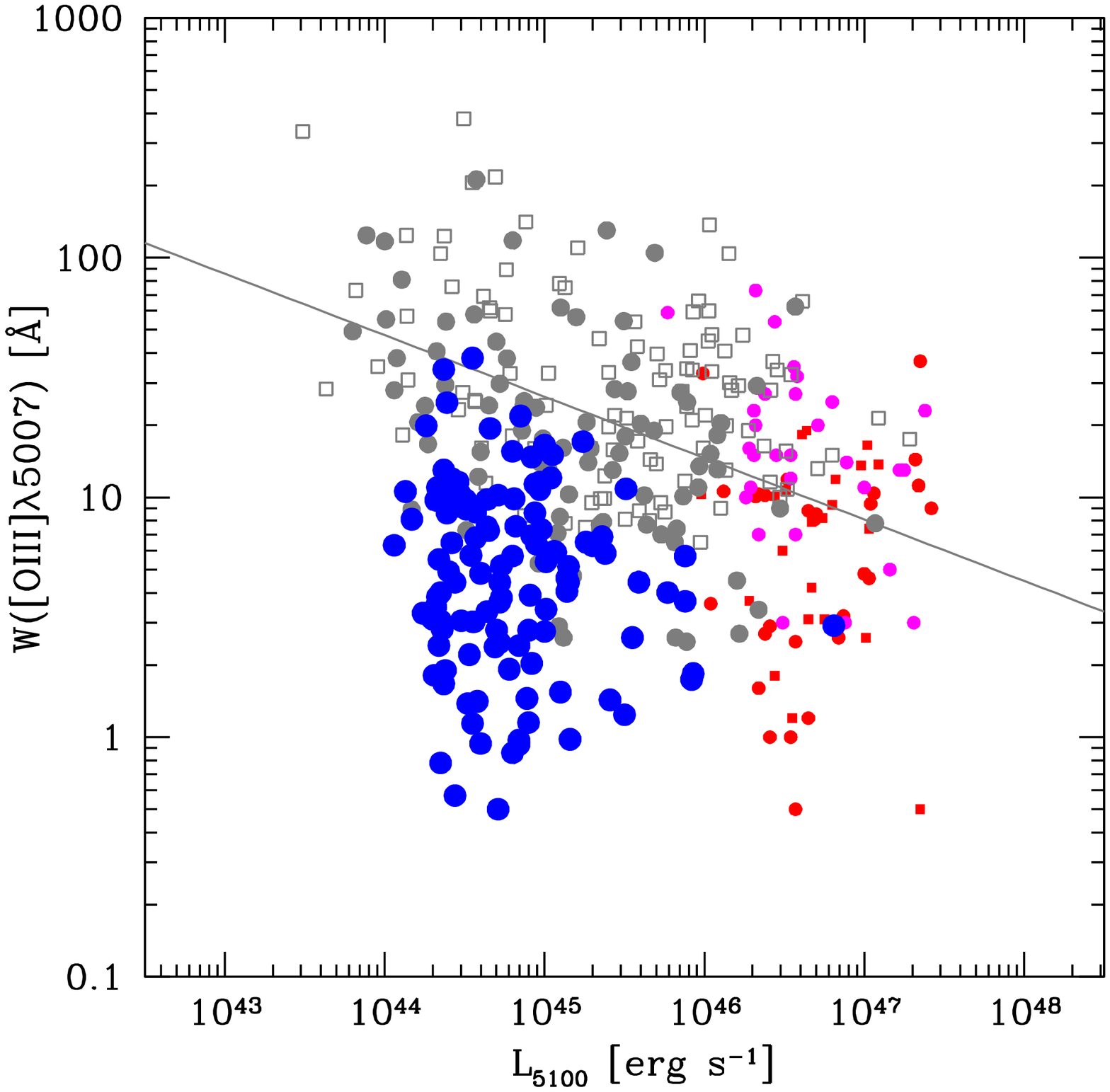}
\caption{Relation between W \oiii\  and 5100 \AA\  luminosity for the cosmo sample and the total \oiii\ SB+NC emission. Red color: \citealt{sulenticetal17}; purple: \citealt{netzeretal04}; gray: low-$z$ sample \citep{marzianietal03a}. Open squares are Pop. B, filled circles Pop. A. The cosmo sample data points are shown as filled blue circles. 
\label{fig:lewnew} }
\end{figure}

In addition, Fig. \ref{fig:EWO3SB_Lbol} shows that there is no well-defined trend between luminosity and W(\oiii), that is, there is no Baldwin effect. As in the case of \civ, a clear Baldwin effect becomes detectable only when relatively large samples are considered \cite[e.g., ][]{baskinlaor05c,zhangetal11,zhangetal13}. The xA sources belong to a particular class  whose \oiiiopt\ emission is known to be of low EW. In Fig. \ref{fig:lewnew} we show the cosmo sample sources along with a large sample of low-$z$ quasars, and  the samples of \citet{netzeretal04} and \citet{sulenticetal17} of high-$z$, high-$L$\ quasars. The vast majority of the xA cosmo sample sources have $W \lesssim 10$ \AA, a property that locates them  in the space of Pop. A sources at high-$z$ in terms of EW (albeit the cosmo sample has a much lower luminosity). The significant correlation in Fig. \ref{fig:lewnew} with slope $a \approx -0.256 \pm 0.027$\  (without including the xA sources of the present work) arises because of the detection of Pop. B sources at low-$z$. Sources from Pop B show systematically higher W(\oiii), frequently reaching values of several tens of \AA, up to $\sim $ 100 \AA. Sources with W(\oiii)$\gtrsim 30$ \ almost disappear in high-$z$, high-$L$ samples. The origin of this trend is not fully clear, but two main effects may concur in its reinforcement: (1) a selection effect, disfavoring the discovery of Pop. B sources that are the weaker sources for a given mass \citep[a similar mechanism is expected to operate for \civ; ][]{sulenticetal14}; and (2) a luminosity effect that can arise if the \oiii\ luminosity is upper bounded because of limits in the physical extension of the NLR \citep{netzeretal04,bennertetal06}. 

The $W$(\oiii) -- $L$ trend disappears if the xA sources of the cosmo sample are added (the correlation coefficient becomes $0.11$, and the slope $a \approx -0.06$, emphasizing the effect of the sample selection. A proper analysis should consider the $W$(\oiii) -- $L$  trend for the same spectral types, but this  goes beyond the aim of the present paper.

\subsection{xA sources: implications for feedback}

The xA sources observed at high luminosity yield the most powerful radiative and mechanical feedback per unit black hole mass \citep{martinez-aldamaetal18}.  The typical luminosity of the \citet{martinez-aldamaetal18} sample is, even before correcting for intrinsic absorption, a factor of ten higher than the typical luminosities in the present sample, but already yields a mechanical output that is larger or at least comparable to that of the most luminous quasars studied by \citet{sulenticetal17}.  What is the feedback level due to the black hole activity in the present xA sample?

The mass of ionized gas needed to sustain the  \oiii\ semi-broad  component  (the average \oiii\ semi-broad component luminosity  in our sample is $\bar{\log L} \approx 41.6$): $M_\mathrm{ion} \sim 5 \cdot 10^{4} \, L_{\rm [OIII], 42} Z_{10}^{-1} n_{3}^{-1}  M_{\odot}$,   where  the density is assumed to be $10^{3}$ cm$^{-3}$, and the relation  is normalized to  a metal content ten times solar \citep[e.g.,][]{canodiazetal12}.   The computation of the mass outflow rate, of its thrust, and of its kinetic power requires knowledge of the distance of the line emitting gas from the central continuum source.  The mass outflow rate can   be written as: $ \dot{M}_\mathrm{ion}  \sim  0.15  L_{\rm [OIII], 42} v_\mathrm{o,1000} r^{-1}_{\rm [OIII], 1kpc} Z_{10}^{-1} n_{3}^{-1} M_{\odot}$ yr$^{-1}$, where the outflow radial velocity is in units of 1000 \kms\ (close to the average shift of the SB component, $\approx 1060 $ \kms).  The  thrust can be expressed as: $ \dot{M}_{\mathrm{ion}} v_\mathrm{o} \sim 1 \cdot 10^{33}  L_{\rm [OIII], 42}   v^{2}_\mathrm{o,1000} r^{-1}_{\rm [OIII],1kpc}  Z_{10}^{-1} n_{3}^{-1}  $   g cm s$^{-2}$, where we have assumed that the $v_\mathrm{o}$ is the terminal velocity of the outflow. The kinetic power of the outflow is $ \dot{\epsilon}  \sim 5 \cdot 10^{40} L_{\rm [OIII], 42}   v^{3}_\mathrm{o,1000}   r^{-1}_{\rm [OIII], \rm 1 kpc}  Z_{10}^{-1}     n_{3}^{-1}  $\   erg s$^{-1}$. The values of the thrust and kinetic power should be compared with $L/c$ and $L$ ($\log L/c \approx 35.3$\ and $\log L \approx 45.79$), respectively. The  $\dot{\epsilon}$ is below the 0.05 $L$\ value thought to be the minimum mechanical input needed to explain the black hole and bulge mass scaling \citep[e.g.,][and references therein]{zubovasking12,kingpounds15} by almost four orders of magnitude. 

Even if the NLR are spatially extended, we can still define a characteristic distance that may represent an emissivity-weighted radius, $r_\mathrm{[OIII]}$.  In the present context, we can make two independent assumptions: (1)    $r_\mathrm{[OIII]}$ is simply one half the radius of the aperture size; and (2) $r_\mathrm{[OIII]}$ \ follows a scaling law with luminosity \citep{bennertetal02,bennertetal06a}. This second approach is especially risky, as we are dealing with   only an outflowing part of the NLR \citep{zamanovetal02}, as well as  sources whose NLR may be intrinsically underdeveloped.  At a typical redshift of  $z \approx 0.3$, the half width of the SDSS fiber, 1.5 arcsec, would correspond to 6.7 kpc of projected linear distance. Our estimates will be lowered accordingly. The scaling law with luminosity implies $r_\mathrm{[OIII]} \approx 10^{2.8}$ pc for \oiii\ luminosity $\approx 10^{41.6}$ \ergss, consistent with the  above estimates.  More recent work indicates a consistent flux-weighted size $\sim 10^{3}$ pc at $\approx 10^{41}$ \ergss\ \citep{riccietal17}.

Much lower density (10$^{2}$ cm$^{-3}$) and a compact NLR of $\sim$100 pc can increase  the $\dot{\epsilon}$ estimate by a factor of 100. This condition may not be unlikely considering the compact NLR suggested by \citet{zamanovetal02}. Even in this case, and at the highest $\log L_{\rm BLUE} \approx 10^{42.6} $ [erg/s], the value of $\dot{\epsilon}$\ remains below 0.05$L$ by a large factor.  Even  upper limits under reasonable assumptions from the xA sample  imply a limited mechanical feedback effect, as found in nearby AGNs  \citep[c.f. ][]{karouzosetal15,baeetal17}. Since the physical parameters scale with luminosity, it is likely that only high-luminosity sources can reach the energetic limits that may imply a galaxy-wide feedback effect \citep[][]{martinez-aldamaetal18}.

\subsection{The virial luminosity equation: a strong influence of orientation}
\label{sec:cosmo}

\subsubsection{The virial luminosity equation}

The virial luminosity equation derived by \citetalias{marzianisulentic14} can be written in the form: 
\begin{eqnarray}
L(\rm FWHM) & = & {\cal L_{\rm 0}} \cdot  (\rm FWHM)^{4}_{1000} \\ \nonumber
 & = & 7.88 \cdot 10^{44} \left(\frac{L}{L_\mathrm{Edd}}\right)_{,1}^{2} \cdot \\ \nonumber 
& & \frac{ \kappa_{0.5}f_\mathrm{S,2}^2}{h \bar{\nu}_\mathrm{i,100 eV}} \frac{1}{(n_\mathrm{H}U)_{10^{9.6}}} (\rm FWHM)^{4}_{1000} {\rm erg  \, s}^{-1}
\label{eq:vir},
\end{eqnarray}
where the energy value has been normalized to 100 eV ($ \bar{\nu}_{i,100ev} \approx 2.42 \cdot 10^{16}$ Hz), the product ($n_{\rm H}U$) has been normalized to the typical value $10^{9.6} $cm$^{-3}$ \citep{padovanirafanelli88, matsuokaetal08,negreteetal12}  and the FWHM of the \hb\ BC has been normalized to 1000 \kms. The $f_\mathrm{S}$ is scaled to a value of 2 following the determination of \citet{collinetal06}.  The FWHM of \hb\ BC (\hbbc)\  is hereafter adopted as a VBE.

The distance modulus $\mu$  can be written as:

\begin{equation}
\mu   =   2.5 [\log L - C] - 2.5 \log  (\lambda f_{\lambda}) -2.5 \log (4 \pi \delta_\mathrm{10pc}^{2})  
\label{eq:mu0}
,\end{equation}

where the constant  $-2.5 \log (4 \pi \delta_\mathrm{10pc}^{2})$ =-100.19, with $\delta_\mathrm{10pc} \approx 3.08 \cdot 10^{19}$\, the distance of 10pc expressed in cm. The  $ \lambda f_{\lambda} $\ is the flux at 5100 \AA\ in the cosmo sample.  In this case, $L$ is the virial luminosity $L$(FWHM) or the customary $L$($z, H_{0}, \Omega_{M}, \Omega_{\Lambda}$) computed from $z$ and concordance cosmology. The difference between the $\mu$ computed from $L$(FWHM)  and from $L$($z, H_{0}, \Omega_{M}, \Omega_{\Lambda}$) is 

\begin{equation}
\delta \mu = \mu^\mathrm{vir} - \mu^\mathrm{z} = 2.5 \log L({\rm FWHM}) - 2.5 \log  L (z, H_{0}, \Omega_{M}, \Omega_{\Lambda})
.\end{equation}

\subsection{Sources of the scatter in the virial luminosity estimates}

From the previous analysis we infer several main sources of scatter that may be affecting the virial luminosity estimates:

\begin{itemize}
\item The \hb\ BLUE, strongly affecting  the \hb\ line base at 0.1 and 0.25 fractional intensity. 
\item \rfe, which  ranges from 1.2 (by definition) to 3 in the most extreme case. While \rfe\ and \lledd\ are correlated for the sample of 304 sources, they show no correlation if we restrict the attention to the cosmo sample (Fig. \ref{fig:BlueRels}):  
the Pearson correlation coefficient is $r \approx 0.01$. 
\item The \oiii/\hb\ ratio: sources with \oiii/\hb\ $< 1$ should be at the tip of the MS according to the original formulation of \citetalias{borosongreen92}.
\item Orientation effects (discussed in Sect. \ref{sec:orien}).  
\end{itemize}

We hereafter restrict our analysis to the parameters listed above. 

\subsubsection{The   \hb\ blue shifted component and its influence on the virial broadening luminosity estimates}

\begin{figure}
\includegraphics[scale=0.425]{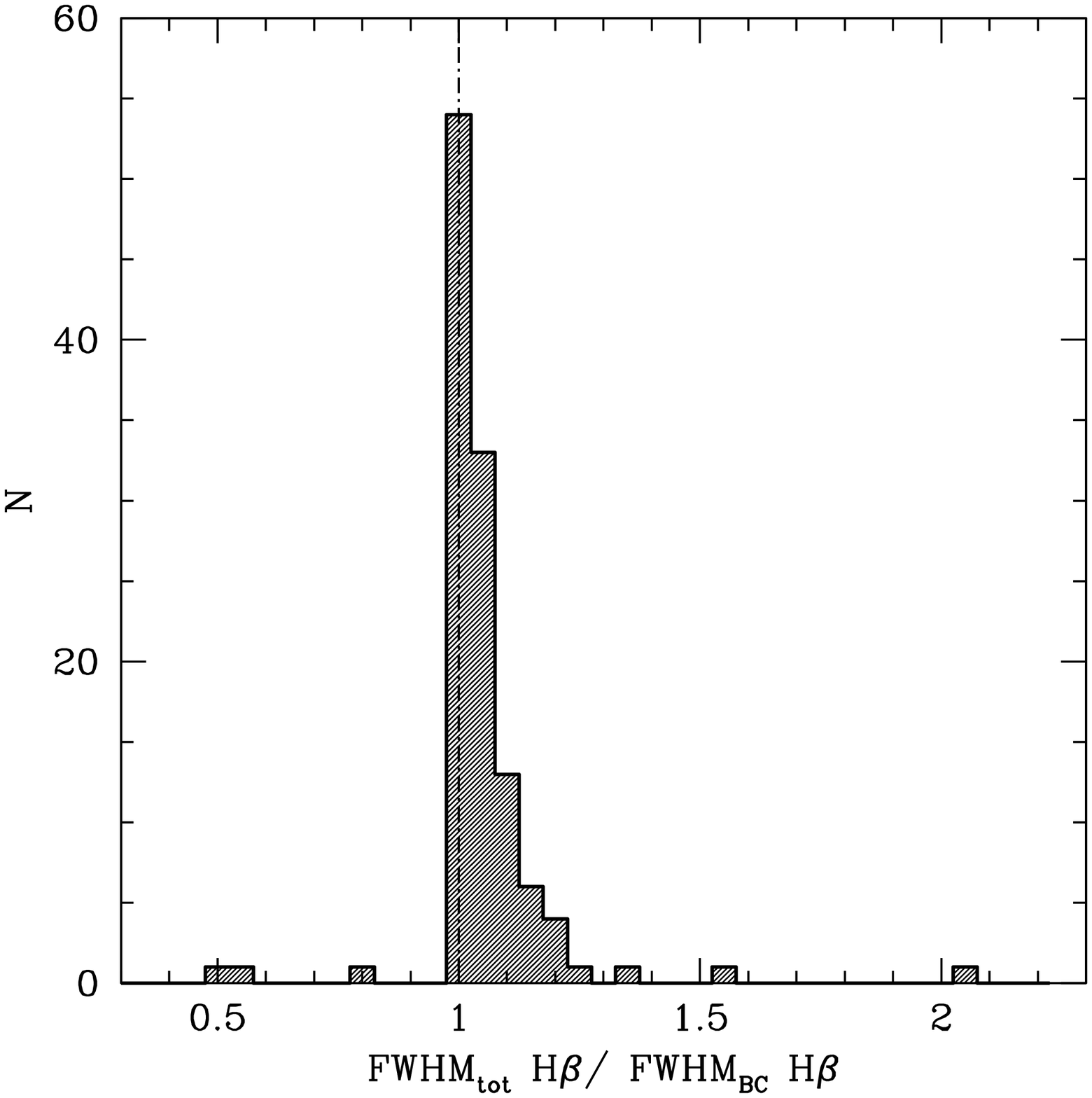}
\caption{Distribution  of the ratio between the FWHM of the full \hb\ profile and the FWHM of the \hb\ BC. 
\label{fig:fratios}}
\end{figure}

\begin{figure}
\includegraphics[scale=0.42]{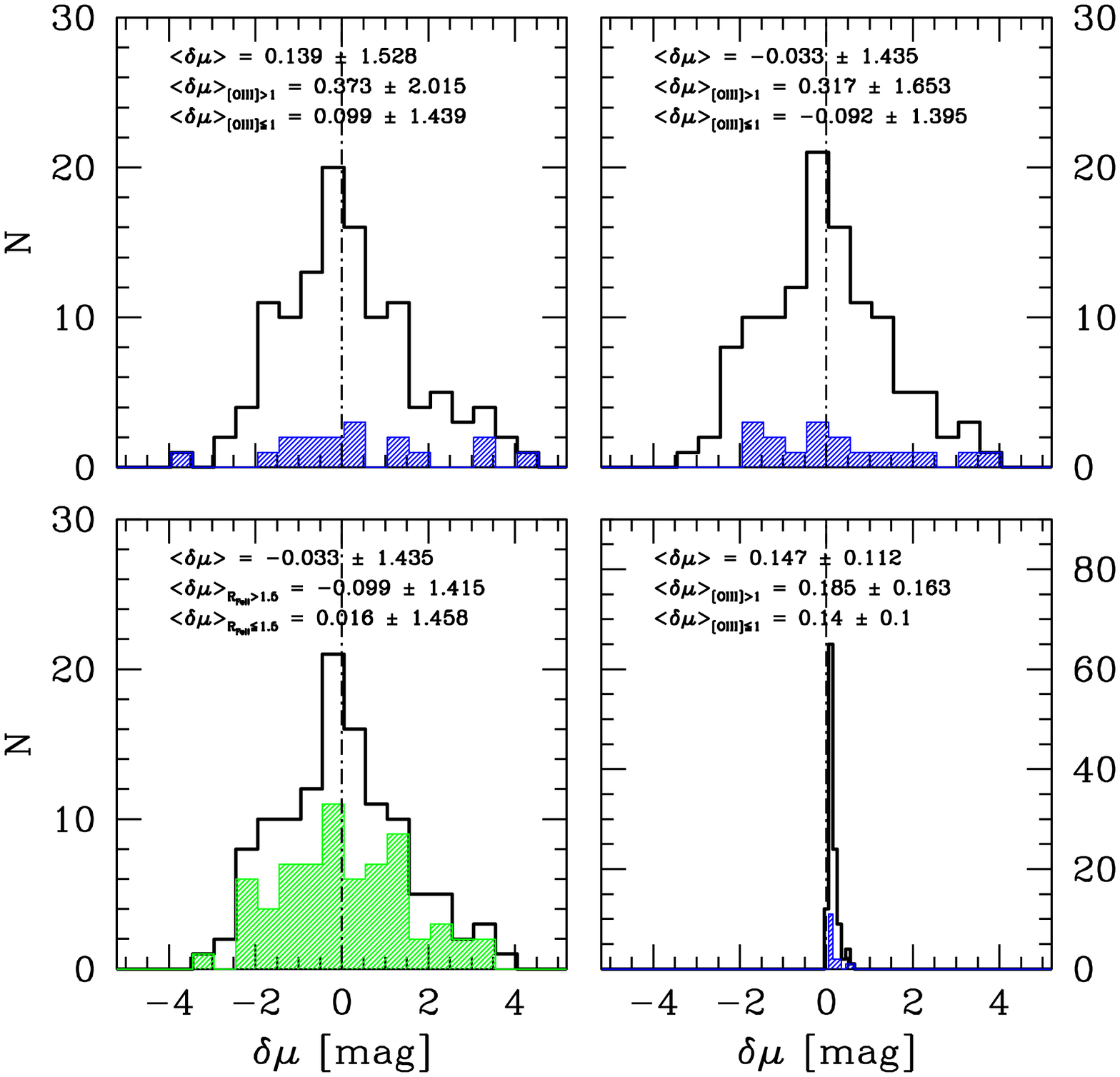}
\caption{Distribution of the  difference $\delta \mu = \mu^\mathrm{vir} - \mu^\mathrm{z} $ between virial luminosity and concordance cosmology estimates of the distance modulus $\mu$. Top left: Full cosmo-sample sources uncorrected for \hb\ BLUE, and sources  with  $R_{\rm [OIII]} > 1$ (blue). Top right: As in top-left  panel, but with \hb\ BLUE excluded.  Bottom left:  As in  top-right panel but distinguishing sources with \rfe $< 1.5$ (green).  Bottom right: Residuals after correction for \hb\ BLUE and orientation effects. The distribution of sources  with  $R_{\rm [OIII]} > 1$ (blue) is also shown, as in the top panels.} \label{fig:fwhm}
\end{figure}

The presence of line emission that is barely resolved and associated with outflows complicates the derivation of the VBE to be used in Eq. 4. 
From the {\tt specfit} \ analysis we derive the FWHM of a symmetric Lorentzian, considered  our VBE. The effect on the line profile of the blueshifted emission is very important  at very low fractional intensity, and is much smaller but not negligible also at half maximum. 

Figure \ref{fig:fratios} shows the distribution of the ratio between the FWHM of the full profile and of the \hb\ BC:
\begin{equation}
\frac{1}{\xi} = \frac{\rm FWHM\, H\beta}{\rm FWHM\, H\beta_{\rm BC}}
,\end{equation}
where $\xi$\ is a correction to virial broadening from the observed profile (in practice the ratio of the full-profile and BC FWHM). 
 Figure \ref{fig:fratios} shows that the average excess broadening is modest; 30\%\ of the sources have $\xi =1.00$ (i.e., they are considered with a symmetric profile and show no evidence of blueshifted emission), 46\%\ are in the range $0.98 \le \frac{1}{\xi} \le 1.02$ and the median $<\frac{1}{\xi}> \approx 1.05 \pm 0.035$.  If the five outliers outside in the range  with $0.85 \le \frac{1}{\xi} \le 1.5$ are excluded, the average is $<{1/\xi}> \approx $\  1.05 $\pm$ 0.06. The distribution of  Fig. \ref{fig:fratios} shows that about two thirds of the sample will be within $1 \le \frac{1}{\xi} \le <{1/\xi}>$. The difference from unity for median and average implies a systematic overestimate of the virial luminosity by $\approx $ 20\%, or $\delta \log L \approx 0.085$.  Multi-component line fittings are  needed: 70\%\ of sources show evidence of asymmetry to some extent, and the dispersion $\approx 0.06$\ implies that  for 14\%\ of sources the luminosity is overestimated  by about 60\%, with $\delta \log L \gtrsim 0.14$; for 7 \%\ of the sample the disagreement reaches a factor two, implying $\delta \log L \approx 0.3$. Therefore, the overall effect on the FWHM is  small, but unfortunately uneven and with a skewed distribution  significantly contributing to the scatter  appreciable for about one third of the sources in the present sample.  This is especially not negligible if we want to achieve the accuracy necessary for meaningful cosmological constraints because the effect may become comparable to root mean square  (rms) values that are conducive to clear results: an $rms \approx$ 0.2 --  0.3 dex could yield, in the absence of   systematic effects,  meaningful constraints on $\Omega_{\rm M}$\ with 400 quasars over the redshift range 0.2 -- 3.0.

Figure \ref{fig:fwhm} (top panel) shows that the inclusion of the \hb\ BLUE in the FWHM produces the worst    scatter of the $\delta \mu$ differences in the cosmo sample, if compared to the cases where only the symmetric Lorentzian FWHM is considered (middle and bottom panels). 

\subsubsection{Role of \rfe\ and \oiiiopt}

Restricting the attention to the \hbbc\ ``clean'' of \hb\ BLUE, Fig. \ref{fig:fwhm} (lower left panel)  shows that the  scatter is not greatly reduced if we consider the A3 spectral type of the cosmo sample, and the 
spectral types A4 and beyond with higher \rfe $\ge$ 1.5.  On the contrary, there is a significant enhancement in the scatter (from rms $\approx$ 1.6 to 1.4) if the $R_{\rm [OIII]}$\ is limited to $> 1$ (upper right panel). This result is somewhat unexpected, as  sources with $R_{\rm [OIII]}> 1$ are present if \rfe $\ge$ 1.5. However, $R_{\rm [OIII]} > 1$ is found in only 16 sources, and this may explain the larger scatter. A larger sample is needed to test whether or not $R_{\rm [OIII]} \le 1$\ can be considered as an additional selection criterion to identify xA quasars.

\subsubsection{Orientation effects on virial luminosity estimates}
\label{sec:orien}

The effect of orientation can be quantified  by assuming that the line broadening is due to an isotropic component plus a flattened component whose velocity field projection along the line of sight is $\propto 1/\sin \theta$: 
\begin{equation}
\delta v^{2}_\mathrm{obs} = \frac{\delta v_\mathrm{iso}^{2}}{3} + \delta v_\mathrm{K}^{2} \sin^{2}\theta.
\label{eq:v}
\end{equation}

The structure factor $f_{\rm S}$ relates the observed velocity dispersion to the real, virial velocity dispersion $\delta v_{\rm K}$. The virial mass equation:

\begin{equation}
M_{\rm BH} = \frac{r \delta v_{\rm K}^{2}}{G}
,\end{equation}

can be of use if we can relate $\delta v_{\rm K}$\ to the observed velocity dispersion, represented here by the FWHM of the line profile: 

\begin{equation}
M_{\rm BH} = f_{\rm S} \frac{r  {\rm FWHM}^{2}}{G}
,\end{equation}

via the structure factor whose definition is given by

\begin{equation}
 \delta v_{\rm K}^{2} = f_{\rm S}{\rm FWHM}^{2}
.\end{equation}

The structure factor  in Eq. 4 
is set to $f_{\rm S} = 2$. If we considered a flattened distribution of clouds with an isotropic and a  velocity component associated with a flat disk, the structure factor appearing in Eq. 4 
can be written as 

\begin{equation}
f_{\rm S}  = \frac{1}{4 \left[ \frac{1}{3}\left(\frac{\delta v_{\rm iso}}{\delta v_{\rm K}}\right)^{2} + \sin^{2} \theta  \right]   } \label{eq:fs}
,\end{equation}

which can reach values $\gtrsim 1$\ if $\kappa = \frac{\delta v_{\rm iso}}{\delta v_{\rm K}} \ll 1$, and if $\theta$ is also small ($\lesssim 30$ deg). The assumption $f_{\rm S} = 2$ implies that we are seeing a highly flattened system (if all parameters in Eq. 4 
are set to their appropriate values): an isotropic velocity field would yield $f_{\rm S} = 0.75$ (i.e., setting $v_{\rm K} = 0$ in Eq \ref{eq:v}).   

The virial luminosity equation may  be rewritten in the form
\begin{eqnarray}
L({\rm FWHM}) &=&    {{\cal L^{\bullet}_{\rm 0}}}  f^{-2}_{\rm S} (\delta v_{\rm K})^{4}   \\ \nonumber
& = &  {   {\cal L^{\bullet}_{\rm 0}}  (\delta v_{\rm K})^{4}  4^{2}    }{\left[ \frac{\kappa^{2}}{3} + \sin^{2} \theta \right]}^{2} =   L_{\rm vir} 4^{2} {\left[ \frac{\kappa^{2}}{3} + \sin^{2} \theta \right]}^{2}, \\ \nonumber
 \end{eqnarray}

where $L_{\rm vir}$ is the true virial  luminosity (which implies $f_{\rm S} = 1$) with   ${\cal L^{\bullet}_{\rm 0} }= \frac{1}{4}{\cal L_{\rm 0}}$, since  ${\cal L_{\rm 0}}$ was scaled to $f_{\rm S} = 2.$

Considering the difference between the observed virial luminosity and the concordance cosmology luminosity computed from the observed fluxes, we can write

\begin{equation}
\delta \log L = \log L({\rm FWHM}) - \log L(z, H_{0}, \Omega s)  
,\end{equation}
which in general can be written as

\begin{eqnarray}
\delta \log L &=& \log { 16 {\cal L^{\bullet}_{\rm 0}}   \delta v_{\rm K}^{4} }{\left[ \frac{\kappa^{2}}{3} + \sin^{2} \theta  \right]^{2}  - }\\  \nonumber
&  & \log \left[ \frac{L_{0}(z, H_{0}, \Omega s)}{2} \cos \theta  \frac{(1 + \beta \cos \theta)}{1+\beta}  +  \frac{L_{0}(z, H_{0}, \Omega s)}{2}\right ],\\ \nonumber 
\label{eq:delta}
\end{eqnarray} 

where we have considered that only half of the luminosity (the anisotropic component) is released by the accretion disk \citep{franketal02}, and  that the accretion disk is a Lambertian radiator subject to limb darkening ($\beta \approx 1.5 - 2$; \citealt{netzer87,netzer13}).

\begin{eqnarray}
\delta \log L  &= &   \log \frac{ 16 \cdot 2 (1+\beta) \left[ \frac{\kappa^{2}}{3} + \sin^{2} \theta   \right]^{2}}{ \left[ \cos \theta   \left( 1 + \beta \cos \theta) + (1+\beta) \right)     \right]}  .\\ \nonumber
&& \cdot \frac{ {\cal L^{\bullet}_{\rm 0}}   \delta v_{\rm K}^{4}}{L_{0}(z, H_{0}, \Omega s)}  \\ \nonumber
\end{eqnarray}

In the ideal situation in which the virial luminosity is a perfect estimator of the face-on quasar luminosity, the factor $\frac{{\cal L^{\bullet}_{\rm 0}}  \delta v_{\rm K}^{4}}{L_{\rm 0}(z, H_{0}, \Omega s) }$\ should be equal to one.

\begin{figure}[h]
\centering
\includegraphics[scale=0.425]{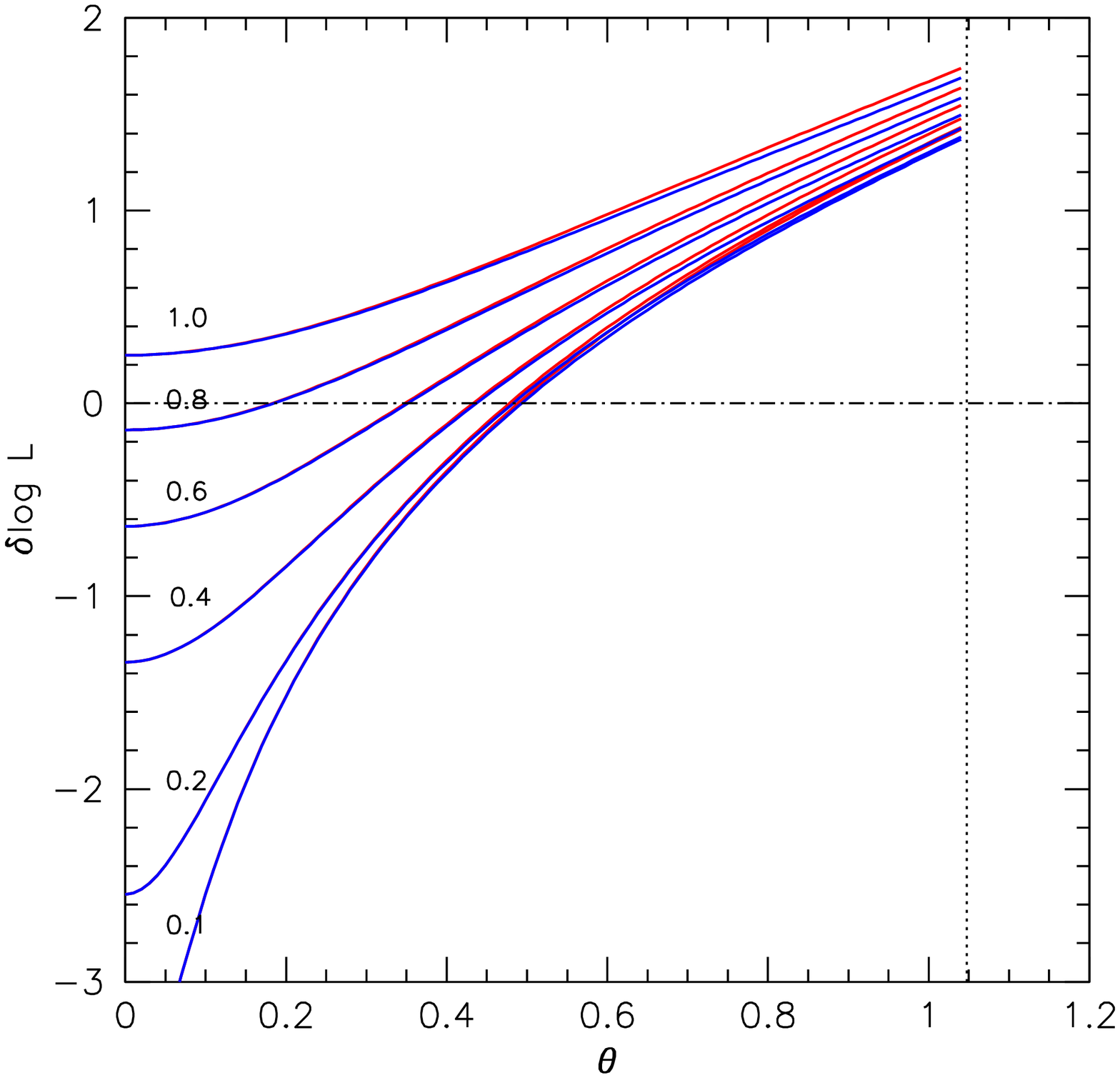}
\caption{Luminosity difference between virial and concordance cosmology determinations, 
$\delta \log L = \log L({\rm FWHM}) - \log L(z,H_{0}, \Omega s) = \log {\cal L^{\bullet}_{\rm 0}}  f^{-2}_{\rm S} (\delta v_{\rm K})^{4} - \log L(z,H_{0}, \Omega s)$, 
for different $\kappa$ values. The blue lines refer to the absence of limb darkening ($\beta = 0$) while the  red ones are computed for $\beta = 2$.  For $\kappa=0.1$ the red line computed for $\beta = 2$\ is fully superimposed on the one for $\beta = 0$.  \label{fig:orienxa}  }
\end{figure}

Figure \ref{fig:orienxa} shows the behavior of $\delta \log L$\ as a function of $\theta$. Large values of $\delta \log L$ are possible for highly flattened systems. The effect of the anisotropy decreases dramatically as $\kappa$ grows to values of approximately one. 

The orientation angle for each individual source can be retrieved from the $\delta \log L$ which is known from the observations: 

\begin{equation}
 \frac{ 16 \cdot 2 (1+\beta) \left[ \frac{\kappa^{2}}{3} + \sin^{2} \theta   \right ]^{2}}{ \left[ \cos \theta  (1 + \beta \cos \theta) + (1+\beta)      \right ]}    = 10^{\delta \log L} = \frac{L({\rm FWHM})}{L(z,H_{0},\Omega s)} 
.\end{equation}

Substituting $x = \cos \ {\theta}$:

\begin{equation}
\frac{ 16 \cdot 2 (1+\beta)    }{10^{\delta \log L}}   = \frac{ \left[x \left( 1 + \beta x) + (1+\beta) \right)\right]}{ \left[ \frac{\kappa^{2}}{3} + 1 -x^{2}   \right ]^{2}} \label{eq:correction}
.\end{equation}

This rational equation yields an estimate of $\theta$ that can be valid  for an individual source as well as for an average. The   $\theta$\ values needed to account for the $\delta \log L$\ are only slightly  dependent on the value of $\kappa$ \ of $0.1 \lesssim \kappa \lesssim 0.3$\ (Fig. \ref{fig:orienxa}).  The minimum scatter is obtained for $\kappa = 0.1$\ and 0.2\ which is the value consistent with the expectations for a flat disk. Several simplifying assumptions can be made; for example, setting $\beta = 0$\ (no limb darkening). Figure \ref{fig:orienxa} shows that the effect of neglecting limb darkening is negligible for $\kappa = 0.1$, and also remains small (few hundreds of dex at most) in the other cases. Another possibility is to assume that the whole quasar luminosity follows a Lambertian dependence on $\theta$, or that $\log L$\ is fully isotropic. In both cases there are no significant changes, as the dominant term setting the orientation effect is the one associated with the $f_{ \rm S}$. 

If the derived correction  is applied to the cosmo sample for a fixed value of $\kappa$, we obtain a significant reduction of the rms scatter. The bottom right of Fig. \ref{fig:fwhm} shows that the rms is reduced from 1.4 to 0.1 mag for $\kappa = 0.1$. The distribution of the residuals   shows a modest bias (0.15 mag) and that for all data  $|\delta \mu| \le 0.6$, and that the rms scatter $\approx 0.1$\ is significantly increased by a minority of data points with larger $\delta \mu$\ that skew the distribution.

Figures \ref{fig:fwhm} and \ref{fig:dl}   provide a confirmation that orientation is a major factor at the origin of large $\delta \mu$ in the Hubble diagram presented by \citet{negreteetal17}, and that it can account for the vast majority of the scatter. However, the scatter distribution becomes skewed after orientation correction.  The skewness may be associated with intrinsic differences in the physical conditions of the line emitting regions that enters into the ``constant'' ${\cal L}_{0}$. A rest-frame ultraviolet (UV) analysis of individual sources is needed to ascertain the origin of the largest deviation after orientation correction.

\subsection{Dependence on FWHM}

\begin{figure}[htp]
\centering
\includegraphics[scale=0.425]{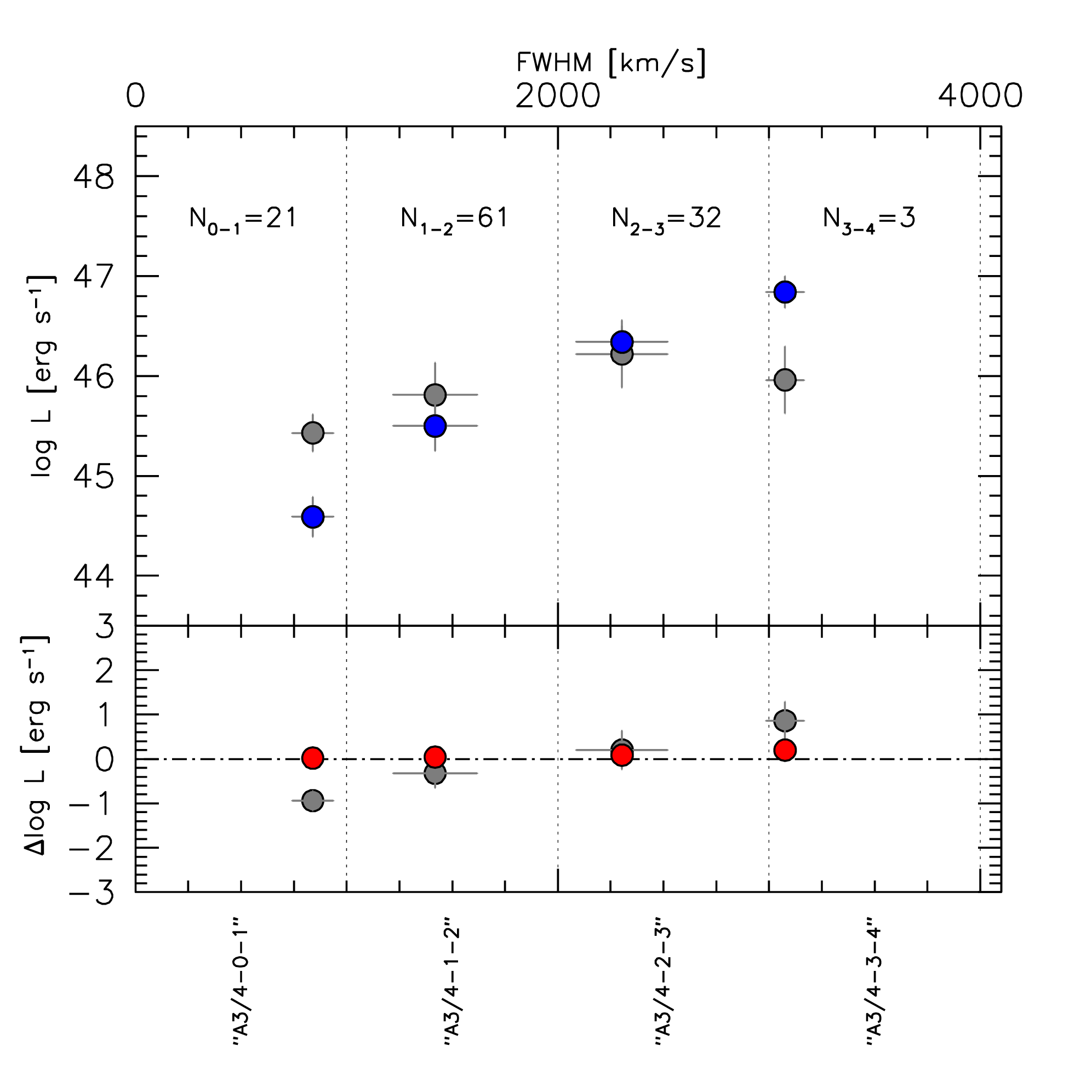}
\caption{Top: Comparison between virial (blue) and redshift-based luminosity (gray) for the A3 and A4 spectral types (the cosmo sample), for  the sub-bins of $\Delta $ FWHM = 1000 \kms\ in order of increasing FWHM.   Bottom: Residuals  before (gray) and after (red) correction for orientation. 
 \label{fig:dl}  }
\end{figure}

We created median composites in sub-bins of width $\Delta $FWHM = 1000 \kms. We computed the virial luminosity using the FWHM of the median composite. Figure \ref{fig:dl} shows a comparison between     median virial luminosity and  median concordance luminosity, computed from the individual sources in the bins as a function of FWHM, for the cosmo sample.  

 We infer two main results:
1) The spectral bins with the largest numbers (between 1000 and 3000 \kms) show very good agreement between the two $L$\ estimates;
and 2) there is a systematic trend in $\Delta \log L$: at low FWHM, the $L_\mathrm{vir}$\ is much lower than $L(z)$, while it becomes significantly larger above 3000 \kms\ (although the 3-4000 \kms\ bin should be viewed with care as there are only 3 objects). The trend is almost fully accounted for by the correction for orientation: $\Delta \log L \approx 0.02$ for the 0-1 bin becomes 0.08 in the 2-3 bin.

Figure \ref{fig:dl}  provides confirmation of the validity of the $L_\mathrm{vir}$ estimates for four fifths  of the cosmo sample sources. The systematic discrepancy can be traced to the effect of orientation, at least in Pop. A, as  the width of the \hb\ is expected to be roughly proportional to ${\sin \theta}$.  If broader sources are considered, the situation is probably more complex: the basic question is whether they are still xA or rather  Pop. B. We defer the issue to an eventual paper.

The residuals    in Fig. \ref{fig:dl} show that the vast majority of data cluster around an average $<\delta \log L> \approx 0$, with a relatively small dispersion.   Therefore,

\begin{eqnarray}
    \sum_{i} \frac{1}{n}\delta \log L_{i} &=&  \log \frac{ 16 \cdot 2 (1+\beta) \left[ \kappa^{2} + \sin^{2} \bar{\theta}   \right ]^{2}}{ \left[ \cos \bar{\theta}   \left( 1 + \beta \cos \bar{\theta}) + (1+\beta) \right)     \right ]}.\\ \nonumber
    & &+\frac{1}{n} \sum_{i=1}^{n}  \log  \frac{{\cal L^{\bullet}_{\rm 0}} \rm \delta v_{\rm K,i}^{4} }{L_{\rm 0,i}(z, H_{0}, \Omega s)   } = 0\\ \nonumber
\end{eqnarray}

Consistently with the assumption of agreement between the two luminosity measurements, we can consider that $\theta_{i} \sim \bar{\theta}$, that is, that the bulk of the sources are observed at the same viewing angle. Therefore,
\begin{eqnarray}
    \sum_{i} \frac{1}{n}\delta \log L_{i} &=& \log \frac{ 16 \cdot 2 (1+\beta) \left[ \kappa^{2} + \sin^{2} \bar{\theta}   \right ]^{2}}{ \left[ \cos \bar{\theta}   \left( 1 + \beta \cos \bar{\theta}) + (1+\beta) \right)     \right ]}.\\ \nonumber
    & &+\frac{1}{n} \sum_{i=1}^{n}  \log  \frac{{\cal L^{\bullet}_{\rm 0}} \rm \delta v_{\rm K,i}^{4} }{L_{\rm 0,i}(z, H_{0}, \Omega s)   } = 0\\ \nonumber
\end{eqnarray}

A constant can be defined as  $A = \frac{1}{n} \sum_{i=1}^{n}  \log  \frac{{\cal L^{\bullet}_{\rm 0}} \rm FWHM_{\rm i,1000}^{4} }{L_{\rm 0,i}(z, H_{0}, \Omega s) }$. As in the case of individual sources, we expect  $A \approx 1$. The value of $\bar{\theta}$\ can then be recovered as in the case of individual sources. Most sources show $\theta \approx 20$, a value slightly lower than the one expected for sources randomly oriented between 0 and 45, but in agreement with the results obtained in \citet{bonetal09a}. Two factors may bias our sample: (1) we restrict the line width to sources narrower than 4000 \kms\ FWHM; and (2) sources with sharper features are identified more easily. This is why the prevalence of sources in the range 3-4000 \kms is just less than 3\%. 

The trend of the residuals as a function of luminosity is reduced but not zeroed after the correction for orientation.  Equation \ref{eq:v} clearly depends on the \mbh\ of each source: our assumptions imply that $r \propto L^{1/2}$, and that $L/$\mbh\ = $const$, so $v_{\rm K} \propto L^{1/4}$. The $v_{\rm  K} \approx const$ applies if all sources have a similar \mbh, which may happen in particular samples. However, this is not true in  general. 
As a consequence it is not directly possible to estimate an orientation angle using the expression for   $v_{\rm K}$.   Using $\delta \log L_{i}$ and subtracting the concordance cosmology luminosity from each virial luminosity estimate we somehow ``eliminate'' different intrinsic luminosities, which is equivalent to saying that, in the case of fixed $L/$\mbh, we eliminate the effect of different masses. This  leaves  the effect of the orientation as a major factor affecting $\delta \log L$: our analysis strongly suggests that the effect of orientation is indeed significant. The origin of the residual scatter and of the trend with FWHM is not clear. In principle the sources detected at FWHM $\rightarrow 4000$\ \kms\ could be the most inclined xAs, as the orientation effects   (Fig. \ref{fig:orienxa}) predict a $\delta \log L \approx $ 1.5 -- 2 at $\theta = \pi/3$  with no strong dependence on $\kappa$. However, a comparison of FWHM makes sense only if the profile remains exactly the same as a function of the FWHM. A detailed analysis of the \hb\ line profile  with increasing FWHM is needed to ascertain whether changes in line profile may also be contributing to $\delta \log L_{i}$.

\subsection{Ongoing work}

Several methods have been proposed for the use of quasars as cosmological standard candles (or better, redshift-independent luminosity indicators), or standard rulers \citep[some early and some recent ones  are reviewed in the Chapter by \citealt{bartelmannetal09} in ][]{donofrioburigana09}.  The advantage of quasars is that we can easily detect them up to $z \sim$ 4.  There are however very serious difficulties. Quasar luminosity  spans over six orders of magnitude, which is the opposite of the standard candle concept. There is a wide spectral diversity  which reflects the existence of intrinsic differences in the physical and kinematic conditions of the region where the broad lines are emitted (the broad line region, BLR). Early efforts to establish correlations between luminosity and one or more parameters  \citep[e.g., the Baldwin effect; ][]{baldwinetal78,bianetal12} did not live up to cosmological expectations. Other developments are  paving the road to the use of quasars as distance indicators.

\citetalias{marzianisulentic14} proposed the use of xA quasars as ``Eddington standard candles'' \citep[see also][]{teerikorpi11}. The fundamental hypothesis for using an object as a cosmological candle is that all of them should have the same intrinsic luminosity. For the case of quasars, as mentioned in the Introduction, theory predicts that the luminosity-per-unit-mass is similar in the sources that have high accretion rates of material onto the supermassive black hole \citep[e.g.,][]{wataraietal00,mineshigeetal00}.    \citetalias{marzianisulentic14}   verified that sources with \rfe $\gtrsim$ 1 are radiating close to the Eddington limit, as also confirmed by \citet[][and references therein]{duetal16a}. 

We will explore a new method to compute cosmological distances in a forthcoming study. The present sample will be considered, along with a high-z (2 $\le$ z $\le$ 2.9) high-S/N xA quasar sample observed with the Gran Telescopio de Canarias \citep[GTC; ][accepted]{martinez-aldamaetal18}.

\section{Conclusion}
\label{sec:conclusion}

In the above sections, we describe the process used to effectively isolate and characterize   extreme quasars.  We analyzed the individual spectra   used to measure the emission lines that allow us to locate quasars in the E1 diagram, and analyzed composite spectra as a function of FWHM, \rfe, and \oiii\ prominence. The main results  concerning the observational properties of xA sources can be summarized as follows:

\begin{itemize}
\item We confirm that the \hb\ profile remains Lorentzian-like up to about 4000 \kms. 
\item We detect a weak but significant blueshifted \hb\ emission, present in all model profile cases we considered. This seems to be a widespread feature, especially in the most extreme xA sources of spectral type A4/B4. 
\item The \oiii\ profiles are also consistent with line broadening partly associated with outflow motions. The correlation between shift and width of Fig. \ref{fig:fwhm_o3shift_histo} (also found by \citealt{komossaetal08} in the case of ``blue outliers,'' sources with large \oiii\ blueshift)  is consistent with that of \civ\ \citep{coatmanetal16,sulenticetal17}, supporting the conclusion  that semi-broad profiles are mainly broadened by non-virial motions.
\item The \oiii\ and \hb\ blueshifts are loosely  related, indicating that in some cases the \hb\ blueshifted component may be in part associated with the \oiii\ outflow. 
\item Since xA sources in general show powerful \civ\ blueshifted emission, the weakness of the \hb\ blueshifted emission is consistent with high-ionization in the outflowing gas from the BLR \citep[e.g.,][]{leighlyetal07,richardsetal11}. 

\end{itemize}

In this paper we found that the \rfe\ and \lledd\ are correlated and we confirm that \rfe\ is $\gtrsim 1$ at the highest \lledd. Regarding the possibility of employing xA sources to estimate the luminosity in a
way that is independent of redshift, we  identify two major  sources of scatter and systematic effects at the origin of the dispersion of the virial luminosity estimates with respect to conventional estimates.

\begin{itemize}
\item  The most relevant one, which was indeed expected, is  due to the viewing angle of the source. This can lead to deviation of several magnitudes in the Hubble diagram. We derived a correction  which cannot, however, be immediately applied to cosmological studies. As a matter of fact  we did assume a cosmological model for computing the $\delta \log L$. Even if the correction followed only  from the assumed viewing angle dependence of the velocity field where the value of $\kappa$\ is chosen a priori  (Eq. \ref{eq:fs}), we minimized  $\delta \mu$ imposing a cosmological model.   
\item The second one, less impressive but not negligible, is associated with the trace of outflowing line-emitting gas in the \hb\ profile. The ``trace'' is feeble, an order of magnitude below the intensities observed in \civ,  acting especially on the centroids below or at one quarter peak intensity, but is affecting the FWHM measurements in a way that is different from object-to-object, and that, in the presence of selection biases, can affect results for cosmology.  
\end{itemize}

These explorative results will be further analyzed in future works aimed at further reducing the   statistical dispersion of virial luminosity estimates and at  analyzing systematic sources of error \citep[e.g.,][]{caietal18}.

\begin{acknowledgements}
D. Dultzin and A. Negrete acknowledge support form grant IN108716 PAPIIT UNAM and CONACyT project CB-221398. M. L. Mart\'inez-Aldama acknowledges a CONACyT postdoctoral fellowship. A. del Olmo,  M. L. Mart\'inez-Aldama and J. W. Sulentic acknowledge financial support from the Spanish Ministry for Economy and Competitiveness through grants AYA2013-42227-P and AYA2016-76682-C3-1-P. J. 
This research is part of projects 176003 ``Gravitation and the large scale structure of the Universe'' and 176001 ``Astrophysical spectroscopy of extragalactic objects'' supported by the Ministry of Education and Science of the Republic of Serbia.
Funding for the SDSS and SDSS-II has been provided by the Alfred P. Sloan Foundation, the Participating Institutions, the National Science Foundation, the U.S. Department of Energy, the National Aeronautics and Space Administration, the Japanese Monbukagakusho, the Max Planck Society, and the Higher Education Funding Council for England. The SDSS website is http://www.sdss.org. The SDSS is managed by the Astrophysical Research Consortium for the Participating Institutions listed at the http://www.sdss.org.
\end{acknowledgements}

\end{document}